\documentclass[%
reprint,
 amsmath,amssymb,
 aps,
floatfix,
]{revtex4-2}
\usepackage{graphicx}
\usepackage{dcolumn}
\usepackage{bm}
\usepackage{color}
\usepackage{xcolor}
\usepackage{soul}
\usepackage{braket}
\usepackage{hyperref}

\begin{document}
\preprint{APS/123-QED}

\title{First-principles determination of the phonon-point defect scattering and thermal transport due to fission products in ThO$_2$ }
\author{Linu Malakkal$^1$}
 \email[]{linu.malakkal@inl.gov}
\author{Ankita Katre$^2$}
 \email{ankitamkatre@gmail.com}
\author{Shuxiang Zhou$^1$}%
\author{Chao Jiang$^1$}%
\author{David H. Hurley$^3$}%
\author{Chris A. Marianetti$^4$}%
\author{Marat Khafizov$^5$}%
 \email{khafizov.1@osu.edu}
\affiliation{
 $^1$Computational Mechanics and Materials Department, Idaho National Laboratory, Idaho Falls, Idaho 83415, USA
}%
\affiliation{
 $^2$Department of Scientific Computing, Modeling and Simulation, SP Pune University, Pune-411007 India}
  \affiliation{
 $^3$Idaho National Laboratory, Idaho Falls, ID, 83415, USA}
 
 \affiliation{
 $^4$Department of Applied Physics and Applied Mathematics, Columbia University, 500 West 120th Street, New York, New York 10027, United States}

 \affiliation{
 $^5$Department of Mechanical and Aerospace Engineering, The Ohio State University, 201 West 19th Ave, Columbus, Ohio 43210, United States
}
\date{\today}

\begin{abstract}
This work presents the first-principles calculations of the lattice thermal conductivity degradation due to point defects in thorium dioxide using an iterative solution of the Peirels-Boltzmann transport equation. We have used the non-perturbative Green’s function methodology to compute the phonon-point defect scattering rates that consider the local distortion around the point defect, including the mass difference changes, interatomic force constants, and structural relaxation near the point defects. The point defects considered in this work include the vacancy of thorium (V$_{Th}$) and oxygen (V$_{O}$), substitutions of helium (He$_{Th}$), krypton (Kr$_{Th}$), zirconium (Zr$_{Th}$), iodine (I$_{Th}$) and xenon (Xe$_{Th}$) in the thorium site, and the three different configurations of the Schottky defects. The results of the phonon-defect scattering rate reveal that, among all the considered intrinsic defects, the thorium vacancy and helium substitution in the thorium site scatter the phonon most due to the substantial changes in the force constant and structural distortions. The scattering of phonons due to the substitutional defects unveils that the zirconium atom scatters phonons the least, followed by xenon, iodine, krypton, and helium. This is contrary to the intuition that the scattering strength follows He$_{Th}$ $>$ Kr$_{Th}$ $>$ Zr$_{Th}$ $>$ I$_{Th}$ $>$ Xe$_{Th}$ based on the mass difference. This striking difference in the zirconium phonon scattering is due to the local chemical environment changes. Zirconium is an electropositive element with valency similar to thorium and, therefore, can bond with the oxygen atoms, thus creating less force constant variance compared to iodine, an electronegative element, and the noble gases helium, xenon, and krypton. These results can serve as a benchmark for analytical models and help the engineering-scale modeling effort for nuclear design.
\end{abstract}

\maketitle


\section{\label{sec:level1}Introduction \protect}
Thorium dioxide (ThO$_2$) is expected to play a vital role as a material for advanced nuclear fuel cycles in future energy needs and is considered a better and safer alternative to the currently used uranium dioxide \cite{iaea2005}. Since the thermal conductivity of nuclear fuel determines the efficiency and safety of a nuclear reactor, it is imperative to understand the thermal transport of ThO$_2$ in all possible reactor conditions \cite{hurley2022thermal}. It is well-known that point defects are ubiquitous in irradiated materials and can scatter phonons, causing a significant reduction in the material's thermal conductivity, especially at low temperatures or high defect concentrations \cite{DENNETT2021}. Enormous efforts are being undertaken to accurately model and ultimately control the thermal transport in nuclear fuels. A detailed description of the new computational and experimental tools for the same is detailed in a recent review article \cite{hurley2022thermal}. However, a fundamental understanding of how different fission products scatter the phonons in nuclear fuels is still lacking. Therefore, an accurate prediction of the phonon-point defect scattering rate by first-principles calculations is imperative to shed light not only on the fundamental physics of phonon transport in irradiated fuels but also in modeling the thermal transport of ThO$_2$ during regular operation and accident conditions in a nuclear reactor.

The scattering of phonons by point defects is described as a perturbation to harmonic lattice dynamics. The responsible mechanisms include the mass difference, harmonic force constant changes, and strain field surrounding the defect, often captured by ionic  radius difference \cite{ramya2020,Deskins_2021}. So far, the widely used description of the phonon-point defect scattering is based on the seminal work by Klemens \cite{Klemens_1955}. In this model, Klemens derived the analytical expression for the phonon elastic scattering cross section by point defects, dislocations, and the grain boundary using second-order perturbation theory. However, the Klemens model is restricted to small perturbations due to mass variance, and the equations are derived assuming a single atom unit cell and linear phonon dispersion \cite{Asheghi_2002,Zou_2002,Fletcher_1987,Malekpour_2016}. Another popular mass difference model used to describe phonon scattering by point defects utilizing realistic phonon dispersion was implemented by Tamura \cite{Tamura_1983}. The main difference between the Klemens and Tamura models is that the mass difference term in the Tamura model is weighted by the eigenvector components corresponding to the atom in the incident and final vibrational modes. The Born approximation is another standard method that treats the defect using lowest-order perturbation theory \cite{sakurai1967advanced}. These models become questionable for point defects with large force constant variance and structural changes \cite{Polanco_InN} (i.e., for conditions beyond the low-order perturbation regime). Hence, defining the phonon-defect scattering by a nonperturbative method that includes the scattering due to large variance in force constants becomes vital. Recently, an exact calculation of the phonon-defect scattering rate using Green’s function approach, which is nonperturbative and includes the interatomic force constants (IFCs), has been proposed \cite{Mingo_2010,Kundu_2011}. To date, this method has been successfully used to describe the phonon scattering by point defects in various materials, such as diamond \cite{Katcho_2014}, boron arsenide \cite{protik_2016}, silicon carbide \cite{Ankita_2017}, gallium nitride \cite{Ankita_2018}, graphene \cite{Polanco_g}, and indium nitride \cite{Polanco_InN}. These calculations have already identified shortcomings of the simplified perturbative methods. They have been instrumental in providing fundamental insights into the phonon-defect scattering of the respective materials.

 In ThO$_2$, where there is a colossal mass difference between the constituent atoms and the possibility of sizeable interatomic force constant differences with point defects, it becomes essential to describe the phonon scattering by point defects with Green’s function methodology. To our knowledge, Green’s-function-based methods have not been employed to study thermal transport in any actinide materials. In the literature, a few first-principles investigations of the thermal conductivity by iteratively solving the Pierels-Boltzmann transport equation of the defect-free ThO$_2$ are available \cite{LIU201811, MALAKKAL2019507, Jin_2021, Enda_2022_prb}. In 2018, Liu \textit{et al.} \cite{LIU201811}, using the generalized gradient approximation (GGA) pseudopotential, computed the second-order force constant from the finite difference method and predicted the thermal conductivity of ThO$_2$ at 300 K to be 12.4 W/mK, which was underpredicted by $\approx$40\% in comparison with experimental results (20.4 W/mK at 300 K) for a single crystal \cite{Mann_2010}. The following year, using the GGA pseudopotential, Malakkal \textit{et al.} \cite{MALAKKAL2019507} reported a thermal conductivity of 15.4 W/mK at 300 K (underpredicted by $\approx$25\%) by computing the second-order force constant using the density functional perturbation theory. Using the local density approximation (LDA) pseudopotential, Jin \textit{et al.} \cite{Jin_2021} predicted a thermal conductivity of 18.6 W/mK (underpredicted by $\approx$9\%) at room temperature. Recently, Xiao \textit{et al.} \cite{Enda_2022_prb} calculated the phonon lifetimes and thermal conductivity of ThO$_2$ with the strongly constrained and approximately normed functional \cite{Sun_2015_SCAN} with excellent agreement to experiments, indicating that, over the years, researchers were able to provide a good description of the thermal transport in the pristine ThO$_2$ using the first-principles simulations by refining the accuracy of calculated interatomic forces. 
 
In practical application as nuclear fuel,a variety of lattice defects are formed including the point defects, small defect clusters and other extended defects such as dislocation loops, inert gas bubbles, grain subdivision induced grain boundaries, and defect segregation at grain boundaries, which all scatters the phonons and hinders the thermal transport \cite{hurley2022thermal}.Until recently, the impact of many of these defects on thermal transport have been treated empirically. To predict temperature distribution within the fuel elements, it is imperative to have a fundamental understanding of the role played by each of these defect types on thermal conductivity\cite{FERRIGNO_2023}.However, to isolate the impact of different defects on thermal transport, it is more practical to consider the low temperature thermal conductivity. At high temperature intrinsic three-phonon scattering dominates reduction of thermal conductivity. At low cryogenic temperatures, the three-phonon scattering is weaker and defects scattering become dominant. As a step towards systematically understanding the influence of various defects on the thermal transport, we begin by focusing on the role played by the irradiation-induced small-scale point defects because at the early stages of damage accumulation these small-scale lattice defects significantly impact the thermal conductivity \cite{Khafizov_2016_investigation,Dennett_2020}. Although the main irradiation-induced point defects in nuclear fuels includes vacancies, interstitials, fission product acting as substitutions, Frenkel pairs and the Schottky defects, this work focus on the impact of vacancies, substitution of fission products and three different configurations of the Schottky defects. The role of interstitials and Frenkel pairs requires additional code development and will be the subject of future work.

Multiple researchers have already applied various computational techniques to understand the effect of point defects in the thermal transport of ThO$_2$. For instance, Cooper \textit{et al.} \cite{COOPER201529} reported the degradation of thermal conductivity due to the nonuniform cation lattice of (U$_x$Th$_{1-x}$)O$_2$ solid solutions, using the non-equilibrium molecular dynamics method (NEMD), from 300 to 2000 K. Park \textit{et al.} \cite{PARK2018198} used reverse non-equilibrium molecular dynamics (r-NEMD) to investigate the effect of vacancy and uranium substitutional defects in the thermal transport of ThO$_2$. Rahman \textit{et al.} \cite{RAHMAN2020152050} studied the dependence of thermal conductivity on fission-generated products (xenon and krypton) and vacancies in ThO$_2$ within 300-1500 K using NEMD simulations. Recently, Jin \textit{et al.} \cite{JIN2022Impact}, using NEMD simulations, reported the phonon-defect scattering cross section for a range of defects to be used with reduced order analytical models. In all these studies, the quantum effects are ignored. The phonons follow the Boltzmann distribution instead of the physically correct Bose-Einstein distribution \cite{Zhou_2018} and, therefore, are restricted to temperatures above the Debye temperature. But phonon-point defect scattering is more dominant at low temperatures; thus, molecular dynamics (MD) based simulations are not ideal for describing phonon-point defect scattering. The other challenge for MD-based simulation is that the results depend on the accuracy of the interatomic potential. The work by Jin \textit{et al.} \cite{Jin_2021} suggests that the empirical potential used previously needs to be further optimized for the robust prediction of thermal conductivity of ThO$_2$, both in perfect crystals and in the presence of complex defects. Apart from MD simulations, Ryan \textit{et al.} \cite{Deskins_2021} studied the effect of point defects on the thermal conductivity of ThO$_2$ using the Klemens model for phonon relaxation times that result from the change in mass and induced lattice strain associated with point defects. This model computes the difference in force constants and atomic radii using MD simulation in ThO$_2$. Again, this method is also limited to small perturbations and does not consider the complete scattering to all orders.

Therefore, for the first time, the phonon scattering due to point defects and their impact on the thermal conductivity of ThO$_2$ is described by an \textit {ab initio} calculation utilizing the Green-function T-matrix method. In the current study, we focus on the thermal conductivity degradation in ThO$_2$ due to a selected set of point defects, including the substitutions of transmutation and radioactive decay product helium (He$_{Th}$) and fission products, such as  krypton (Kr$_{Th}$), zirconium (Zr$_{Th}$), iodine (I$_{Th}$), and xenon (Xe$_{Th}$) in thorium sites, along with other point defects, such as the single vacancy of thorium (V$_{Th}$) and oxygen (V$_{O}$) and three configurations of Schottky defects (SD$_{100}$, SD$_{110}$, SD$_{111}$).

\section{Methodology}
The lattice thermal conductivity (\textit{$k_{L}$}) of ThO$_{2}$ with dilute point defects sites is calculated by summing over contributions from all phonons as
\begin{eqnarray}
{k_L={\frac{1}{3\Omega}\sum_{j\textbf{q}} c_{j\textbf{q}} v^2_{j\textbf{q}}\tau_{j\textbf{q}}},}
\end{eqnarray}

\noindent where $\Omega$ is unit cell volume, $c_{j\textbf{q}}$, $v_{j\textbf{q}}$,and $\tau_{j\textbf{q}}$ are the specific heat capacity, group velocity, and relaxation time for the phonon in mode $j\textbf{q}$, respectively, and $j$ and $\textbf{q}$ represent the phonon branch index and wave vector, respectively. The inverse of the phonon relaxation time $\tau^{-1}_{j\textbf{q}}$ is called the phonon scattering rate, and the total phonon scattering rate can be written as the sum of the contribution from different scattering mechanisms; in this work, we have considered contributions from three major scattering phenomena, including the intrinsic three-phonon scattering ($\tau^{-1}_{j\textbf{q},anh}$) \cite{Ziman_2001}, phonon-isotope scattering ($\tau^{-1}_{j\textbf{q},iso}$), and phonon-defect scattering ($\tau^{-1}_{j\textbf{q},def}$) \cite{Mingo_2010}: 
\begin{eqnarray}
{\tau^{-1}_{j\textbf{q}} = \tau^{-1}_{j\textbf{q},anh}+\tau^{-1}_{j\textbf{q},iso}+\tau^{-1}_{j\textbf{q},def}.}
\end{eqnarray}
 For brevity, the equation governing the scattering rate for the intrinsic anharmonic scattering and the isotopic scattering rate is not provided here and can be found in Ref. \cite{Ziman_2001} and \cite{Tamura_1983}, respectively. However, we provide a detailed discussion on the phonon-defect scattering. The phonon-defect scattering is computed using Green’s function methodology using \textbf{T} matrix \cite{Mingo_2010}.The advantages of using this approach is that, firstly, it captures the scattering due to variance in both mass and force constants, with the latter incorporating structural relaxation effects; secondly, it also accounts for complete scattering to all orders to capture the effect of resonances that would otherwise be missing in the lower order perturbative methods. This approach is valid under the assumption of a random and dilute concentration of defects, which means the phonon scatters independently from different defects at a different location. The scattering cross-section $\sigma$, an essential parameter in the phonon-defect scattering using Green’s function approach, is explained in detail in Ref. \cite{Kundu_2011} and in the supplementary materials of the Ref. \cite{Ankita_2017}. The scattering rate due to all possible elastic scattering processes of phonons by a random and dilute distribution is given by Equation (3):
\begin{align}
{\tau^{-1}_{j\textbf{q},def}={\pi\chi_{def}\frac{\Omega}{V_{def}}}{\frac{1}{{\omega_{j\textbf{\textbf{q}}}}}}\sum_{j^{'}\textbf{q}^{'}}{{|\bra{j^{'}\textbf{q}^{'}}\textbf{T}\ket{j\textbf{q}}|}^2}}{\delta(\omega^{2}_{j^{'}\textbf{q}^{'}}{-}\omega^{2}_{j\textbf{q}}),} \notag \\
  & \label{eq:1}
\end{align}
where $\chi_{def}$ is the number fraction of defects, $V_{def}$ is the volume of a defected supercell, $\ket{j\textbf{q}}$ and $\ket{j^{'}\textbf{q}^{'}}$ represent the incident and scattered phonon eigenstates, respectively, $\omega$ is the phonon angular frequency, and $\textbf{T}$ is the matrix associated with the point defect scatterer. 
The $\textbf{T}$ matrix is defined in terms of the perturbation matrix ${(\textbf{V})}$ and the retarded Green's function \cite{Kundu_2011} for the perfect structure ${(\textbf{g}^+)}$ as: 
\begin{eqnarray}
{\textbf{T}=(\textbf{I}-\textbf{V}\textbf{g}^+)^{-1}\textbf{V},}
\end{eqnarray}
where ${(\textbf{I})}$ is the identity matrix. Here, the perturbation matrix includes the mass ${(\textbf{V}_{M})}$ and harmonic force constant ${(\textbf{V}_{K})}$ differences between the perfect and defective systems, as shown below: 
\begin{eqnarray}
{\textbf{V}=(\textbf{V}_{M}+\textbf{V}_{K}).}
\end{eqnarray}

The mass term is a diagonal matrix, with nonzero matrix elements corresponding to defect site given as {${{\textbf{V}_{M,i,i}}}$=$-{(\frac{{\text{M}^{'}_{i}}-\text{M}_{i}}{\text{M}_{i}})\omega^{2}}$},
where $\text{M}^{'}_{i}$ is the mass of the defect atom, $\text{M}_{i}$ is the mass of the original atom at the $i^{th}$ site, and $\omega$ is the angular frequency of the incoming phonon \cite{Ankita_2017}. Similarly, the presence of a point defect will also change the harmonic force constant around the defect, and this is included in the $\textbf{V}_{\text{K}}$ matrix, which is computed from the difference between the IFCs for the defective $\text{K}^{'}$ and perfect {K} structures as: 
\begin{eqnarray}
{\textbf{V}_{K,i\alpha,k\beta}={\frac{{K^{'}_{i\alpha,k\beta}}-K_{i\alpha,k\beta}}{\sqrt{M_{i}M_{k}}}}.}
\end{eqnarray}
Here ${i}$ and ${k}$ are atom indices and ${\alpha}$ and ${\beta}$ are the cartesian axes. Born approximation is achieved by Taylor series expansion of equation (4) and approximation $\textbf{T} \approx \textbf{V}$.

It is worth noting that ${\textbf{V}_{M}}$ is proportional to the square of phonon frequency, and hence is smaller near the zero frequency, whereas ${\textbf{V}_{K}}$ is independent of the phonon frequency, and therefore, a significant change in the harmonic force constants cannot be considered small at any frequency. Moreover, the IFCs are calculated at zero Kelvin, hence, the dependence of temperature on the scattering rate is not included. But as the phonon defect scattering rates are frequency dependent, and the phonon population are both frequency and temperature dependent, the scattering rate indirectly influences the thermal conductivity as a function of temperature \cite{Polanco_InN}.

In this approach, the perturbation matrix is defined in real space, and to limit the size of the matrix, a cutoff radius ${r_{cut}}$ needs to be enforced for the structural distortion. A large ${r_{cut}}$ corresponding to the fifth neighbour shell in ThO$_{2}$ is chosen for all the defects. Past this cutoff the IFC changes are found to be negligible. Within ${r_{cut}}$, the changes in IFCs up to the second nearest neighbors are considered. ${\textbf{V}_{K}}$ must obey the translational invariance rule, and when computed numerically, this symmetry must be enforced. This can be done by adding a minor correction to the harmonic IFCs of the perturbed system following a procedure outlined in \cite{Ankita_2017}. 

We simplified the calculation represented by equation (3) with the use of the optical theorem. The total elastic scattering of phonons by defects can be efficiently computed from the imaginary part of the diagonal element of the \textbf{T}-matrix, given by: 
\begin{eqnarray}
{\tau^{-1}_{j\textbf{q},def}={{-}\chi_{def}\frac{\Omega}{V_{def}}}{\frac{1}{{\omega_{j\textbf{q}}}}}{\text{Im}\{{{\bra{j\textbf{q}}\textbf{T}\ket{j\textbf{q}}}\}}}.}
\end{eqnarray}
When using the optical theorem, the Born approximation requires inclusion of higher order terms in the Taylor series expansion $\textbf{T} \approx \textbf{V}+\textbf{V}\textbf{g}^+\textbf{V}$. 

\section{Simulation details}
All density functional theory \cite{Kohn_1965} calculations in this work were performed using the projector-augmented-wave \cite{blochl_1994} method implemented in the plane wave software package, Vienna Ab initio Simulation Package (VASP) \cite{Kresse_1996}, with the LDA \cite{Perdew_1981} pseudopotential for the exchange and correlation. Also, it must be noted that no Hubbard U correction \cite{Dudarev_1988_HubbardU} were included in this work because it was previously shown that the Hubbard U correction offered no improvement to the property predictions of ThO$_2$ \cite{SHIELDS201699,Lu_2012}. Moreover, ThO$_{2}$ is specifically used as a model material to UO$_{2}$ to avoid implementation of this correction. We performed the geometry optimization of ThO$_{2}$ (space group $Fm\bar{3}m$) on a primitive unit cell by minimizing the total energy with respect to changes in both cell parameters and atom positions using the conjugate gradient method. We obtained the energy convergence of ThO$_{2}$, using an electron wave vector grid and plane wave energy cutoff of a 12 × 12 × 12 mesh and 550 eV, respectively. The criteria for the electronic energy convergence was set at $10^{-8}$ eV. The lattice parameter for the obtained relaxed structure of ThO$_{2}$ at 0 K is 5.529 Å, which is consistent with the previously reported value \cite{Jin_2021} for the LDA functional from VASP and is comparable with the experimental lattice constant value of 5.60 Å \cite{idiri_2004}. However, it is worth noting that the GGA functional predicted the lattice constant better than the LDA, however, the thermal conductivity predicted by the GGA functional was found to significantly lower. Hence the functional based on the LDA was used in this work. A 5 × 5 × 5 supercell with 375 atoms and kpoints of 2 × 2 × 2 were used to evaluate the harmonic force constants using the finite displacement method as implemented in PHONOPY \cite{TOGO2015} and the calculated phonon dispersion spectra was compared with the experiment done by Bryan et al. \cite{Bryan2020}, as shown in supplementary information (SI) [xx]. The third-order force constants (anharmonic force constants) were also calculated using a  5 × 5 × 5 supercell at the gamma point using Thirdorder.py \cite{shengBTE_2014}, and the force cutoff distance was set to the fifth nearest neighboring atoms. In addition, because ThO$_{2}$ is a polar material, the non-analytical contribution was considered, and the Born charges and dielectric constant required to evaluate the non-analytical correction \cite{Wang2016} were calculated using density functional perturbation theory \cite{Baroni_2001_Phonons}. The phonon-phonon scattering processes are evaluated using Fermi's golden rule from the cubic force constants. Finally, \textit{$k_{L}$} was calculated using the iterative solutions of the BTE as implemented in 
AlmaBTE \cite{CARRETE2017}. All the \textit{$k_{L}$} presented in this work are the fully iterative solutions of the Peirels-Boltzmann equation. The converged \textit{$k_{L}$} values were obtained using a cubic force constant, considering the fifth nearest-neighbor interaction. The number of grid planes along each axis in the reciprocal space for solving the BTE was 24 × 24 × 24.

All point defects considered in this work were created in a 5 × 5 × 5 supercell. For vacancy defects, we removed a Th or O atom from their lattice position. We built the substitutional defects by replacing a Th atom by the corresponding fission products. The Schottky defects were formed by removing a unit of ThO$_{2}$ in the three-bound configuration of the SD, the SD$_{100}$, SD$_{110}$, and SD$_{111}$ \cite{Misako_2009,Dorado2009}. Each defected supercell was relaxed by keeping the cell volume constant. For the substitutional defects, a slight displacement of the defect atom or its nearest neighbors is introduced before relaxation to avoid the system getting trapped in a saddle point of the potential energy surface. We confirmed the dynamic stability of the relaxed structures by ensuring that the phonon band structure did not have any modes with imaginary frequency. The IFCs of the system with point defects were calculated using the finite-displacement method. The space group symmetry of the supercell containing the point defect of V$_{Th}$ and V$_{O}$ was face centered cubic, while the space group symmetry of the supercell containing the substitutional defects of He$_{Th}$, Kr$_{Th}$, Zr$_{Th}$, I$_{Th}$, and Xe$_{Th}$ were orthorhombic, and the space group symmetry of the supercell with the Schottky defects were orthorhombic for SD$_{100}$ and SD$_{110}$ and trigonal for SD$_{111}$. Space group symmetries were determined using spglib as implemented in the PHONOPY code. Finally, the phonon-defect scattering rates were calculated on a uniformly spaced 24 × 24 × 24 q-point mesh with an 18 × 18 × 18 grid for the Green’s function, using the tetrahedron method to integrate over the Brillouin zone. Finally, to understand the charge transfer of the substituted defects to the neighbouring atoms, the charge density difference plot was obtained using the VESTA software package \cite{Momma:ko5060} using the expression: ${\Delta\rho}_{\text{diff}}$ = ${\rho[\text{ThO}_{2}}+\text{defect}]$ - ${\rho[\text{ThO}_{2}]}$ - ${\rho[\text{defect}]}$ where ${\rho[\text{ThO}_{2}}+\text{defect}]$, ${\rho[\text{ThO}_{2}]}$, and ${\rho[\text{defect}]}$ are the charge density of ThO$_2$ with point defect, the pristine ThO$_2$, and isolated atom corresponding to that particular defect, respectively.

\section{Results and Discussion}
\begin{figure}[h]
    \centering
    \includegraphics [width=0.95\columnwidth]{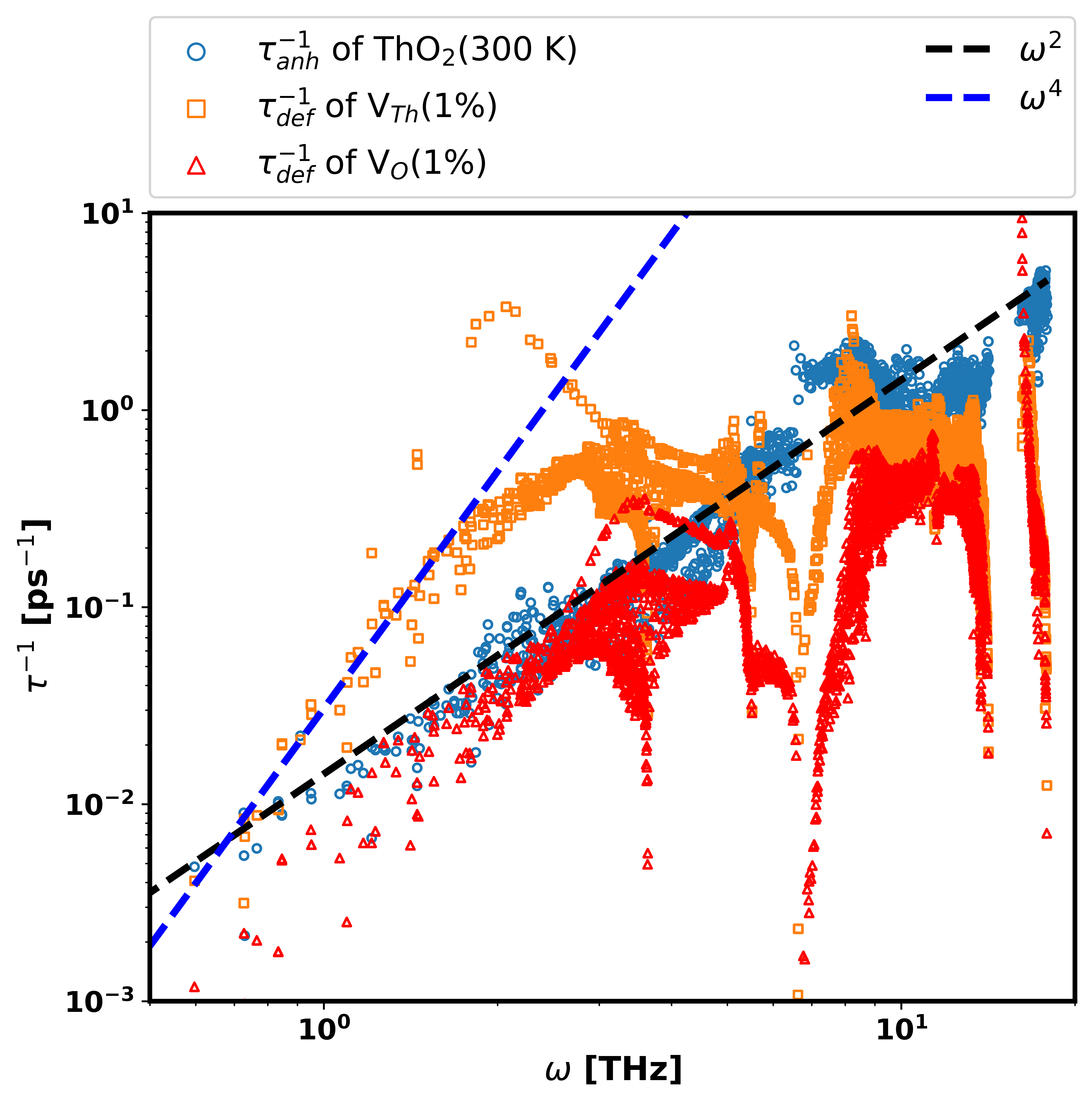}
    \caption{Comparison of $\tau^{-1}_{def}$ of the V$_{Th}$ (orange square) and V$_{O}$ (red triangles) as a function of phonon frequency $(\omega)$ from the Green-function T-matrix method compared to $\tau^{-1}$ predicted by Klemens model (anharmonic scattering [black dashed lines] and defect scattering [blue dashed lines]). Hollow blue dots represent the anharmonic scattering ($\tau^{-1}_{anh}$) of the pristine ThO$_{2}$ due to three phonon interactions at 300 K.}
    \label{Fig1.png}
\end{figure}

We plotted the scattering rates ($\tau^{-1}$) of each point defect type at a 1{\%} concentration as a function of phonon frequency $(\omega)$ to provide insights into phonon scattering by various defects in ThO$_{2}$. Fig.~\ref{Fig1.png} compares the $\tau^{-1}$  due to vacancy point defects to the anharmonic scattering rate of pristine ThO$_{2}$ at 300 K. The analysis of the phonon-defect scattering rate suggests that a thorium vacancy scatters the phonons more strongly than an oxygen vacancy. In vacancy defects, where the host atom is absent, the contribution to $\tau^{-1}$ comes from the IFC differences. To understand the extent of the force constant changes due to single vacancies of thorium or oxygen, we computed the Frobenius norm ratio of the IFC matrix. The values of the Frobenius norm ratio of the IFC matrix for the thorium vacancy (0.17) and oxygen vacancy (0.05) defects suggest that a thorium vacancy causes significantly more changes to the force constants than the oxygen vacancy, which explains why the thorium vacancy scatters the phonons more than the oxygen vacancy. 
Additionally, Fig.~\ref{Fig1.png} also compares the frequency trends suggested by Klemens' model \cite{Klemens_1955} for anharmonic scattering and point defects. According to Klemens' model, the anharmonic scattering rate follows an $\omega^2$ dependence (black dashed lines) and the defect scattering has an $\omega^4$ dependence (blue dashed lines). The latter is generally referred to as Rayleigh scattering. Klemens' model captures the anharmonic scattering in ThO$_2$ in agreement with the \textit {ab initio} predictions. In contrast, Klemens' expression for point defect scattering, while correctly capturing the low frequency trends is unable to reproduce the high frequency phonons, especially the optical modes which appear to have much weaker dependence on frequency. One can argue that this is due to an oversimplified description of phonon density of states employed in the Klemens expression not valid for high frequency modes. Nevertheless, the Tamura model, which uses exact phonon dispersions is unable to capture high frequency trends either \cite{Tamura_1983}. In fact the Tamura model makes the situation even worse. As the density of phonons increases at high frequency, it predicts scattering rates that are even larger than one expects from Klemens' expressions \cite{ramya2020,Deskins_2021}. The high frequency overestimates of the Klemens' and Tamura models is in fact due to their consideration of only the mass term which has strong frequency dependence, whereas the perturbation due to IFC is frequency independent, suggesting that IFC term dominates at high frequencies. 
\begin{figure}
    \centering
    \includegraphics [width=0.95\columnwidth]{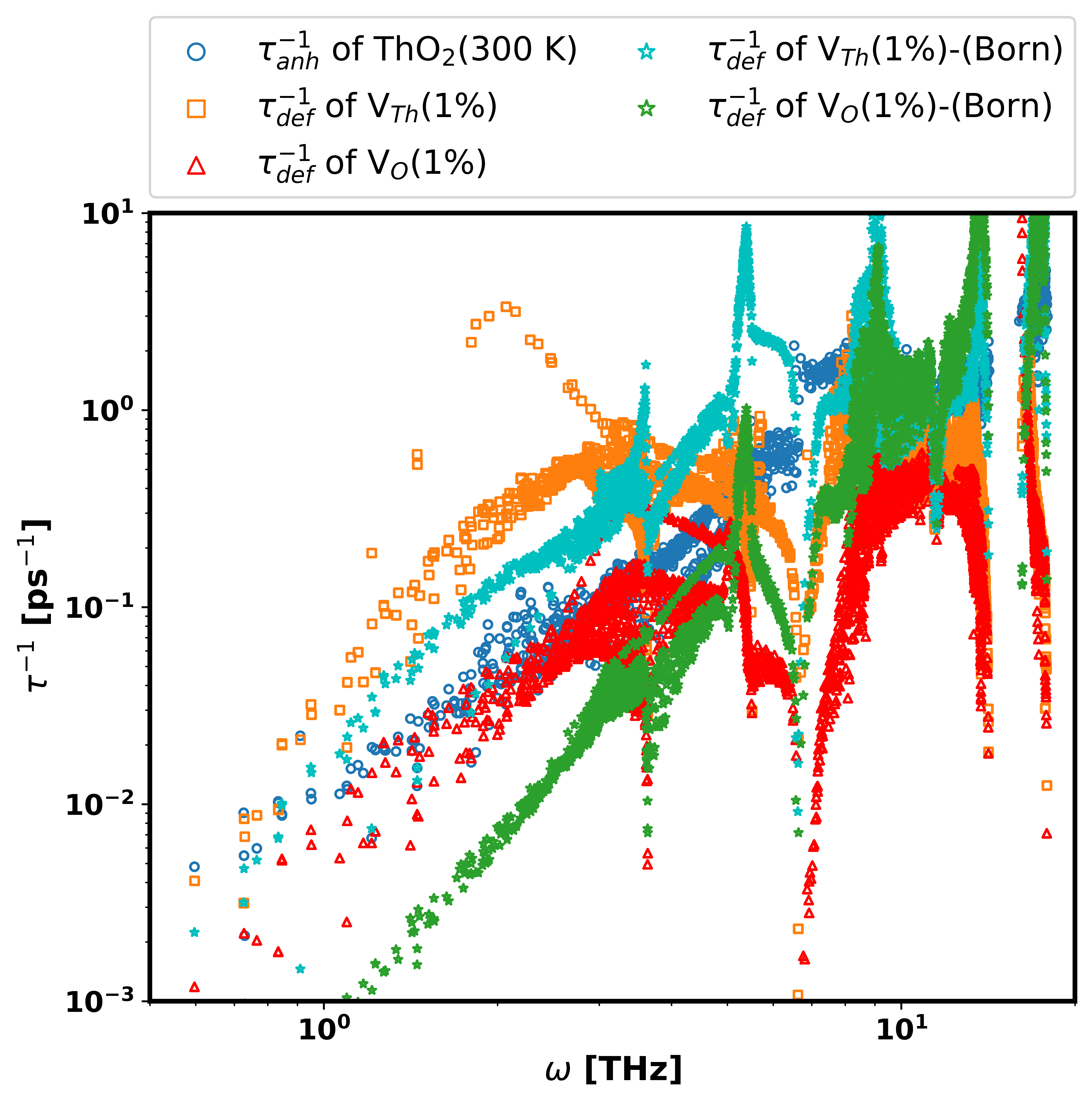} \caption{Comparison of $\tau^{-1}_{def}$ of the V$_{Th}$ (orange square) and V$_{O}$ (red triangles) as a function of phonon frequency $(\omega)$ from the Green-function T-matrix method compared to the Born approximation for V$_{Th}$ (cyan star) and V$_{O}$ (green star). Hollow blue dots represent the anharmonic scattering ($\tau^{-1}_{anh}$) of pristine ThO$_{2}$ due to three phonon interactions at 300 K.}
    \label{Fig2.png}
\end{figure}
Compared to the Green-function T-matrix methods, our calculations show that single-atom vacancy scattering centers are not well described by the Born approximation (Fig.~\ref{Fig2.png}), with a significant underestimation of $\tau^{-1}$  at low phonon frequencies and overestimation at high phonon frequencies for both thorium and oxygen vacancies. It is worth noting that there are prominent peaks observed at 2 Thz for thorium vacancies and 3 THz for oxygen vacancies. Further modal analysis reveals that these peaks are associated with longitudinal acoustic (LA) waves and appear at the location where the transverse acoustic  (TA) mode has large density of states. Under the Born approximation the scattering of LA modes to TA is not possible as the momentum is not conserved. Previously, such behavior in silicon carbide and gallium arsenide has been attributed to resonant scattering \cite{Ankita_2017,kundu2019}. The Green's function approach considers full scattering to all orders, thus capturing the effect that leads to enhanced phonon scattering due to emergence of localized models. 
Apart from vacancy defects, neutron irradiation will introduce substitutional defects. To discern the phonon scattering due to the substitutional atoms in ThO$_{2}$, we considered some common fission products such as Kr, Zr, I, and Xe as well as He in the thorium sites. Our reason for only considering thorium sites and not oxygen sites is that the Th atom’s vibration governs the acoustic modes and contributes $\approx$70{\%} of the thermal conductivity \cite{MALAKKAL2019507}. Hence, defects on the Th sites, particularly those that have a large mass or IFC variance, can degrade the thermal conductivity more significantly than those of the light O atoms. Unlike vacancy defects, the scattering from substitutional atoms arises from the mass difference between the substitutional and host atom and the IFCs variations around the substituted site due to the structural relaxation and modified chemical environment. 

\begin{figure}[h]
   \centering
   \includegraphics [width=0.95\columnwidth]{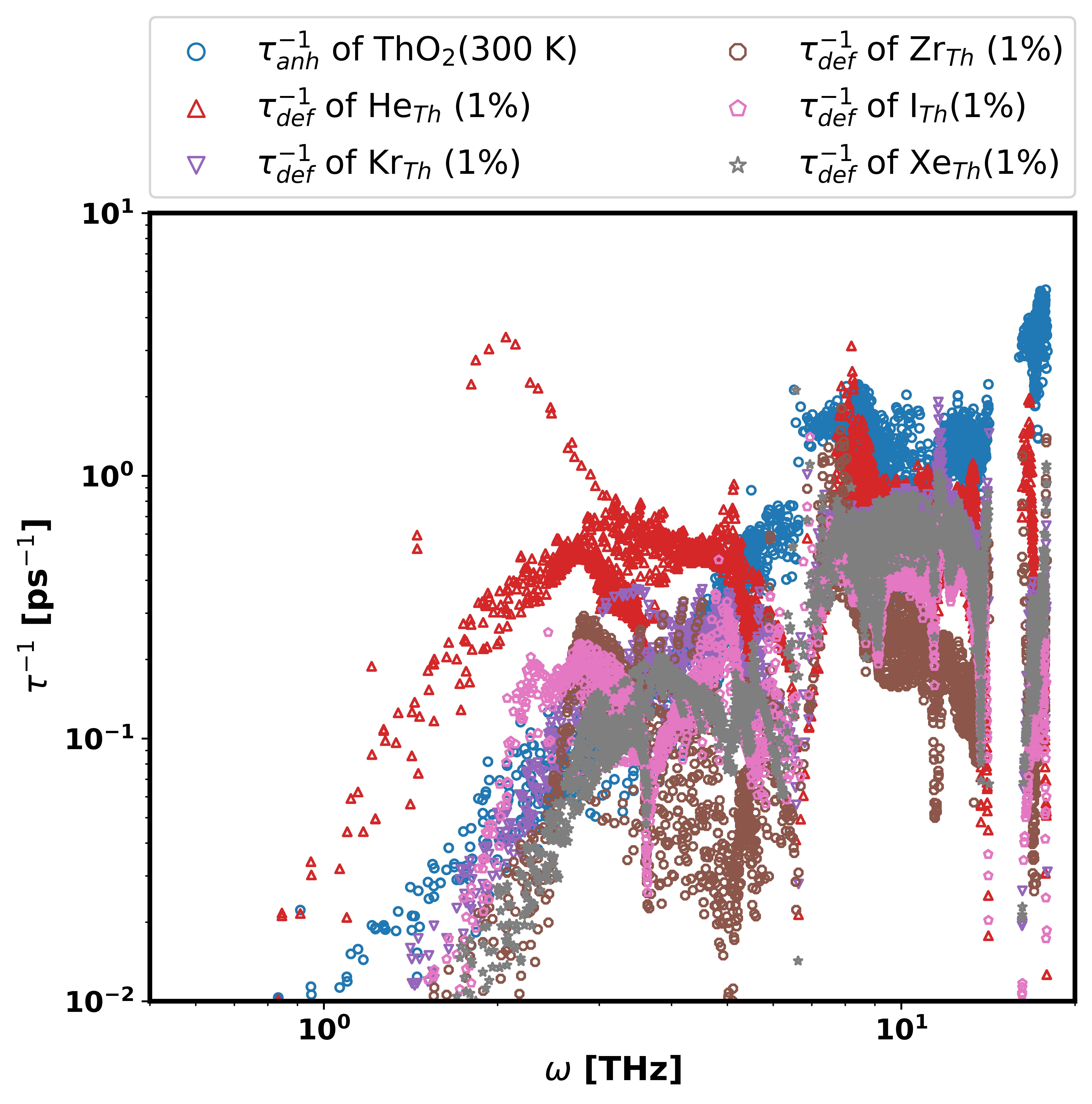}
   \caption{Scattering rates ($\tau^{-1}_{def}$) of phonons as a function of phonon frequency $(\omega)$ due to substitutional point defects of He$_{Th}$ (up-pointing crimson red triangle), Kr$_{Th}$ (down-pointing purple triangle), Zr$_{Th}$ (brown octagon), I$_{Th}$ (pink pentagon), and Xe$_{Th}$ (gray star) in thorium sites compared to $\tau^{-1}_{anh}$ of the pristine ThO$_{2}$ due to three phonon interactions at 300 K (blue dots).}
   \label{Fig3.png} 
\end{figure}

The previous study using same methodology has shown that, in general, substitutional atoms scatter phonons less effectively than vacancies \cite{Polanco_InN}. Our ThO$_2$ results as shown in Fig.~\ref{Fig3.png} indicate that a substitution by a significantly lighter He atom results in phonon scattering with a strength comparable to thorium vacancy, to the extent that they overlap. The primary reason for He substitution to strongly scatter phonons is because He as a light chemically inert atom induces a force constant variance very similar to that of the thorium vacancy(SI: Table 1 [xx]). As expected, a similar resonant peak as seen in a thorium vacancy is also observed for a helium substitution at the same frequency. 

Among the other substitutional atoms, intuitively based on mass differences, one expects the scattering strength to be He$_{Th}$ $>$ Kr$_{Th}$ $>$ Zr$_{Th}$ $>$ I$_{Th}$ $>$ Xe$_{Th}$. However, contrary to this intuition, the predicted scattering strengths of the fission products were He$_{Th}$ $>$ Kr$_{Th}$ $>$  I$_{Th}$ $>$ Xe$_{Th}$  $>$ Zr$_{Th}$, where the Zr$_{Th}$ substitution scattered phonons the least. To understand why the Zr$_{Th}$ substitution showed minimal phonon scattering, we computed the Frobenius norm ratio of the IFC matrix (SI: Table 1) and found that the force constant perturbations were the lowest when Zr was in thorium sites. This can be attributed to the fact that, unlike noble gas elements (He, Xe and Kr) and electronegative elements (I),  Zr is an electropositive element, and Zr in a cation lattice site can donate electrons to the oxygen and form bonds. The charge transfer in the substitutional defects was quantified (SI: Fig. 3 [xx]) to provide insights into the redistribution of charge and bond strength. In the case of a Zr substitution, the charge transfer takes place from the Zr atom out towards the surrounding O atoms. Also, the charge accumulated around the Zr atom is negligible since Th and Zr have similar ionic charges. This observation clearly indicates that local chemical modification can significantly impact force constant perturbations and, thus, phonon scattering.

\begin{figure}[h]
    \centering
    \includegraphics [width=0.95\columnwidth]{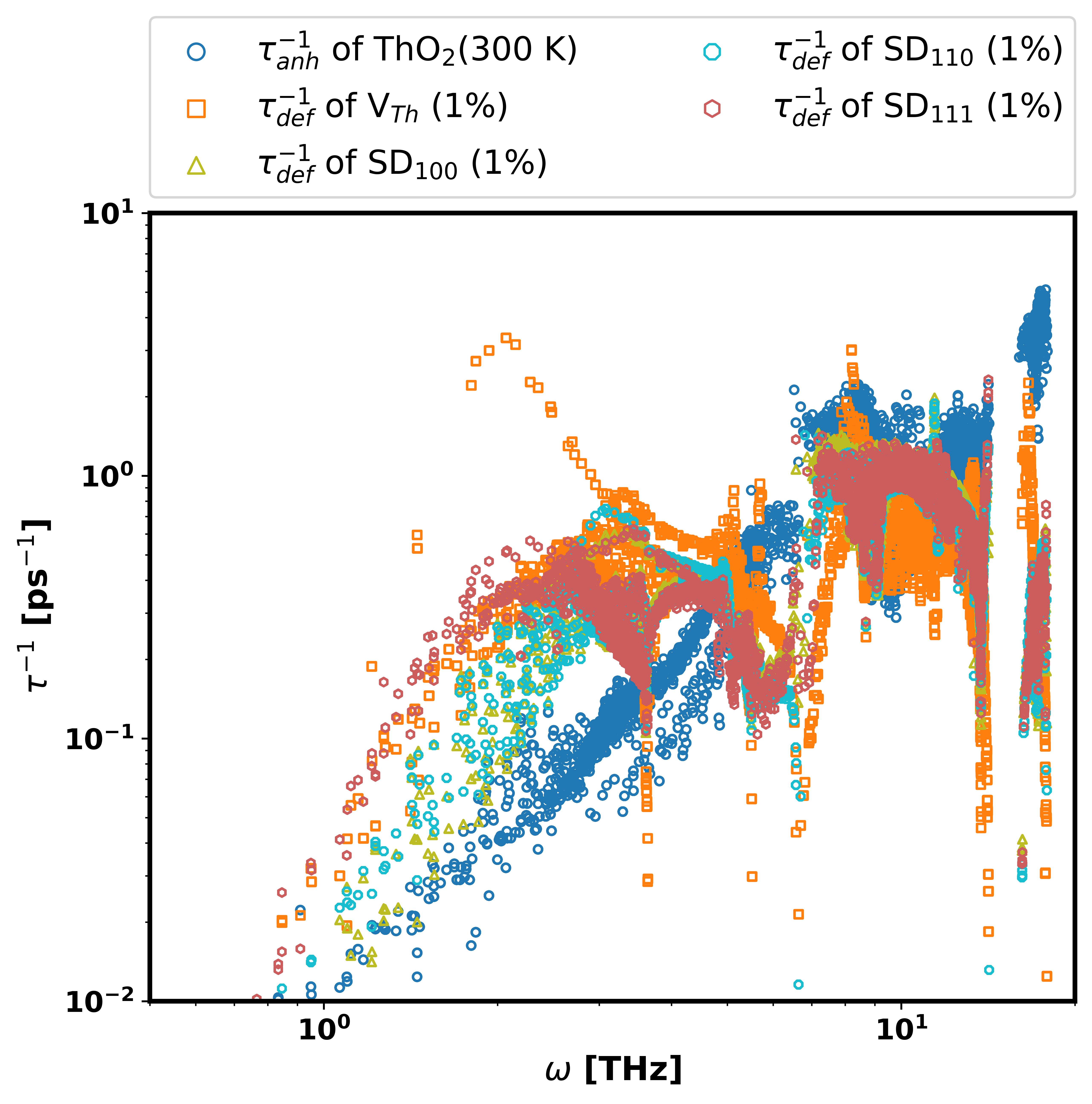}
    \caption{Scattering rates ($\tau^{-1}_{def}$) of phonons by three different configurations of the Schottky defects (SD$_{100}$ [up-pointing green triangle], SD$_{110}$ [blue octagon], SD$_{111}$ [red pentagon]) in ThO$_{2}$ compared to the $\tau^{-1}_{def}$ of the V$_{Th}$ and the $\tau^{-1}_{anh}$ of the pristine ThO$_{2}$ due to three phonon interactions  at 300 K.}
    \label{Fig4.png}
\end{figure}

Finally as an example of larger defects, we investigated the role of Schottky defects in ThO$_{2}$ in scattering the phonons, as depicted in Fig.~\ref{Fig4.png}. Among the three different Schottky defect configurations, the scattering strengths are SD$_{111}$ $>$ SD$_{110}$ $>$ SD$_{100}$, which indicates that ThO$_{2}$ with an SD$_{111}$ defect scatters phonons the most.

\begin{figure}[h]
   \centering
   \includegraphics [width=0.95\columnwidth]{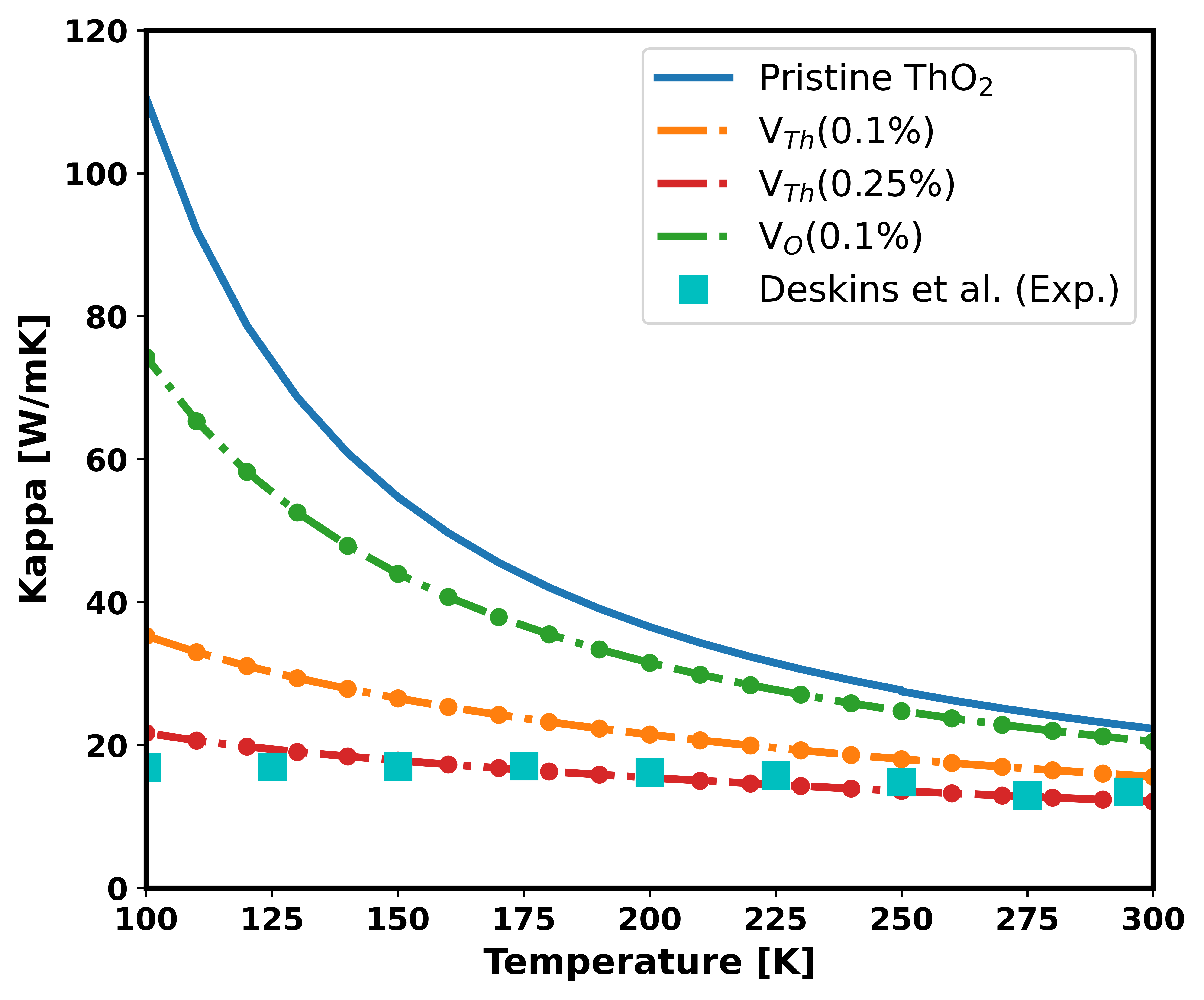}
   \caption{The calculated \textit{$k_{L}$} of ThO$_{2}$ with vacancy defects (V$_{Th}$ (0.1$\%$ and 0.25$\%$) , and V$_{O}$ (0.1$\%$)) at a lower temperature range of 100--300 K. The experiments are from Deskins \textit{et al.} \cite{DESKINS2022} corresponding to ThO$_{2}$ single crystals irradiated with 2 MeV protons to a dose of 0.004 displacements per atom at room temperature.}
   \label{Fig5.png} 
\end{figure}

Using the calculated scattering rates, including the anharmonic, isotropic, and phonon-defect scattering, the {$k_{L}$} of ThO$_{2}$ was computed using an iterative solution of the BTE. Fig.~\ref{Fig5.png} shows the \textit{$k_{L}$} of pristine ThO$_{2}$ and ThO$_2$ with vacancy defects as a function of temperature. Thorium vacancies caused the most significant reduction in thermal conductivity, consistent with the observation made with the phonon-point defect scattering rate (Fig.~\ref{Fig1.png}). Oxygen vacancies at 0.001 concentration caused a marginal decrease in thermal conductivity. The theoretical results are compared with the recent experimental measurement done on a proton-irradiated single crystal
of ThO$_{2}$ at a small dose of 0.004 displacements per atom \cite{DESKINS2022}. At these low dose and room temperature irradiation conditions,thorium vacancies and interstitials are the most likely defects, as predicted by the rate theory calculation presented in Dennett et al. \cite{DENNETT2021}. However, current T-matrix formalism can not be applied to interstitial defects located at octahedral sites \cite{JIANG2022}, therefore to account for this type of defects some assumptions need to be made. The scattering cross sections of the vacancies and interstitials are similar based on the NEMD simulations by Jin et al. \cite{JIN2022Impact}. Based on these arguments, we estimate that 0.004 dpa corresponds to an equivalent concentration of V$_{Th}$ to be  $\approx $ 0.25 $\%$. Using the concentration of 0.25 $\%$ V$_{Th}$ yields an excellent agreement between the predicted and the experimental thermal conductivity.
\begin{figure}[h]
   \centering
   \includegraphics [width=0.95\columnwidth]{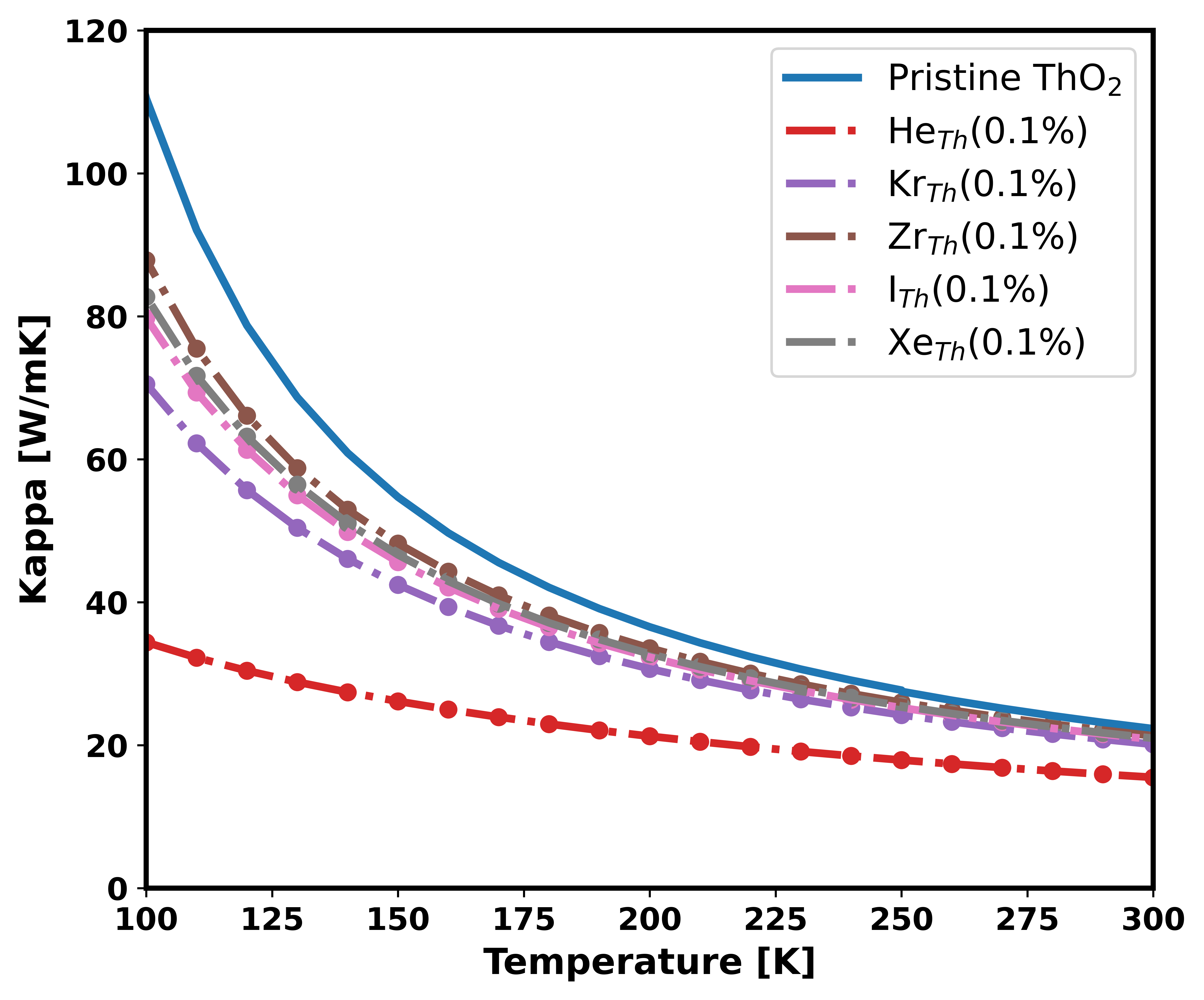}
   \caption{The calculated \textit{$k_{L}$} of ThO$_{2}$ with 0.1$\%$ fission products of He$_{Th}$, Kr$_{Th}$, Zr$_{Th}$, I$_{Th}$, Xe$_{Th}$ at a lower temperature range of 100--300 K.}
   \label{Fig6.png} 
\end{figure}
Fig.~\ref{Fig6.png} shows the degradation of thermal conductivity due to substitutional defects. Among the considered elements, He with the largest perturbation caused the largest reduction in thermal conductivity followed by Kr$_{Th}$ $>$ I$_{Th}$ $>$ Xe$_{Th}$ $>$ Zr$_{Th}$. Fig.~\ref{Fig7.png} shows \textit{$k_{L}$} as a function of defect concentration at 300 K. The results indicate that, for point defects to impact thermal transport, the concentration has to be on the order of  10$^{-2}$ atomic $\%$. For brevity, the \textit{$k_{L}$} degradation for 1\% defect concentrations for all the vacancies and substitutional defects, the {$k_{L}$} reduction due to the Schottky defects for two different concentrations of 0.1\% and 1\%, and the thermal conductivity degradation at high temperatures are provided in SI (Fig S1-S2)[xx]. The results for Schottky defects suggest that, among the three different configurations, the SD$_{111}$ impacted thermal transport the most, followed by SD$_{110}$ and SD$_{100}$. All these observations shed light on the effect of point defects on thermal transport in thorium dioxide and help in understanding the behavior of nuclear reactors \cite{hurley2022thermal}.
Lastly, we should point out that experimental measurements of thermal conductivity in irradiated ThO$_2$ and (Th,U)O$_2$ alloys provided evidence of the presence of phonon resonant scattering as the temperature dependence deviated from a 1/T-trend at low temperatures \cite{DESKINS2022,hua2023thermal}. Our phonon scattering rates shown in Figs.~\ref{Fig1.png} and ~\ref{Fig3.png} provide some indication of the resonances, which for other materials have resulted in the calculated thermal conductivity departing from a 1/T-trend \cite{kundu2019}. However, our calculated thermal conductivity values reported Figs.~\ref{Fig5.png} and ~\ref{Fig6.png} do not reflect this. This observation suggests that further work is needed to improve estimation of low temperature thermal conductivity.
\begin{figure}[h]
    \centering
    \includegraphics [width=0.95\columnwidth]{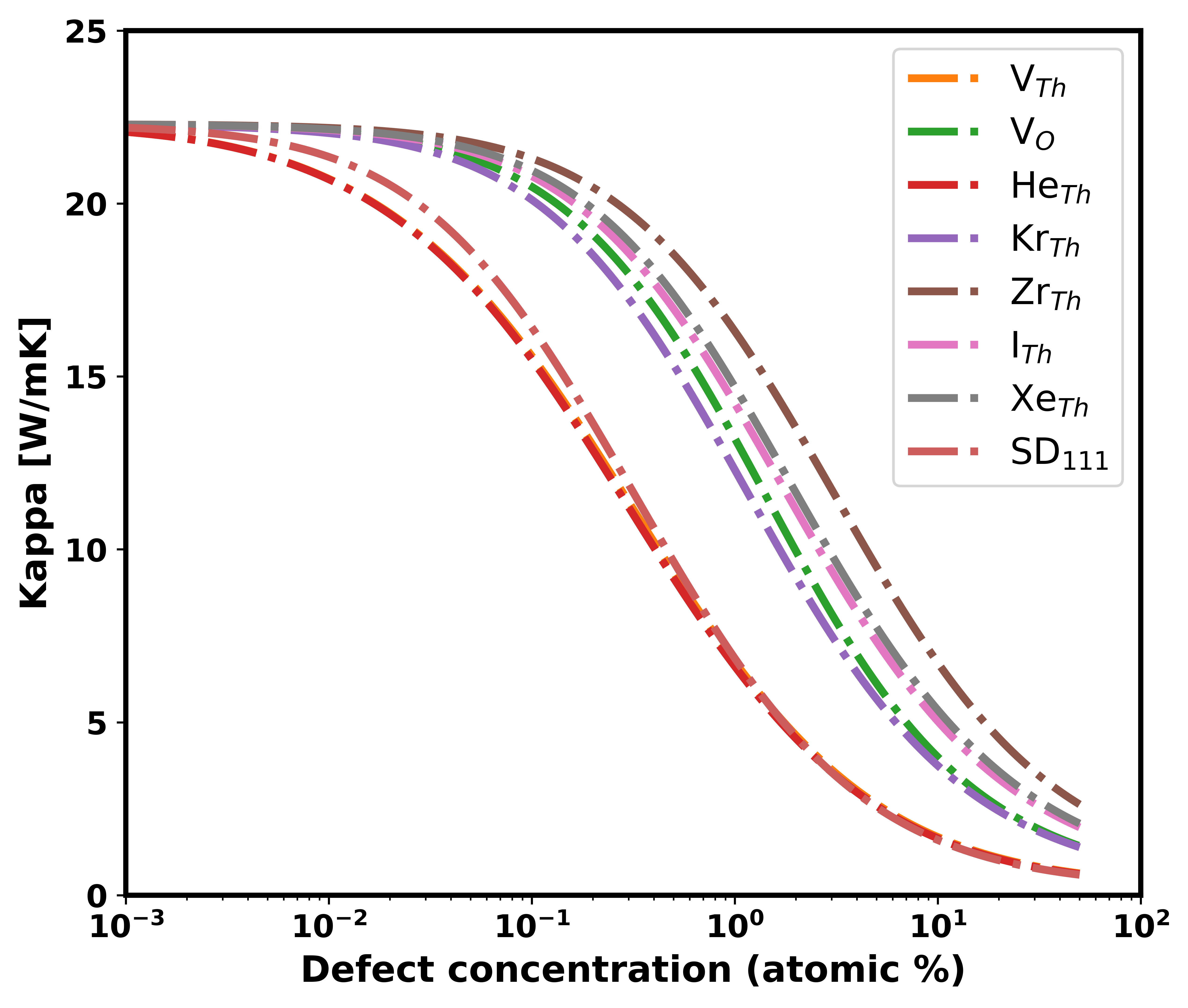}
    \caption{ThO$_{2}$ \textit{$k_{L}$} variation with defect concentration for V$_{Th}$, V$_{O}$, He$_{Th}$, Kr$_{Th}$, Zr$_{Th}$, I$_{Th}$, Xe$_{Th}$, and SD$_{111}$ at 300 K.}
    \label{Fig7.png}
\end{figure}

\section{Conclusion}
For the first time, using a fully \textit {ab initio} Green’s function \textbf{T}-matrix method, we have presented an extensive analysis of the phonon-defect scattering rate of ThO$_{2}$ with point defects. The phonon-point defect scattering rates were unified with the Peierls-Boltzmann transport equation to determine the effects of defects on phonon thermal conductivity in ThO$_{2}$. The results revealed that, among all the intrinsic defects considered in this work, the thorium vacancy and helium substitution in the thorium site caused the most significant reduction in the thermal conductivity of ThO$_{2}$. The similar behaviours of the thorium vacancy and helium substitution was due to the comparable force constant difference induced by these two defects. Also, a large peak, which is a signature of resonant scattering was observed in the case of a thorium vacancy and helium substitution.  Among the substitutional defects, the zirconium substitution had the smallest reduction in thermal conductivity despite having a significant mass difference compared to iodine and xenon. The primary reason for the relatively weaker scattering by zirconium substitution is its similar valency to thorium, thus enabling bond formation with oxygen, inducing the least force constant differences. Unlike zirconium, the chemical environment due to helium, krypton, iodine, and xenon creates a larger force constant difference, leading to higher scattering rates. This study revealed that the modification of the local chemical environment and the associated structure and force variations caused by substitutional defects should be considered while studying phonon-defect scattering. This work enhances our understanding of the phonon-defect scattering due to various fission products in ThO$_{2}$, which will serve as a benchmark for analytical models and help improve the engineering-scale modeling predictions for nuclear reactor design.

\section{ACKNOWLEDGMENTS}
This work was supported by the Center for Thermal Energy Transport under Irradiation (TETI), an Energy Frontier Research Center (EFRC) funded by the U.S. Department of Energy, Office of Science, and Office of Basic Energy Sciences. The authors also acknowledge that this research made use of the resources of the High Performance Computing Center at Idaho National Laboratory, which is supported by the Office of Nuclear Energy of the U.S. Department of Energy and the Nuclear Science User Facilities under Contract No. DE-AC07-05ID14517. This manuscript has been authored by Battelle Energy Alliance, LLC under Contract No. DE-AC07-05ID14517 with the U.S. Department of Energy. The United States Government retains and the publisher, by accepting the article for publication, acknowledges that the U.S. Government retains a nonexclusive, paid-up, irrevocable, world-wide license to publish or reproduce the published form of this manuscript, or allow others to do so, for U.S. Government purposes.

\bibliography{final.bib}

\providecommand{\noopsort}[1]{}\providecommand{\singleletter}[1]{#1}%
\begin{thebibliography}{3}%
\makeatletter
\providecommand \@ifxundefined [1]{%
 \@ifx{#1\undefined}
}%
\providecommand \@ifnum [1]{%
 \ifnum #1\expandafter \@firstoftwo
 \else \expandafter \@secondoftwo
 \fi
}%
\providecommand \@ifx [1]{%
 \ifx #1\expandafter \@firstoftwo
 \else \expandafter \@secondoftwo
 \fi
}%
\providecommand \natexlab [1]{#1}%
\providecommand \enquote  [1]{``#1''}%
\providecommand \bibnamefont  [1]{#1}%
\providecommand \bibfnamefont [1]{#1}%
\providecommand \citenamefont [1]{#1}%
\providecommand \href@noop [0]{\@secondoftwo}%
\providecommand \href [0]{\begingroup \@sanitize@url \@href}%
\providecommand \@href[1]{\@@startlink{#1}\@@href}%
\providecommand \@@href[1]{\endgroup#1\@@endlink}%
\providecommand \@sanitize@url [0]{\catcode `\\12\catcode `\$12\catcode
  `\&12\catcode `\#12\catcode `\^12\catcode `\_12\catcode `\%12\relax}%
\providecommand \@@startlink[1]{}%
\providecommand \@@endlink[0]{}%
\providecommand \url  [0]{\begingroup\@sanitize@url \@url }%
\providecommand \@url [1]{\endgroup\@href {#1}{\urlprefix }}%
\providecommand \urlprefix  [0]{URL }%
\providecommand \Eprint [0]{\href }%
\providecommand \doibase [0]{https://doi.org/}%
\providecommand \selectlanguage [0]{\@gobble}%
\providecommand \bibinfo  [0]{\@secondoftwo}%
\providecommand \bibfield  [0]{\@secondoftwo}%
\providecommand \translation [1]{[#1]}%
\providecommand \BibitemOpen [0]{}%
\providecommand \bibitemStop [0]{}%
\providecommand \bibitemNoStop [0]{.\EOS\space}%
\providecommand \EOS [0]{\spacefactor3000\relax}%
\providecommand \BibitemShut  [1]{\csname bibitem#1\endcsname}%
\let\auto@bib@innerbib\@empty
\bibitem [{\citenamefont {Deskins}\ \emph {et~al.}(2022)\citenamefont
  {Deskins}, \citenamefont {Khanolkar}, \citenamefont {Mazumder}, \citenamefont
  {Dennett}, \citenamefont {Bawane}, \citenamefont {Hua}, \citenamefont
  {Ferrigno}, \citenamefont {He}, \citenamefont {Mann}, \citenamefont
  {Khafizov}, \citenamefont {Hurley},\ and\ \citenamefont
  {El-Azab}}]{DESKINS2022}%
  \BibitemOpen
  \bibfield  {author} {\bibinfo {author} {\bibfnamefont {W.~R.}\ \bibnamefont
  {Deskins}}, \bibinfo {author} {\bibfnamefont {A.}~\bibnamefont {Khanolkar}},
  \bibinfo {author} {\bibfnamefont {S.}~\bibnamefont {Mazumder}}, \bibinfo
  {author} {\bibfnamefont {C.~A.}\ \bibnamefont {Dennett}}, \bibinfo {author}
  {\bibfnamefont {K.}~\bibnamefont {Bawane}}, \bibinfo {author} {\bibfnamefont
  {Z.}~\bibnamefont {Hua}}, \bibinfo {author} {\bibfnamefont {J.}~\bibnamefont
  {Ferrigno}}, \bibinfo {author} {\bibfnamefont {L.}~\bibnamefont {He}},
  \bibinfo {author} {\bibfnamefont {J.~M.}\ \bibnamefont {Mann}}, \bibinfo
  {author} {\bibfnamefont {M.}~\bibnamefont {Khafizov}}, \bibinfo {author}
  {\bibfnamefont {D.~H.}\ \bibnamefont {Hurley}},\ and\ \bibinfo {author}
  {\bibfnamefont {A.}~\bibnamefont {El-Azab}},\ }\href
  {https://doi.org/https://doi.org/10.1016/j.actamat.2022.118379} {\bibfield
  {journal} {\bibinfo  {journal} {Acta Materialia}\ }\textbf {\bibinfo {volume}
  {241}},\ \bibinfo {pages} {118379} (\bibinfo {year} {2022})}\BibitemShut
  {NoStop}%
\bibitem [{\citenamefont {Mann}\ \emph {et~al.}(2010)\citenamefont {Mann},
  \citenamefont {Thompson}, \citenamefont {Serivalsatit}, \citenamefont
  {Tritt}, \citenamefont {Ballato},\ and\ \citenamefont {Kolis}}]{Mann_2010}%
  \BibitemOpen
  \bibfield  {author} {\bibinfo {author} {\bibfnamefont {M.}~\bibnamefont
  {Mann}}, \bibinfo {author} {\bibfnamefont {D.}~\bibnamefont {Thompson}},
  \bibinfo {author} {\bibfnamefont {K.}~\bibnamefont {Serivalsatit}}, \bibinfo
  {author} {\bibfnamefont {T.~M.}\ \bibnamefont {Tritt}}, \bibinfo {author}
  {\bibfnamefont {J.}~\bibnamefont {Ballato}},\ and\ \bibinfo {author}
  {\bibfnamefont {J.}~\bibnamefont {Kolis}},\ }\href
  {https://doi.org/10.1021/cg901308f} {\bibfield  {journal} {\bibinfo
  {journal} {Crystal Growth \& Design}\ }\textbf {\bibinfo {volume} {10}},\
  \bibinfo {pages} {2146} (\bibinfo {year} {2010})}\BibitemShut {NoStop}%
\bibitem [{\citenamefont {Matthew}\ \emph {et~al.}(2020)\citenamefont
  {Matthew}, \citenamefont {Lyuwen}, \citenamefont {Rickert}, \citenamefont
  {David}, \citenamefont {Prusnick}, \citenamefont {Matthew}, \citenamefont
  {Douglas}, \citenamefont {Chris},\ and\ \citenamefont {Michael}}]{Bryan2020}%
  \BibitemOpen
  \bibfield  {author} {\bibinfo {author} {\bibfnamefont {B.}~\bibnamefont
  {Matthew}}, \bibinfo {author} {\bibfnamefont {F.}~\bibnamefont {Lyuwen}},
  \bibinfo {author} {\bibfnamefont {K.}~\bibnamefont {Rickert}}, \bibinfo
  {author} {\bibfnamefont {T.}~\bibnamefont {David}}, \bibinfo {author}
  {\bibfnamefont {T.~A.}\ \bibnamefont {Prusnick}}, \bibinfo {author}
  {\bibfnamefont {M.}~\bibnamefont {Matthew}}, \bibinfo {author} {\bibfnamefont
  {A.}~\bibnamefont {Douglas}}, \bibinfo {author} {\bibfnamefont
  {M.}~\bibnamefont {Chris}},\ and\ \bibinfo {author} {\bibfnamefont
  {M.}~\bibnamefont {Michael}},\ }\href
  {https://doi.org/10.1038/s42005-020-00483-2} {\bibfield  {journal} {\bibinfo
  {journal} {Communications Physics}\ }\textbf {\bibinfo {volume} {3}},\
  \bibinfo {pages} {217} (\bibinfo {year} {2020})}\BibitemShut {NoStop}%
\end{thebibliography}%


\providecommand{\noopsort}[1]{}\providecommand{\singleletter}[1]{#1}%
\begin{thebibliography}{56}%
\makeatletter
\providecommand \@ifxundefined [1]{%
 \@ifx{#1\undefined}
}%
\providecommand \@ifnum [1]{%
 \ifnum #1\expandafter \@firstoftwo
 \else \expandafter \@secondoftwo
 \fi
}%
\providecommand \@ifx [1]{%
 \ifx #1\expandafter \@firstoftwo
 \else \expandafter \@secondoftwo
 \fi
}%
\providecommand \natexlab [1]{#1}%
\providecommand \enquote  [1]{``#1''}%
\providecommand \bibnamefont  [1]{#1}%
\providecommand \bibfnamefont [1]{#1}%
\providecommand \citenamefont [1]{#1}%
\providecommand \href@noop [0]{\@secondoftwo}%
\providecommand \href [0]{\begingroup \@sanitize@url \@href}%
\providecommand \@href[1]{\@@startlink{#1}\@@href}%
\providecommand \@@href[1]{\endgroup#1\@@endlink}%
\providecommand \@sanitize@url [0]{\catcode `\\12\catcode `\$12\catcode
  `\&12\catcode `\#12\catcode `\^12\catcode `\_12\catcode `\%12\relax}%
\providecommand \@@startlink[1]{}%
\providecommand \@@endlink[0]{}%
\providecommand \url  [0]{\begingroup\@sanitize@url \@url }%
\providecommand \@url [1]{\endgroup\@href {#1}{\urlprefix }}%
\providecommand \urlprefix  [0]{URL }%
\providecommand \Eprint [0]{\href }%
\providecommand \doibase [0]{https://doi.org/}%
\providecommand \selectlanguage [0]{\@gobble}%
\providecommand \bibinfo  [0]{\@secondoftwo}%
\providecommand \bibfield  [0]{\@secondoftwo}%
\providecommand \translation [1]{[#1]}%
\providecommand \BibitemOpen [0]{}%
\providecommand \bibitemStop [0]{}%
\providecommand \bibitemNoStop [0]{.\EOS\space}%
\providecommand \EOS [0]{\spacefactor3000\relax}%
\providecommand \BibitemShut  [1]{\csname bibitem#1\endcsname}%
\let\auto@bib@innerbib\@empty
\bibitem [{iae(2005)}]{iaea2005}%
  \BibitemOpen
  \href
  {https://www.iaea.org/publications/7192/thorium-fuel-cycle-potential-benefits-and-challenges}
  {\emph {\bibinfo {title} {Thorium Fuel Cycle - Potential Benefits and
  Challenges}}},\ \bibinfo {series} {TECDOC Series}\ No.\ \bibinfo {number}
  {1450}\ (\bibinfo  {publisher} {INTERNATIONAL ATOMIC ENERGY AGENCY},\
  \bibinfo {address} {Vienna},\ \bibinfo {year} {2005})\BibitemShut {NoStop}%
\bibitem [{\citenamefont {Hurley}\ \emph {et~al.}(2022)\citenamefont {Hurley},
  \citenamefont {El-Azab}, \citenamefont {Bryan}, \citenamefont {Cooper},
  \citenamefont {Dennett}, \citenamefont {Gofryk}, \citenamefont {He},
  \citenamefont {Khafizov}, \citenamefont {Lander}, \citenamefont {Manley},
  \citenamefont {Mann}, \citenamefont {Marianetti}, \citenamefont {Rickert},
  \citenamefont {Selim}, \citenamefont {Tonks},\ and\ \citenamefont
  {Wharry}}]{hurley2022thermal}%
  \BibitemOpen
  \bibfield  {author} {\bibinfo {author} {\bibfnamefont {D.~H.}\ \bibnamefont
  {Hurley}}, \bibinfo {author} {\bibfnamefont {A.}~\bibnamefont {El-Azab}},
  \bibinfo {author} {\bibfnamefont {M.~S.}\ \bibnamefont {Bryan}}, \bibinfo
  {author} {\bibfnamefont {M.~W.~D.}\ \bibnamefont {Cooper}}, \bibinfo {author}
  {\bibfnamefont {C.~A.}\ \bibnamefont {Dennett}}, \bibinfo {author}
  {\bibfnamefont {K.}~\bibnamefont {Gofryk}}, \bibinfo {author} {\bibfnamefont
  {L.}~\bibnamefont {He}}, \bibinfo {author} {\bibfnamefont {M.}~\bibnamefont
  {Khafizov}}, \bibinfo {author} {\bibfnamefont {G.~H.}\ \bibnamefont
  {Lander}}, \bibinfo {author} {\bibfnamefont {M.~E.}\ \bibnamefont {Manley}},
  \bibinfo {author} {\bibfnamefont {J.~M.}\ \bibnamefont {Mann}}, \bibinfo
  {author} {\bibfnamefont {C.~A.}\ \bibnamefont {Marianetti}}, \bibinfo
  {author} {\bibfnamefont {K.}~\bibnamefont {Rickert}}, \bibinfo {author}
  {\bibfnamefont {F.~A.}\ \bibnamefont {Selim}}, \bibinfo {author}
  {\bibfnamefont {M.~R.}\ \bibnamefont {Tonks}},\ and\ \bibinfo {author}
  {\bibfnamefont {J.~P.}\ \bibnamefont {Wharry}},\ }\bibfield  {title}
  {\bibinfo {title} {Thermal energy transport in oxide nuclear fuel},\ }\href
  {https://doi.org/10.1021/acs.chemrev.1c00262} {\bibfield  {journal} {\bibinfo
   {journal} {Chemical Reviews}\ }\textbf {\bibinfo {volume} {122}},\ \bibinfo
  {pages} {3711} (\bibinfo {year} {2022})}\BibitemShut {NoStop}%
\bibitem [{\citenamefont {Dennett}\ \emph {et~al.}(2021)\citenamefont
  {Dennett}, \citenamefont {Deskins}, \citenamefont {Khafizov}, \citenamefont
  {Hua}, \citenamefont {Khanolkar}, \citenamefont {Bawane}, \citenamefont {Fu},
  \citenamefont {Mann}, \citenamefont {Marianetti}, \citenamefont {He},
  \citenamefont {Hurley},\ and\ \citenamefont {El-Azab}}]{DENNETT2021}%
  \BibitemOpen
  \bibfield  {author} {\bibinfo {author} {\bibfnamefont {C.~A.}\ \bibnamefont
  {Dennett}}, \bibinfo {author} {\bibfnamefont {W.~R.}\ \bibnamefont
  {Deskins}}, \bibinfo {author} {\bibfnamefont {M.}~\bibnamefont {Khafizov}},
  \bibinfo {author} {\bibfnamefont {Z.}~\bibnamefont {Hua}}, \bibinfo {author}
  {\bibfnamefont {A.}~\bibnamefont {Khanolkar}}, \bibinfo {author}
  {\bibfnamefont {K.}~\bibnamefont {Bawane}}, \bibinfo {author} {\bibfnamefont
  {L.}~\bibnamefont {Fu}}, \bibinfo {author} {\bibfnamefont {J.~M.}\
  \bibnamefont {Mann}}, \bibinfo {author} {\bibfnamefont {C.~A.}\ \bibnamefont
  {Marianetti}}, \bibinfo {author} {\bibfnamefont {L.}~\bibnamefont {He}},
  \bibinfo {author} {\bibfnamefont {D.~H.}\ \bibnamefont {Hurley}},\ and\
  \bibinfo {author} {\bibfnamefont {A.}~\bibnamefont {El-Azab}},\ }\bibfield
  {title} {\bibinfo {title} {An integrated experimental and computational
  investigation of defect and microstructural effects on thermal transport in
  thorium dioxide},\ }\href
  {https://doi.org/https://doi.org/10.1016/j.actamat.2021.116934} {\bibfield
  {journal} {\bibinfo  {journal} {Acta Materialia}\ }\textbf {\bibinfo {volume}
  {213}},\ \bibinfo {pages} {116934} (\bibinfo {year} {2021})}\BibitemShut
  {NoStop}%
\bibitem [{\citenamefont {Gurunathan}\ \emph {et~al.}(2020)\citenamefont
  {Gurunathan}, \citenamefont {Hanus}, \citenamefont {Dylla}, \citenamefont
  {Katre},\ and\ \citenamefont {Snyder}}]{ramya2020}%
  \BibitemOpen
  \bibfield  {author} {\bibinfo {author} {\bibfnamefont {R.}~\bibnamefont
  {Gurunathan}}, \bibinfo {author} {\bibfnamefont {R.}~\bibnamefont {Hanus}},
  \bibinfo {author} {\bibfnamefont {M.}~\bibnamefont {Dylla}}, \bibinfo
  {author} {\bibfnamefont {A.}~\bibnamefont {Katre}},\ and\ \bibinfo {author}
  {\bibfnamefont {G.~J.}\ \bibnamefont {Snyder}},\ }\bibfield  {title}
  {\bibinfo {title} {Analytical models of phonon--point-defect scattering},\
  }\href {https://doi.org/10.1103/PhysRevApplied.13.034011} {\bibfield
  {journal} {\bibinfo  {journal} {Phys. Rev. Applied}\ }\textbf {\bibinfo
  {volume} {13}},\ \bibinfo {pages} {034011} (\bibinfo {year}
  {2020})}\BibitemShut {NoStop}%
\bibitem [{\citenamefont {Deskins}\ \emph {et~al.}(2021)\citenamefont
  {Deskins}, \citenamefont {Hamed}, \citenamefont {Kumagai}, \citenamefont
  {Dennett}, \citenamefont {Peng}, \citenamefont {Khafizov}, \citenamefont
  {Hurley},\ and\ \citenamefont {El-Azab}}]{Deskins_2021}%
  \BibitemOpen
  \bibfield  {author} {\bibinfo {author} {\bibfnamefont {W.~R.}\ \bibnamefont
  {Deskins}}, \bibinfo {author} {\bibfnamefont {A.}~\bibnamefont {Hamed}},
  \bibinfo {author} {\bibfnamefont {T.}~\bibnamefont {Kumagai}}, \bibinfo
  {author} {\bibfnamefont {C.~A.}\ \bibnamefont {Dennett}}, \bibinfo {author}
  {\bibfnamefont {J.}~\bibnamefont {Peng}}, \bibinfo {author} {\bibfnamefont
  {M.}~\bibnamefont {Khafizov}}, \bibinfo {author} {\bibfnamefont
  {D.}~\bibnamefont {Hurley}},\ and\ \bibinfo {author} {\bibfnamefont
  {A.}~\bibnamefont {El-Azab}},\ }\bibfield  {title} {\bibinfo {title} {Thermal
  conductivity of \uppercase{T}h\uppercase{O}$_2$: Effect of point defect
  disorder},\ }\href {https://doi.org/10.1063/5.0038117} {\bibfield  {journal}
  {\bibinfo  {journal} {Journal of Applied Physics}\ }\textbf {\bibinfo
  {volume} {129}},\ \bibinfo {pages} {075102} (\bibinfo {year}
  {2021})}\BibitemShut {NoStop}%
\bibitem [{\citenamefont {Klemens}(1955)}]{Klemens_1955}%
  \BibitemOpen
  \bibfield  {author} {\bibinfo {author} {\bibfnamefont {P.~G.}\ \bibnamefont
  {Klemens}},\ }\bibfield  {title} {\bibinfo {title} {The scattering of
  low-frequency lattice waves by static imperfections},\ }\href
  {https://doi.org/10.1088/0370-1298/68/12/303} {\bibfield  {journal} {\bibinfo
   {journal} {Proceedings of the Physical Society. Section A}\ }\textbf
  {\bibinfo {volume} {68}},\ \bibinfo {pages} {1113} (\bibinfo {year}
  {1955})}\BibitemShut {NoStop}%
\bibitem [{\citenamefont {Asheghi}\ \emph {et~al.}(2002)\citenamefont
  {Asheghi}, \citenamefont {Kurabayashi}, \citenamefont {Kasnavi},\ and\
  \citenamefont {Goodson}}]{Asheghi_2002}%
  \BibitemOpen
  \bibfield  {author} {\bibinfo {author} {\bibfnamefont {M.}~\bibnamefont
  {Asheghi}}, \bibinfo {author} {\bibfnamefont {K.}~\bibnamefont
  {Kurabayashi}}, \bibinfo {author} {\bibfnamefont {R.}~\bibnamefont
  {Kasnavi}},\ and\ \bibinfo {author} {\bibfnamefont {K.~E.}\ \bibnamefont
  {Goodson}},\ }\bibfield  {title} {\bibinfo {title} {Thermal conduction in
  doped single-crystal silicon films},\ }\href
  {https://doi.org/10.1063/1.1458057} {\bibfield  {journal} {\bibinfo
  {journal} {Journal of Applied Physics}\ }\textbf {\bibinfo {volume} {91}},\
  \bibinfo {pages} {5079} (\bibinfo {year} {2002})}\BibitemShut {NoStop}%
\bibitem [{\citenamefont {Zou}\ \emph {et~al.}(2002)\citenamefont {Zou},
  \citenamefont {Kotchetkov}, \citenamefont {Balandin}, \citenamefont
  {Florescu},\ and\ \citenamefont {Pollak}}]{Zou_2002}%
  \BibitemOpen
  \bibfield  {author} {\bibinfo {author} {\bibfnamefont {J.}~\bibnamefont
  {Zou}}, \bibinfo {author} {\bibfnamefont {D.}~\bibnamefont {Kotchetkov}},
  \bibinfo {author} {\bibfnamefont {A.~A.}\ \bibnamefont {Balandin}}, \bibinfo
  {author} {\bibfnamefont {D.~I.}\ \bibnamefont {Florescu}},\ and\ \bibinfo
  {author} {\bibfnamefont {F.~H.}\ \bibnamefont {Pollak}},\ }\bibfield  {title}
  {\bibinfo {title} {Thermal conductivity of \uppercase{G}a\uppercase{N} films:
  Effects of impurities and dislocations},\ }\href
  {https://doi.org/10.1063/1.1497704} {\bibfield  {journal} {\bibinfo
  {journal} {Journal of Applied Physics}\ }\textbf {\bibinfo {volume} {92}},\
  \bibinfo {pages} {2534} (\bibinfo {year} {2002})}\BibitemShut {NoStop}%
\bibitem [{\citenamefont {Fletcher}(1987)}]{Fletcher_1987}%
  \BibitemOpen
  \bibfield  {author} {\bibinfo {author} {\bibfnamefont {R.}~\bibnamefont
  {Fletcher}},\ }\bibfield  {title} {\bibinfo {title} {Scattering of phonons by
  dislocations in potassium},\ }\href
  {https://doi.org/10.1103/PhysRevB.36.3042} {\bibfield  {journal} {\bibinfo
  {journal} {Phys. Rev. B}\ }\textbf {\bibinfo {volume} {36}},\ \bibinfo
  {pages} {3042} (\bibinfo {year} {1987})}\BibitemShut {NoStop}%
\bibitem [{\citenamefont {Malekpour}\ \emph {et~al.}(2016)\citenamefont
  {Malekpour}, \citenamefont {Ramnani}, \citenamefont {Srinivasan},
  \citenamefont {Balasubramanian}, \citenamefont {Nika}, \citenamefont
  {Mulchandani}, \citenamefont {Lake},\ and\ \citenamefont
  {Balandin}}]{Malekpour_2016}%
  \BibitemOpen
  \bibfield  {author} {\bibinfo {author} {\bibfnamefont {H.}~\bibnamefont
  {Malekpour}}, \bibinfo {author} {\bibfnamefont {P.}~\bibnamefont {Ramnani}},
  \bibinfo {author} {\bibfnamefont {S.}~\bibnamefont {Srinivasan}}, \bibinfo
  {author} {\bibfnamefont {G.}~\bibnamefont {Balasubramanian}}, \bibinfo
  {author} {\bibfnamefont {D.~L.}\ \bibnamefont {Nika}}, \bibinfo {author}
  {\bibfnamefont {A.}~\bibnamefont {Mulchandani}}, \bibinfo {author}
  {\bibfnamefont {R.~K.}\ \bibnamefont {Lake}},\ and\ \bibinfo {author}
  {\bibfnamefont {A.~A.}\ \bibnamefont {Balandin}},\ }\bibfield  {title}
  {\bibinfo {title} {Thermal conductivity of graphene with defects induced by
  electron beam irradiation},\ }\href {https://doi.org/10.1039/C6NR03470E}
  {\bibfield  {journal} {\bibinfo  {journal} {Nanoscale}\ }\textbf {\bibinfo
  {volume} {8}},\ \bibinfo {pages} {14608} (\bibinfo {year}
  {2016})}\BibitemShut {NoStop}%
\bibitem [{\citenamefont {Tamura}(1983)}]{Tamura_1983}%
  \BibitemOpen
  \bibfield  {author} {\bibinfo {author} {\bibfnamefont {S.}~\bibnamefont
  {Tamura}},\ }\bibfield  {title} {\bibinfo {title} {Isotope scattering of
  dispersive phonons in \uppercase{G}e},\ }\href
  {https://doi.org/10.1103/PhysRevB.27.858} {\bibfield  {journal} {\bibinfo
  {journal} {Phys. Rev. B}\ }\textbf {\bibinfo {volume} {27}},\ \bibinfo
  {pages} {858} (\bibinfo {year} {1983})}\BibitemShut {NoStop}%
\bibitem [{\citenamefont {Sakurai}(1967)}]{sakurai1967advanced}%
  \BibitemOpen
  \bibfield  {author} {\bibinfo {author} {\bibfnamefont {J.}~\bibnamefont
  {Sakurai}},\ }\href {https://books.google.com/books?id=lvmSZkzDFt0C} {\emph
  {\bibinfo {title} {Advanced Quantum Mechanics}}},\ Always learning\ (\bibinfo
   {publisher} {Pearson Education, Incorporated},\ \bibinfo {year}
  {1967})\BibitemShut {NoStop}%
\bibitem [{\citenamefont {Polanco}\ and\ \citenamefont
  {Lindsay}(2018{\natexlab{a}})}]{Polanco_InN}%
  \BibitemOpen
  \bibfield  {author} {\bibinfo {author} {\bibfnamefont {C.~A.}\ \bibnamefont
  {Polanco}}\ and\ \bibinfo {author} {\bibfnamefont {L.}~\bibnamefont
  {Lindsay}},\ }\bibfield  {title} {\bibinfo {title} {Thermal conductivity of
  \uppercase{I}n\uppercase{N} with point defects from first principles},\
  }\href {https://doi.org/10.1103/PhysRevB.98.014306} {\bibfield  {journal}
  {\bibinfo  {journal} {Phys. Rev. B}\ }\textbf {\bibinfo {volume} {98}},\
  \bibinfo {pages} {014306} (\bibinfo {year} {2018}{\natexlab{a}})}\BibitemShut
  {NoStop}%
\bibitem [{\citenamefont {Mingo}\ \emph {et~al.}(2010)\citenamefont {Mingo},
  \citenamefont {Esfarjani}, \citenamefont {Broido},\ and\ \citenamefont
  {Stewart}}]{Mingo_2010}%
  \BibitemOpen
  \bibfield  {author} {\bibinfo {author} {\bibfnamefont {N.}~\bibnamefont
  {Mingo}}, \bibinfo {author} {\bibfnamefont {K.}~\bibnamefont {Esfarjani}},
  \bibinfo {author} {\bibfnamefont {D.~A.}\ \bibnamefont {Broido}},\ and\
  \bibinfo {author} {\bibfnamefont {D.~A.}\ \bibnamefont {Stewart}},\
  }\bibfield  {title} {\bibinfo {title} {Cluster scattering effects on phonon
  conduction in graphene},\ }\href {https://doi.org/10.1103/PhysRevB.81.045408}
  {\bibfield  {journal} {\bibinfo  {journal} {Phys. Rev. B}\ }\textbf {\bibinfo
  {volume} {81}},\ \bibinfo {pages} {045408} (\bibinfo {year}
  {2010})}\BibitemShut {NoStop}%
\bibitem [{\citenamefont {Kundu}\ \emph {et~al.}(2011)\citenamefont {Kundu},
  \citenamefont {Mingo}, \citenamefont {Broido},\ and\ \citenamefont
  {Stewart}}]{Kundu_2011}%
  \BibitemOpen
  \bibfield  {author} {\bibinfo {author} {\bibfnamefont {A.}~\bibnamefont
  {Kundu}}, \bibinfo {author} {\bibfnamefont {N.}~\bibnamefont {Mingo}},
  \bibinfo {author} {\bibfnamefont {D.~A.}\ \bibnamefont {Broido}},\ and\
  \bibinfo {author} {\bibfnamefont {D.~A.}\ \bibnamefont {Stewart}},\
  }\bibfield  {title} {\bibinfo {title} {Role of light and heavy embedded
  nanoparticles on the thermal conductivity of \uppercase{S}i\uppercase{G}e
  alloys},\ }\href {https://doi.org/10.1103/PhysRevB.84.125426} {\bibfield
  {journal} {\bibinfo  {journal} {Phys. Rev. B}\ }\textbf {\bibinfo {volume}
  {84}},\ \bibinfo {pages} {125426} (\bibinfo {year} {2011})}\BibitemShut
  {NoStop}%
\bibitem [{\citenamefont {Katcho}\ \emph {et~al.}(2014)\citenamefont {Katcho},
  \citenamefont {Carrete}, \citenamefont {Li},\ and\ \citenamefont
  {Mingo}}]{Katcho_2014}%
  \BibitemOpen
  \bibfield  {author} {\bibinfo {author} {\bibfnamefont {N.~A.}\ \bibnamefont
  {Katcho}}, \bibinfo {author} {\bibfnamefont {J.}~\bibnamefont {Carrete}},
  \bibinfo {author} {\bibfnamefont {W.}~\bibnamefont {Li}},\ and\ \bibinfo
  {author} {\bibfnamefont {N.}~\bibnamefont {Mingo}},\ }\bibfield  {title}
  {\bibinfo {title} {Effect of nitrogen and vacancy defects on the thermal
  conductivity of diamond: An ab initio green's function approach},\ }\href
  {https://doi.org/10.1103/PhysRevB.90.094117} {\bibfield  {journal} {\bibinfo
  {journal} {Phys. Rev. B}\ }\textbf {\bibinfo {volume} {90}},\ \bibinfo
  {pages} {094117} (\bibinfo {year} {2014})}\BibitemShut {NoStop}%
\bibitem [{\citenamefont {Protik}\ \emph {et~al.}(2016)\citenamefont {Protik},
  \citenamefont {Carrete}, \citenamefont {Katcho}, \citenamefont {Mingo},\ and\
  \citenamefont {Broido}}]{protik_2016}%
  \BibitemOpen
  \bibfield  {author} {\bibinfo {author} {\bibfnamefont {N.~H.}\ \bibnamefont
  {Protik}}, \bibinfo {author} {\bibfnamefont {J.}~\bibnamefont {Carrete}},
  \bibinfo {author} {\bibfnamefont {N.~A.}\ \bibnamefont {Katcho}}, \bibinfo
  {author} {\bibfnamefont {N.}~\bibnamefont {Mingo}},\ and\ \bibinfo {author}
  {\bibfnamefont {D.}~\bibnamefont {Broido}},\ }\bibfield  {title} {\bibinfo
  {title} {Ab initio study of the effect of vacancies on the thermal
  conductivity of boron arsenide},\ }\href
  {https://doi.org/10.1103/PhysRevB.94.045207} {\bibfield  {journal} {\bibinfo
  {journal} {Phys. Rev. B}\ }\textbf {\bibinfo {volume} {94}},\ \bibinfo
  {pages} {045207} (\bibinfo {year} {2016})}\BibitemShut {NoStop}%
\bibitem [{\citenamefont {Katre}\ \emph {et~al.}(2017)\citenamefont {Katre},
  \citenamefont {Carrete}, \citenamefont {Dongre}, \citenamefont {Madsen},\
  and\ \citenamefont {Mingo}}]{Ankita_2017}%
  \BibitemOpen
  \bibfield  {author} {\bibinfo {author} {\bibfnamefont {A.}~\bibnamefont
  {Katre}}, \bibinfo {author} {\bibfnamefont {J.}~\bibnamefont {Carrete}},
  \bibinfo {author} {\bibfnamefont {B.}~\bibnamefont {Dongre}}, \bibinfo
  {author} {\bibfnamefont {G.~K.~H.}\ \bibnamefont {Madsen}},\ and\ \bibinfo
  {author} {\bibfnamefont {N.}~\bibnamefont {Mingo}},\ }\bibfield  {title}
  {\bibinfo {title} {Exceptionally strong phonon scattering by \uppercase{B}
  substitution in cubic \uppercase{S}i\uppercase{C}},\ }\href
  {https://doi.org/10.1103/PhysRevLett.119.075902} {\bibfield  {journal}
  {\bibinfo  {journal} {Phys. Rev. Lett.}\ }\textbf {\bibinfo {volume} {119}},\
  \bibinfo {pages} {075902} (\bibinfo {year} {2017})}\BibitemShut {NoStop}%
\bibitem [{\citenamefont {Katre}\ \emph {et~al.}(2018)\citenamefont {Katre},
  \citenamefont {Carrete}, \citenamefont {Wang}, \citenamefont {Madsen},\ and\
  \citenamefont {Mingo}}]{Ankita_2018}%
  \BibitemOpen
  \bibfield  {author} {\bibinfo {author} {\bibfnamefont {A.}~\bibnamefont
  {Katre}}, \bibinfo {author} {\bibfnamefont {J.}~\bibnamefont {Carrete}},
  \bibinfo {author} {\bibfnamefont {T.}~\bibnamefont {Wang}}, \bibinfo {author}
  {\bibfnamefont {G.~K.~H.}\ \bibnamefont {Madsen}},\ and\ \bibinfo {author}
  {\bibfnamefont {N.}~\bibnamefont {Mingo}},\ }\bibfield  {title} {\bibinfo
  {title} {Phonon transport unveils the prevalent point defects in
  \uppercase{G}a\uppercase{N}},\ }\href
  {https://doi.org/10.1103/PhysRevMaterials.2.050602} {\bibfield  {journal}
  {\bibinfo  {journal} {Phys. Rev. Materials}\ }\textbf {\bibinfo {volume}
  {2}},\ \bibinfo {pages} {050602} (\bibinfo {year} {2018})}\BibitemShut
  {NoStop}%
\bibitem [{\citenamefont {Polanco}\ and\ \citenamefont
  {Lindsay}(2018{\natexlab{b}})}]{Polanco_g}%
  \BibitemOpen
  \bibfield  {author} {\bibinfo {author} {\bibfnamefont {C.~A.}\ \bibnamefont
  {Polanco}}\ and\ \bibinfo {author} {\bibfnamefont {L.}~\bibnamefont
  {Lindsay}},\ }\bibfield  {title} {\bibinfo {title} {Ab initio phonon point
  defect scattering and thermal transport in graphene},\ }\href
  {https://doi.org/10.1103/PhysRevB.97.014303} {\bibfield  {journal} {\bibinfo
  {journal} {Phys. Rev. B}\ }\textbf {\bibinfo {volume} {97}},\ \bibinfo
  {pages} {014303} (\bibinfo {year} {2018}{\natexlab{b}})}\BibitemShut
  {NoStop}%
\bibitem [{\citenamefont {Liu}\ \emph {et~al.}(2018)\citenamefont {Liu},
  \citenamefont {Dai}, \citenamefont {Yang}, \citenamefont {Zhao},\ and\
  \citenamefont {Meng}}]{LIU201811}%
  \BibitemOpen
  \bibfield  {author} {\bibinfo {author} {\bibfnamefont {J.}~\bibnamefont
  {Liu}}, \bibinfo {author} {\bibfnamefont {Z.}~\bibnamefont {Dai}}, \bibinfo
  {author} {\bibfnamefont {X.}~\bibnamefont {Yang}}, \bibinfo {author}
  {\bibfnamefont {Y.}~\bibnamefont {Zhao}},\ and\ \bibinfo {author}
  {\bibfnamefont {S.}~\bibnamefont {Meng}},\ }\bibfield  {title} {\bibinfo
  {title} {Lattice thermodynamic behavior in nuclear fuel
  \uppercase{T}h\uppercase{O}$_2$ from first principles},\ }\href
  {https://doi.org/https://doi.org/10.1016/j.jnucmat.2018.08.054} {\bibfield
  {journal} {\bibinfo  {journal} {Journal of Nuclear Materials}\ }\textbf
  {\bibinfo {volume} {511}},\ \bibinfo {pages} {11} (\bibinfo {year} {2018})},\
  \bibinfo {note} {special Section on "18th International Conference on Fusion
  Reactor Materials"}\BibitemShut {NoStop}%
\bibitem [{\citenamefont {Malakkal}\ \emph {et~al.}(2019)\citenamefont
  {Malakkal}, \citenamefont {Prasad}, \citenamefont {Jossou}, \citenamefont
  {Ranasinghe}, \citenamefont {Szpunar}, \citenamefont {Bichler},\ and\
  \citenamefont {Szpunar}}]{MALAKKAL2019507}%
  \BibitemOpen
  \bibfield  {author} {\bibinfo {author} {\bibfnamefont {L.}~\bibnamefont
  {Malakkal}}, \bibinfo {author} {\bibfnamefont {A.}~\bibnamefont {Prasad}},
  \bibinfo {author} {\bibfnamefont {E.}~\bibnamefont {Jossou}}, \bibinfo
  {author} {\bibfnamefont {J.}~\bibnamefont {Ranasinghe}}, \bibinfo {author}
  {\bibfnamefont {B.}~\bibnamefont {Szpunar}}, \bibinfo {author} {\bibfnamefont
  {L.}~\bibnamefont {Bichler}},\ and\ \bibinfo {author} {\bibfnamefont
  {J.}~\bibnamefont {Szpunar}},\ }\bibfield  {title} {\bibinfo {title} {Thermal
  conductivity of bulk and porous \uppercase{T}h\uppercase{O}$_2$: Atomistic
  and experimental study},\ }\href
  {https://doi.org/https://doi.org/10.1016/j.jallcom.2019.05.274} {\bibfield
  {journal} {\bibinfo  {journal} {Journal of Alloys and Compounds}\ }\textbf
  {\bibinfo {volume} {798}},\ \bibinfo {pages} {507} (\bibinfo {year}
  {2019})}\BibitemShut {NoStop}%
\bibitem [{\citenamefont {Jin}\ \emph {et~al.}(2021)\citenamefont {Jin},
  \citenamefont {Khafizov}, \citenamefont {Jiang}, \citenamefont {Zhou},
  \citenamefont {Marianetti}, \citenamefont {Bryan}, \citenamefont {Manley},\
  and\ \citenamefont {Hurley}}]{Jin_2021}%
  \BibitemOpen
  \bibfield  {author} {\bibinfo {author} {\bibfnamefont {M.}~\bibnamefont
  {Jin}}, \bibinfo {author} {\bibfnamefont {M.}~\bibnamefont {Khafizov}},
  \bibinfo {author} {\bibfnamefont {C.}~\bibnamefont {Jiang}}, \bibinfo
  {author} {\bibfnamefont {S.}~\bibnamefont {Zhou}}, \bibinfo {author}
  {\bibfnamefont {C.~A.}\ \bibnamefont {Marianetti}}, \bibinfo {author}
  {\bibfnamefont {M.~S.}\ \bibnamefont {Bryan}}, \bibinfo {author}
  {\bibfnamefont {M.~E.}\ \bibnamefont {Manley}},\ and\ \bibinfo {author}
  {\bibfnamefont {D.~H.}\ \bibnamefont {Hurley}},\ }\bibfield  {title}
  {\bibinfo {title} {Assessment of empirical interatomic potential to predict
  thermal conductivity in \uppercase{T}h\uppercase{O}$_2$ and
  \uppercase{U}\uppercase{O}$_2$},\ }\href
  {https://doi.org/10.1088/1361-648x/abdc8f} {\bibfield  {journal} {\bibinfo
  {journal} {Journal of Physics: Condensed Matter}\ }\textbf {\bibinfo {volume}
  {33}},\ \bibinfo {pages} {275402} (\bibinfo {year} {2021})}\BibitemShut
  {NoStop}%
\bibitem [{\citenamefont {Xiao}\ \emph {et~al.}(2022)\citenamefont {Xiao},
  \citenamefont {Ma}, \citenamefont {Bryan}, \citenamefont {Fu}, \citenamefont
  {Mann}, \citenamefont {Winn}, \citenamefont {Abernathy}, \citenamefont
  {Hermann}, \citenamefont {Khanolkar}, \citenamefont {Dennett}, \citenamefont
  {Hurley}, \citenamefont {Manley},\ and\ \citenamefont
  {Marianetti}}]{Enda_2022_prb}%
  \BibitemOpen
  \bibfield  {author} {\bibinfo {author} {\bibfnamefont {E.}~\bibnamefont
  {Xiao}}, \bibinfo {author} {\bibfnamefont {H.}~\bibnamefont {Ma}}, \bibinfo
  {author} {\bibfnamefont {M.~S.}\ \bibnamefont {Bryan}}, \bibinfo {author}
  {\bibfnamefont {L.}~\bibnamefont {Fu}}, \bibinfo {author} {\bibfnamefont
  {J.~M.}\ \bibnamefont {Mann}}, \bibinfo {author} {\bibfnamefont
  {B.}~\bibnamefont {Winn}}, \bibinfo {author} {\bibfnamefont {D.~L.}\
  \bibnamefont {Abernathy}}, \bibinfo {author} {\bibfnamefont {R.~P.}\
  \bibnamefont {Hermann}}, \bibinfo {author} {\bibfnamefont {A.~R.}\
  \bibnamefont {Khanolkar}}, \bibinfo {author} {\bibfnamefont {C.~A.}\
  \bibnamefont {Dennett}}, \bibinfo {author} {\bibfnamefont {D.~H.}\
  \bibnamefont {Hurley}}, \bibinfo {author} {\bibfnamefont {M.~E.}\
  \bibnamefont {Manley}},\ and\ \bibinfo {author} {\bibfnamefont {C.~A.}\
  \bibnamefont {Marianetti}},\ }\bibfield  {title} {\bibinfo {title}
  {Validating first-principles phonon lifetimes via inelastic neutron
  scattering},\ }\href {https://doi.org/10.1103/PhysRevB.106.144310} {\bibfield
   {journal} {\bibinfo  {journal} {Phys. Rev. B}\ }\textbf {\bibinfo {volume}
  {106}},\ \bibinfo {pages} {144310} (\bibinfo {year} {2022})}\BibitemShut
  {NoStop}%
\bibitem [{\citenamefont {Mann}\ \emph {et~al.}(2010)\citenamefont {Mann},
  \citenamefont {Thompson}, \citenamefont {Serivalsatit}, \citenamefont
  {Tritt}, \citenamefont {Ballato},\ and\ \citenamefont {Kolis}}]{Mann_2010}%
  \BibitemOpen
  \bibfield  {author} {\bibinfo {author} {\bibfnamefont {M.}~\bibnamefont
  {Mann}}, \bibinfo {author} {\bibfnamefont {D.}~\bibnamefont {Thompson}},
  \bibinfo {author} {\bibfnamefont {K.}~\bibnamefont {Serivalsatit}}, \bibinfo
  {author} {\bibfnamefont {T.~M.}\ \bibnamefont {Tritt}}, \bibinfo {author}
  {\bibfnamefont {J.}~\bibnamefont {Ballato}},\ and\ \bibinfo {author}
  {\bibfnamefont {J.}~\bibnamefont {Kolis}},\ }\bibfield  {title} {\bibinfo
  {title} {Hydrothermal growth and thermal property characterization of
  \uppercase{T}h\uppercase{O}$_2$ single crystals},\ }\href
  {https://doi.org/10.1021/cg901308f} {\bibfield  {journal} {\bibinfo
  {journal} {Crystal Growth \& Design}\ }\textbf {\bibinfo {volume} {10}},\
  \bibinfo {pages} {2146} (\bibinfo {year} {2010})}\BibitemShut {NoStop}%
\bibitem [{\citenamefont {Sun}\ \emph {et~al.}(2015)\citenamefont {Sun},
  \citenamefont {Ruzsinszky},\ and\ \citenamefont {Perdew}}]{Sun_2015_SCAN}%
  \BibitemOpen
  \bibfield  {author} {\bibinfo {author} {\bibfnamefont {J.}~\bibnamefont
  {Sun}}, \bibinfo {author} {\bibfnamefont {A.}~\bibnamefont {Ruzsinszky}},\
  and\ \bibinfo {author} {\bibfnamefont {J.~P.}\ \bibnamefont {Perdew}},\
  }\bibfield  {title} {\bibinfo {title} {Strongly constrained and appropriately
  normed semilocal density functional},\ }\href
  {https://doi.org/10.1103/PhysRevLett.115.036402} {\bibfield  {journal}
  {\bibinfo  {journal} {Phys. Rev. Lett.}\ }\textbf {\bibinfo {volume} {115}},\
  \bibinfo {pages} {036402} (\bibinfo {year} {2015})}\BibitemShut {NoStop}%
\bibitem [{\citenamefont {Ferrigno}\ \emph {et~al.}(2023)\citenamefont
  {Ferrigno}, \citenamefont {Adnan},\ and\ \citenamefont
  {Khafizov}}]{FERRIGNO_2023}%
  \BibitemOpen
  \bibfield  {author} {\bibinfo {author} {\bibfnamefont {J.}~\bibnamefont
  {Ferrigno}}, \bibinfo {author} {\bibfnamefont {S.}~\bibnamefont {Adnan}},\
  and\ \bibinfo {author} {\bibfnamefont {M.}~\bibnamefont {Khafizov}},\
  }\bibfield  {title} {\bibinfo {title} {Influence of point defect accumulation
  on in-pile thermal conductivity degradation: Fuel rod defect distribution and
  deviation between in-pile and post irradiation thermal conductivity},\ }\href
  {https://doi.org/https://doi.org/10.1016/j.jnucmat.2022.154108} {\bibfield
  {journal} {\bibinfo  {journal} {Journal of Nuclear Materials}\ }\textbf
  {\bibinfo {volume} {573}},\ \bibinfo {pages} {154108} (\bibinfo {year}
  {2023})}\BibitemShut {NoStop}%
\bibitem [{\citenamefont {Khafizov}\ \emph {et~al.}(2017)\citenamefont
  {Khafizov}, \citenamefont {Chauhan}, \citenamefont {Wang}, \citenamefont
  {Riyad}, \citenamefont {Hang},\ and\ \citenamefont
  {Hurley}}]{Khafizov_2016_investigation}%
  \BibitemOpen
  \bibfield  {author} {\bibinfo {author} {\bibfnamefont {M.}~\bibnamefont
  {Khafizov}}, \bibinfo {author} {\bibfnamefont {V.}~\bibnamefont {Chauhan}},
  \bibinfo {author} {\bibfnamefont {Y.}~\bibnamefont {Wang}}, \bibinfo {author}
  {\bibfnamefont {F.}~\bibnamefont {Riyad}}, \bibinfo {author} {\bibfnamefont
  {N.}~\bibnamefont {Hang}},\ and\ \bibinfo {author} {\bibfnamefont
  {D.}~\bibnamefont {Hurley}},\ }\bibfield  {title} {\bibinfo {title}
  {Investigation of thermal transport in composites and ion beam irradiated
  materials for nuclear energy applications},\ }\href
  {https://doi.org/10.1557/jmr.2016.421} {\bibfield  {journal} {\bibinfo
  {journal} {Journal of Materials Research}\ }\textbf {\bibinfo {volume}
  {32}},\ \bibinfo {pages} {204} (\bibinfo {year} {2017})}\BibitemShut
  {NoStop}%
\bibitem [{\citenamefont {Dennett}\ \emph {et~al.}(2020)\citenamefont
  {Dennett}, \citenamefont {Hua}, \citenamefont {Khanolkar}, \citenamefont
  {Yao}, \citenamefont {Morgan}, \citenamefont {Prusnick}, \citenamefont
  {Poudel}, \citenamefont {French}, \citenamefont {Gofryk}, \citenamefont {He},
  \citenamefont {Shao}, \citenamefont {Khafizov}, \citenamefont {Turner},
  \citenamefont {Mann},\ and\ \citenamefont {Hurley}}]{Dennett_2020}%
  \BibitemOpen
  \bibfield  {author} {\bibinfo {author} {\bibfnamefont {C.~A.}\ \bibnamefont
  {Dennett}}, \bibinfo {author} {\bibfnamefont {Z.}~\bibnamefont {Hua}},
  \bibinfo {author} {\bibfnamefont {A.}~\bibnamefont {Khanolkar}}, \bibinfo
  {author} {\bibfnamefont {T.}~\bibnamefont {Yao}}, \bibinfo {author}
  {\bibfnamefont {P.~K.}\ \bibnamefont {Morgan}}, \bibinfo {author}
  {\bibfnamefont {T.~A.}\ \bibnamefont {Prusnick}}, \bibinfo {author}
  {\bibfnamefont {N.}~\bibnamefont {Poudel}}, \bibinfo {author} {\bibfnamefont
  {A.}~\bibnamefont {French}}, \bibinfo {author} {\bibfnamefont
  {K.}~\bibnamefont {Gofryk}}, \bibinfo {author} {\bibfnamefont
  {L.}~\bibnamefont {He}}, \bibinfo {author} {\bibfnamefont {L.}~\bibnamefont
  {Shao}}, \bibinfo {author} {\bibfnamefont {M.}~\bibnamefont {Khafizov}},
  \bibinfo {author} {\bibfnamefont {D.~B.}\ \bibnamefont {Turner}}, \bibinfo
  {author} {\bibfnamefont {J.~M.}\ \bibnamefont {Mann}},\ and\ \bibinfo
  {author} {\bibfnamefont {D.~H.}\ \bibnamefont {Hurley}},\ }\bibfield  {title}
  {\bibinfo {title} {{The influence of lattice defects, recombination, and
  clustering on thermal transport in single crystal thorium dioxide}},\ }\href
  {https://doi.org/10.1063/5.0025384} {\bibfield  {journal} {\bibinfo
  {journal} {APL Materials}\ }\textbf {\bibinfo {volume} {8}},\ \bibinfo
  {pages} {111103} (\bibinfo {year} {2020})},\ \Eprint
  {https://arxiv.org/abs/https://pubs.aip.org/aip/apm/article-pdf/doi/10.1063/5.0025384/14564694/111103\_1\_online.pdf}
  {https://pubs.aip.org/aip/apm/article-pdf/doi/10.1063/5.0025384/14564694/111103\_1\_online.pdf}
  \BibitemShut {NoStop}%
\bibitem [{\citenamefont {Cooper}\ \emph {et~al.}(2015)\citenamefont {Cooper},
  \citenamefont {Middleburgh},\ and\ \citenamefont {Grimes}}]{COOPER201529}%
  \BibitemOpen
  \bibfield  {author} {\bibinfo {author} {\bibfnamefont {M.}~\bibnamefont
  {Cooper}}, \bibinfo {author} {\bibfnamefont {S.}~\bibnamefont
  {Middleburgh}},\ and\ \bibinfo {author} {\bibfnamefont {R.}~\bibnamefont
  {Grimes}},\ }\bibfield  {title} {\bibinfo {title} {Modelling the thermal
  conductivity of \uppercase{U}$_x$\uppercase{T}h$_{1-x}$\uppercase{O}$_2$ and
  \uppercase{U}$_x$\uppercase{P}u$_{1-x}$\uppercase{O}$_2$},\ }\href
  {https://doi.org/https://doi.org/10.1016/j.jnucmat.2015.07.022} {\bibfield
  {journal} {\bibinfo  {journal} {Journal of Nuclear Materials}\ }\textbf
  {\bibinfo {volume} {466}},\ \bibinfo {pages} {29} (\bibinfo {year}
  {2015})}\BibitemShut {NoStop}%
\bibitem [{\citenamefont {Park}\ \emph {et~al.}(2018)\citenamefont {Park},
  \citenamefont {Farfán}, \citenamefont {Mitchell}, \citenamefont {Resnick},
  \citenamefont {Enriquez},\ and\ \citenamefont {Yee}}]{PARK2018198}%
  \BibitemOpen
  \bibfield  {author} {\bibinfo {author} {\bibfnamefont {J.}~\bibnamefont
  {Park}}, \bibinfo {author} {\bibfnamefont {E.~B.}\ \bibnamefont {Farfán}},
  \bibinfo {author} {\bibfnamefont {K.}~\bibnamefont {Mitchell}}, \bibinfo
  {author} {\bibfnamefont {A.}~\bibnamefont {Resnick}}, \bibinfo {author}
  {\bibfnamefont {C.}~\bibnamefont {Enriquez}},\ and\ \bibinfo {author}
  {\bibfnamefont {T.}~\bibnamefont {Yee}},\ }\bibfield  {title} {\bibinfo
  {title} {Sensitivity of thermal transport in thorium dioxide to defects},\
  }\href {https://doi.org/https://doi.org/10.1016/j.jnucmat.2018.03.043}
  {\bibfield  {journal} {\bibinfo  {journal} {Journal of Nuclear Materials}\
  }\textbf {\bibinfo {volume} {504}},\ \bibinfo {pages} {198} (\bibinfo {year}
  {2018})}\BibitemShut {NoStop}%
\bibitem [{\citenamefont {Rahman}\ \emph {et~al.}(2020)\citenamefont {Rahman},
  \citenamefont {Szpunar},\ and\ \citenamefont {Szpunar}}]{RAHMAN2020152050}%
  \BibitemOpen
  \bibfield  {author} {\bibinfo {author} {\bibfnamefont {M.}~\bibnamefont
  {Rahman}}, \bibinfo {author} {\bibfnamefont {B.}~\bibnamefont {Szpunar}},\
  and\ \bibinfo {author} {\bibfnamefont {J.}~\bibnamefont {Szpunar}},\
  }\bibfield  {title} {\bibinfo {title} {Dependence of thermal conductivity on
  fission-product defects and vacancy concentration in thorium dioxide},\
  }\href {https://doi.org/https://doi.org/10.1016/j.jnucmat.2020.152050}
  {\bibfield  {journal} {\bibinfo  {journal} {Journal of Nuclear Materials}\
  }\textbf {\bibinfo {volume} {532}},\ \bibinfo {pages} {152050} (\bibinfo
  {year} {2020})}\BibitemShut {NoStop}%
\bibitem [{\citenamefont {Jin}\ \emph {et~al.}(2022)\citenamefont {Jin},
  \citenamefont {Dennett}, \citenamefont {Hurley},\ and\ \citenamefont
  {Khafizov}}]{JIN2022Impact}%
  \BibitemOpen
  \bibfield  {author} {\bibinfo {author} {\bibfnamefont {M.}~\bibnamefont
  {Jin}}, \bibinfo {author} {\bibfnamefont {C.~A.}\ \bibnamefont {Dennett}},
  \bibinfo {author} {\bibfnamefont {D.~H.}\ \bibnamefont {Hurley}},\ and\
  \bibinfo {author} {\bibfnamefont {M.}~\bibnamefont {Khafizov}},\ }\bibfield
  {title} {\bibinfo {title} {Impact of small defects and dislocation loops on
  phonon scattering and thermal transport in \uppercase{T}h\uppercase{O}$_2$},\
  }\href {https://doi.org/https://doi.org/10.1016/j.jnucmat.2022.153758}
  {\bibfield  {journal} {\bibinfo  {journal} {Journal of Nuclear Materials}\ ,\
  \bibinfo {pages} {153758}} (\bibinfo {year} {2022})}\BibitemShut {NoStop}%
\bibitem [{\citenamefont {Zhou}\ \emph {et~al.}(2018)\citenamefont {Zhou},
  \citenamefont {Fan}, \citenamefont {Qin}, \citenamefont {Yang}, \citenamefont
  {Ouyang},\ and\ \citenamefont {Hu}}]{Zhou_2018}%
  \BibitemOpen
  \bibfield  {author} {\bibinfo {author} {\bibfnamefont {Y.}~\bibnamefont
  {Zhou}}, \bibinfo {author} {\bibfnamefont {Z.}~\bibnamefont {Fan}}, \bibinfo
  {author} {\bibfnamefont {G.}~\bibnamefont {Qin}}, \bibinfo {author}
  {\bibfnamefont {J.-Y.}\ \bibnamefont {Yang}}, \bibinfo {author}
  {\bibfnamefont {T.}~\bibnamefont {Ouyang}},\ and\ \bibinfo {author}
  {\bibfnamefont {M.}~\bibnamefont {Hu}},\ }\bibfield  {title} {\bibinfo
  {title} {Methodology perspective of computing thermal transport in
  low-dimensional materials and nanostructures: The old and the new},\ }\href
  {https://doi.org/10.1021/acsomega.7b01594} {\bibfield  {journal} {\bibinfo
  {journal} {ACS Omega}\ }\textbf {\bibinfo {volume} {3}},\ \bibinfo {pages}
  {3278} (\bibinfo {year} {2018})}\BibitemShut {NoStop}%
\bibitem [{Zim(2001)}]{Ziman_2001}%
  \BibitemOpen
  \href {https://doi.org/10.1093/acprof:oso/9780198507796.001.0001} {\emph
  {\bibinfo {title} {Electrons and Phonons: The Theory of Transport Phenomena
  in Solids.}}},\ Oxford University Press\ (\bibinfo  {publisher} {Oxford
  University Press},\ \bibinfo {address} {University of Bristol},\ \bibinfo
  {year} {2001})\BibitemShut {NoStop}%
\bibitem [{\citenamefont {Kohn}\ and\ \citenamefont {Sham}(1965)}]{Kohn_1965}%
  \BibitemOpen
  \bibfield  {author} {\bibinfo {author} {\bibfnamefont {W.}~\bibnamefont
  {Kohn}}\ and\ \bibinfo {author} {\bibfnamefont {L.~J.}\ \bibnamefont
  {Sham}},\ }\bibfield  {title} {\bibinfo {title} {Self-consistent equations
  including exchange and correlation effects},\ }\href
  {https://doi.org/10.1103/PhysRev.140.A1133} {\bibfield  {journal} {\bibinfo
  {journal} {Phys. Rev.}\ }\textbf {\bibinfo {volume} {140}},\ \bibinfo {pages}
  {A1133} (\bibinfo {year} {1965})}\BibitemShut {NoStop}%
\bibitem [{\citenamefont {Bl\"ochl}(1994)}]{blochl_1994}%
  \BibitemOpen
  \bibfield  {author} {\bibinfo {author} {\bibfnamefont {P.~E.}\ \bibnamefont
  {Bl\"ochl}},\ }\bibfield  {title} {\bibinfo {title} {Projector augmented-wave
  method},\ }\href {https://doi.org/10.1103/PhysRevB.50.17953} {\bibfield
  {journal} {\bibinfo  {journal} {Phys. Rev. B}\ }\textbf {\bibinfo {volume}
  {50}},\ \bibinfo {pages} {17953} (\bibinfo {year} {1994})}\BibitemShut
  {NoStop}%
\bibitem [{\citenamefont {Kresse}\ and\ \citenamefont
  {Furthm\"uller}(1996)}]{Kresse_1996}%
  \BibitemOpen
  \bibfield  {author} {\bibinfo {author} {\bibfnamefont {G.}~\bibnamefont
  {Kresse}}\ and\ \bibinfo {author} {\bibfnamefont {J.}~\bibnamefont
  {Furthm\"uller}},\ }\bibfield  {title} {\bibinfo {title} {Efficient iterative
  schemes for ab initio total-energy calculations using a plane-wave basis
  set},\ }\href {https://doi.org/10.1103/PhysRevB.54.11169} {\bibfield
  {journal} {\bibinfo  {journal} {Phys. Rev. B}\ }\textbf {\bibinfo {volume}
  {54}},\ \bibinfo {pages} {11169} (\bibinfo {year} {1996})}\BibitemShut
  {NoStop}%
\bibitem [{\citenamefont {Perdew}\ and\ \citenamefont
  {Zunger}(1981)}]{Perdew_1981}%
  \BibitemOpen
  \bibfield  {author} {\bibinfo {author} {\bibfnamefont {J.~P.}\ \bibnamefont
  {Perdew}}\ and\ \bibinfo {author} {\bibfnamefont {A.}~\bibnamefont
  {Zunger}},\ }\bibfield  {title} {\bibinfo {title} {Self-interaction
  correction to density-functional approximations for many-electron systems},\
  }\href {https://doi.org/10.1103/PhysRevB.23.5048} {\bibfield  {journal}
  {\bibinfo  {journal} {Phys. Rev. B}\ }\textbf {\bibinfo {volume} {23}},\
  \bibinfo {pages} {5048} (\bibinfo {year} {1981})}\BibitemShut {NoStop}%
\bibitem [{\citenamefont {Dudarev}\ \emph {et~al.}(1998)\citenamefont
  {Dudarev}, \citenamefont {Botton}, \citenamefont {Savrasov}, \citenamefont
  {Humphreys},\ and\ \citenamefont {Sutton}}]{Dudarev_1988_HubbardU}%
  \BibitemOpen
  \bibfield  {author} {\bibinfo {author} {\bibfnamefont {S.~L.}\ \bibnamefont
  {Dudarev}}, \bibinfo {author} {\bibfnamefont {G.~A.}\ \bibnamefont {Botton}},
  \bibinfo {author} {\bibfnamefont {S.~Y.}\ \bibnamefont {Savrasov}}, \bibinfo
  {author} {\bibfnamefont {C.~J.}\ \bibnamefont {Humphreys}},\ and\ \bibinfo
  {author} {\bibfnamefont {A.~P.}\ \bibnamefont {Sutton}},\ }\bibfield  {title}
  {\bibinfo {title} {Electron-energy-loss spectra and the structural stability
  of nickel oxide: An lsda+u study},\ }\href
  {https://doi.org/10.1103/PhysRevB.57.1505} {\bibfield  {journal} {\bibinfo
  {journal} {Phys. Rev. B}\ }\textbf {\bibinfo {volume} {57}},\ \bibinfo
  {pages} {1505} (\bibinfo {year} {1998})}\BibitemShut {NoStop}%
\bibitem [{\citenamefont {Shields}\ \emph {et~al.}(2016)\citenamefont
  {Shields}, \citenamefont {Santos-Carballal},\ and\ \citenamefont {{de
  Leeuw}}}]{SHIELDS201699}%
  \BibitemOpen
  \bibfield  {author} {\bibinfo {author} {\bibfnamefont {A.~E.}\ \bibnamefont
  {Shields}}, \bibinfo {author} {\bibfnamefont {D.}~\bibnamefont
  {Santos-Carballal}},\ and\ \bibinfo {author} {\bibfnamefont {N.~H.}\
  \bibnamefont {{de Leeuw}}},\ }\bibfield  {title} {\bibinfo {title} {A density
  functional theory study of uranium-doped thoria and uranium adatoms on the
  major surfaces of thorium dioxide},\ }\href
  {https://doi.org/https://doi.org/10.1016/j.jnucmat.2016.02.009} {\bibfield
  {journal} {\bibinfo  {journal} {Journal of Nuclear Materials}\ }\textbf
  {\bibinfo {volume} {473}},\ \bibinfo {pages} {99} (\bibinfo {year}
  {2016})}\BibitemShut {NoStop}%
\bibitem [{\citenamefont {Lu}\ \emph {et~al.}(2012)\citenamefont {Lu},
  \citenamefont {Yang},\ and\ \citenamefont {Zhang}}]{Lu_2012}%
  \BibitemOpen
  \bibfield  {author} {\bibinfo {author} {\bibfnamefont {Y.}~\bibnamefont
  {Lu}}, \bibinfo {author} {\bibfnamefont {Y.}~\bibnamefont {Yang}},\ and\
  \bibinfo {author} {\bibfnamefont {P.}~\bibnamefont {Zhang}},\ }\bibfield
  {title} {\bibinfo {title} {Thermodynamic properties and structural stability
  of thorium dioxide},\ }\href {https://doi.org/10.1088/0953-8984/24/22/225801}
  {\bibfield  {journal} {\bibinfo  {journal} {Journal of Physics: Condensed
  Matter}\ }\textbf {\bibinfo {volume} {24}},\ \bibinfo {pages} {225801}
  (\bibinfo {year} {2012})}\BibitemShut {NoStop}%
\bibitem [{\citenamefont {Idiri}\ \emph {et~al.}(2004)\citenamefont {Idiri},
  \citenamefont {Le~Bihan}, \citenamefont {Heathman},\ and\ \citenamefont
  {Rebizant}}]{idiri_2004}%
  \BibitemOpen
  \bibfield  {author} {\bibinfo {author} {\bibfnamefont {M.}~\bibnamefont
  {Idiri}}, \bibinfo {author} {\bibfnamefont {T.}~\bibnamefont {Le~Bihan}},
  \bibinfo {author} {\bibfnamefont {S.}~\bibnamefont {Heathman}},\ and\
  \bibinfo {author} {\bibfnamefont {J.}~\bibnamefont {Rebizant}},\ }\bibfield
  {title} {\bibinfo {title} {Behavior of actinide dioxides under pressure:
  \uppercase{U}\uppercase{O}$_2$ and \uppercase{T}h\uppercase{O}$_2$},\ }\href
  {https://doi.org/10.1103/PhysRevB.70.014113} {\bibfield  {journal} {\bibinfo
  {journal} {Phys. Rev. B}\ }\textbf {\bibinfo {volume} {70}},\ \bibinfo
  {pages} {014113} (\bibinfo {year} {2004})}\BibitemShut {NoStop}%
\bibitem [{\citenamefont {Togo}\ and\ \citenamefont {Tanaka}(2015)}]{TOGO2015}%
  \BibitemOpen
  \bibfield  {author} {\bibinfo {author} {\bibfnamefont {A.}~\bibnamefont
  {Togo}}\ and\ \bibinfo {author} {\bibfnamefont {I.}~\bibnamefont {Tanaka}},\
  }\bibfield  {title} {\bibinfo {title} {First principles phonon calculations
  in materials science},\ }\href
  {https://doi.org/https://doi.org/10.1016/j.scriptamat.2015.07.021} {\bibfield
   {journal} {\bibinfo  {journal} {Scripta Materialia}\ }\textbf {\bibinfo
  {volume} {108}},\ \bibinfo {pages} {1} (\bibinfo {year} {2015})}\BibitemShut
  {NoStop}%
\bibitem [{\citenamefont {Matthew}\ \emph {et~al.}(2020)\citenamefont
  {Matthew}, \citenamefont {Lyuwen}, \citenamefont {Rickert}, \citenamefont
  {David}, \citenamefont {Prusnick}, \citenamefont {Matthew}, \citenamefont
  {Douglas}, \citenamefont {Chris},\ and\ \citenamefont {Michael}}]{Bryan2020}%
  \BibitemOpen
  \bibfield  {author} {\bibinfo {author} {\bibfnamefont {B.}~\bibnamefont
  {Matthew}}, \bibinfo {author} {\bibfnamefont {F.}~\bibnamefont {Lyuwen}},
  \bibinfo {author} {\bibfnamefont {K.}~\bibnamefont {Rickert}}, \bibinfo
  {author} {\bibfnamefont {T.}~\bibnamefont {David}}, \bibinfo {author}
  {\bibfnamefont {T.~A.}\ \bibnamefont {Prusnick}}, \bibinfo {author}
  {\bibfnamefont {M.}~\bibnamefont {Matthew}}, \bibinfo {author} {\bibfnamefont
  {A.}~\bibnamefont {Douglas}}, \bibinfo {author} {\bibfnamefont
  {M.}~\bibnamefont {Chris}},\ and\ \bibinfo {author} {\bibfnamefont
  {M.}~\bibnamefont {Michael}},\ }\bibfield  {title} {\bibinfo {title}
  {Nonlinear propagating modes beyond the phonons in fluorite-structured
  crystals},\ }\href {https://doi.org/10.1038/s42005-020-00483-2} {\bibfield
  {journal} {\bibinfo  {journal} {Communications Physics}\ }\textbf {\bibinfo
  {volume} {3}},\ \bibinfo {pages} {217} (\bibinfo {year} {2020})}\BibitemShut
  {NoStop}%
\bibitem [{\citenamefont {Li}\ \emph {et~al.}(2014)\citenamefont {Li},
  \citenamefont {Carrete}, \citenamefont {{A. Katcho}},\ and\ \citenamefont
  {Mingo}}]{shengBTE_2014}%
  \BibitemOpen
  \bibfield  {author} {\bibinfo {author} {\bibfnamefont {W.}~\bibnamefont
  {Li}}, \bibinfo {author} {\bibfnamefont {J.}~\bibnamefont {Carrete}},
  \bibinfo {author} {\bibfnamefont {N.}~\bibnamefont {{A. Katcho}}},\ and\
  \bibinfo {author} {\bibfnamefont {N.}~\bibnamefont {Mingo}},\ }\bibfield
  {title} {\bibinfo {title} {Shengbte: A solver of the boltzmann transport
  equation for phonons},\ }\href
  {https://doi.org/https://doi.org/10.1016/j.cpc.2014.02.015} {\bibfield
  {journal} {\bibinfo  {journal} {Computer Physics Communications}\ }\textbf
  {\bibinfo {volume} {185}},\ \bibinfo {pages} {1747} (\bibinfo {year}
  {2014})}\BibitemShut {NoStop}%
\bibitem [{\citenamefont {Yi}\ \emph {et~al.}(2016)\citenamefont {Yi},
  \citenamefont {Shun-Li}, \citenamefont {Huazhi}, \citenamefont {Zi-Kui},\
  and\ \citenamefont {Qing}}]{Wang2016}%
  \BibitemOpen
  \bibfield  {author} {\bibinfo {author} {\bibfnamefont {W.}~\bibnamefont
  {Yi}}, \bibinfo {author} {\bibfnamefont {S.}~\bibnamefont {Shun-Li}},
  \bibinfo {author} {\bibfnamefont {F.}~\bibnamefont {Huazhi}}, \bibinfo
  {author} {\bibfnamefont {L.}~\bibnamefont {Zi-Kui}},\ and\ \bibinfo {author}
  {\bibfnamefont {C.~L.}\ \bibnamefont {Qing}},\ }\bibfield  {title} {\bibinfo
  {title} {First-principles calculations of lattice dynamics and thermal
  properties of polar solids},\ }\href
  {https://doi.org/10.1038/npjcompumats.2016.6} {\bibfield  {journal} {\bibinfo
   {journal} {npj Computational Materials}\ }\textbf {\bibinfo {volume} {2}},\
  \bibinfo {pages} {16006} (\bibinfo {year} {2016})}\BibitemShut {NoStop}%
\bibitem [{\citenamefont {Baroni}\ \emph {et~al.}(2001)\citenamefont {Baroni},
  \citenamefont {De~Gironcoli}, \citenamefont {Dal~Corso},\ and\ \citenamefont
  {Giannozzi}}]{Baroni_2001_Phonons}%
  \BibitemOpen
  \bibfield  {author} {\bibinfo {author} {\bibfnamefont {S.}~\bibnamefont
  {Baroni}}, \bibinfo {author} {\bibfnamefont {S.}~\bibnamefont
  {De~Gironcoli}}, \bibinfo {author} {\bibfnamefont {A.}~\bibnamefont
  {Dal~Corso}},\ and\ \bibinfo {author} {\bibfnamefont {P.}~\bibnamefont
  {Giannozzi}},\ }\bibfield  {title} {\bibinfo {title} {Phonons and related
  crystal properties from density-functional perturbation theory},\ }\href
  {https://doi.org/10.1103/revmodphys.73.515} {\bibfield  {journal} {\bibinfo
  {journal} {Reviews of Modern Physics}\ }\textbf {\bibinfo {volume} {73}},\
  \bibinfo {pages} {515–562} (\bibinfo {year} {2001})}\BibitemShut {NoStop}%
\bibitem [{\citenamefont {Carrete}\ \emph {et~al.}(2017)\citenamefont
  {Carrete}, \citenamefont {Vermeersch}, \citenamefont {Katre}, \citenamefont
  {{van Roekeghem}}, \citenamefont {Wang}, \citenamefont {Madsen},\ and\
  \citenamefont {Mingo}}]{CARRETE2017}%
  \BibitemOpen
  \bibfield  {author} {\bibinfo {author} {\bibfnamefont {J.}~\bibnamefont
  {Carrete}}, \bibinfo {author} {\bibfnamefont {B.}~\bibnamefont {Vermeersch}},
  \bibinfo {author} {\bibfnamefont {A.}~\bibnamefont {Katre}}, \bibinfo
  {author} {\bibfnamefont {A.}~\bibnamefont {{van Roekeghem}}}, \bibinfo
  {author} {\bibfnamefont {T.}~\bibnamefont {Wang}}, \bibinfo {author}
  {\bibfnamefont {G.~K.}\ \bibnamefont {Madsen}},\ and\ \bibinfo {author}
  {\bibfnamefont {N.}~\bibnamefont {Mingo}},\ }\bibfield  {title} {\bibinfo
  {title} {alma\uppercase{BTE} : A solver of the space–time dependent
  \uppercase{B}oltzmann transport equation for phonons in structured
  materials},\ }\href
  {https://doi.org/https://doi.org/10.1016/j.cpc.2017.06.023} {\bibfield
  {journal} {\bibinfo  {journal} {Computer Physics Communications}\ }\textbf
  {\bibinfo {volume} {220}},\ \bibinfo {pages} {351} (\bibinfo {year}
  {2017})}\BibitemShut {NoStop}%
\bibitem [{\citenamefont {Iwasawa}\ \emph {et~al.}(2009)\citenamefont
  {Iwasawa}, \citenamefont {Ohnuma}, \citenamefont {Chen}, \citenamefont
  {Kaneta}, \citenamefont {Geng}, \citenamefont {Iwase},\ and\ \citenamefont
  {Kinoshita}}]{Misako_2009}%
  \BibitemOpen
  \bibfield  {author} {\bibinfo {author} {\bibfnamefont {M.}~\bibnamefont
  {Iwasawa}}, \bibinfo {author} {\bibfnamefont {T.}~\bibnamefont {Ohnuma}},
  \bibinfo {author} {\bibfnamefont {Y.}~\bibnamefont {Chen}}, \bibinfo {author}
  {\bibfnamefont {Y.}~\bibnamefont {Kaneta}}, \bibinfo {author} {\bibfnamefont
  {H.}~\bibnamefont {Geng}}, \bibinfo {author} {\bibfnamefont {A.}~\bibnamefont
  {Iwase}},\ and\ \bibinfo {author} {\bibfnamefont {M.}~\bibnamefont
  {Kinoshita}},\ }\bibfield  {title} {\bibinfo {title} {First-principles study
  on cerium ion behavior in irradiated cerium dioxide},\ }\href
  {https://doi.org/10.1016/j.jnucmat.2009.06.026} {\bibfield  {journal}
  {\bibinfo  {journal} {Journal of Nuclear Materials}\ }\textbf {\bibinfo
  {volume} {393}},\ \bibinfo {pages} {321} (\bibinfo {year}
  {2009})}\BibitemShut {NoStop}%
\bibitem [{\citenamefont {Dorado}\ \emph {et~al.}(2009)\citenamefont {Dorado},
  \citenamefont {Freyss},\ and\ \citenamefont {Martin}}]{Dorado2009}%
  \BibitemOpen
  \bibfield  {author} {\bibinfo {author} {\bibfnamefont {B.}~\bibnamefont
  {Dorado}}, \bibinfo {author} {\bibfnamefont {M.}~\bibnamefont {Freyss}},\
  and\ \bibinfo {author} {\bibfnamefont {G.}~\bibnamefont {Martin}},\
  }\bibfield  {title} {\bibinfo {title} {\uppercase{GGA+U} study of the
  incorporation of iodine in uranium dioxide},\ }\href
  {https://doi.org/10.1140/epjb/e2009-00145-0} {\bibfield  {journal} {\bibinfo
  {journal} {The European Physical Journal B}\ }\textbf {\bibinfo {volume}
  {69}},\ \bibinfo {pages} {203} (\bibinfo {year} {2009})}\BibitemShut
  {NoStop}%
\bibitem [{\citenamefont {Momma}\ and\ \citenamefont
  {Izumi}(2008)}]{Momma:ko5060}%
  \BibitemOpen
  \bibfield  {author} {\bibinfo {author} {\bibfnamefont {K.}~\bibnamefont
  {Momma}}\ and\ \bibinfo {author} {\bibfnamefont {F.}~\bibnamefont {Izumi}},\
  }\bibfield  {title} {\bibinfo {title} {{{\it VESTA}: a three-dimensional
  visualization system for electronic and structural analysis}},\ }\href
  {https://doi.org/10.1107/S0021889808012016} {\bibfield  {journal} {\bibinfo
  {journal} {Journal of Applied Crystallography}\ }\textbf {\bibinfo {volume}
  {41}},\ \bibinfo {pages} {653} (\bibinfo {year} {2008})}\BibitemShut
  {NoStop}%
\bibitem [{\citenamefont {Kundu}\ \emph {et~al.}(2019)\citenamefont {Kundu},
  \citenamefont {Otte}, \citenamefont {Carrete}, \citenamefont {Erhart},
  \citenamefont {Li}, \citenamefont {Mingo},\ and\ \citenamefont
  {Madsen}}]{kundu2019}%
  \BibitemOpen
  \bibfield  {author} {\bibinfo {author} {\bibfnamefont {A.}~\bibnamefont
  {Kundu}}, \bibinfo {author} {\bibfnamefont {F.}~\bibnamefont {Otte}},
  \bibinfo {author} {\bibfnamefont {J.}~\bibnamefont {Carrete}}, \bibinfo
  {author} {\bibfnamefont {P.}~\bibnamefont {Erhart}}, \bibinfo {author}
  {\bibfnamefont {W.}~\bibnamefont {Li}}, \bibinfo {author} {\bibfnamefont
  {N.}~\bibnamefont {Mingo}},\ and\ \bibinfo {author} {\bibfnamefont
  {G.~K.~H.}\ \bibnamefont {Madsen}},\ }\bibfield  {title} {\bibinfo {title}
  {Effect of local chemistry and structure on thermal transport in doped
  gaas},\ }\href {https://doi.org/10.1103/PhysRevMaterials.3.094602} {\bibfield
   {journal} {\bibinfo  {journal} {Phys. Rev. Mater.}\ }\textbf {\bibinfo
  {volume} {3}},\ \bibinfo {pages} {094602} (\bibinfo {year}
  {2019})}\BibitemShut {NoStop}%
\bibitem [{\citenamefont {Deskins}\ \emph {et~al.}(2022)\citenamefont
  {Deskins}, \citenamefont {Khanolkar}, \citenamefont {Mazumder}, \citenamefont
  {Dennett}, \citenamefont {Bawane}, \citenamefont {Hua}, \citenamefont
  {Ferrigno}, \citenamefont {He}, \citenamefont {Mann}, \citenamefont
  {Khafizov}, \citenamefont {Hurley},\ and\ \citenamefont
  {El-Azab}}]{DESKINS2022}%
  \BibitemOpen
  \bibfield  {author} {\bibinfo {author} {\bibfnamefont {W.~R.}\ \bibnamefont
  {Deskins}}, \bibinfo {author} {\bibfnamefont {A.}~\bibnamefont {Khanolkar}},
  \bibinfo {author} {\bibfnamefont {S.}~\bibnamefont {Mazumder}}, \bibinfo
  {author} {\bibfnamefont {C.~A.}\ \bibnamefont {Dennett}}, \bibinfo {author}
  {\bibfnamefont {K.}~\bibnamefont {Bawane}}, \bibinfo {author} {\bibfnamefont
  {Z.}~\bibnamefont {Hua}}, \bibinfo {author} {\bibfnamefont {J.}~\bibnamefont
  {Ferrigno}}, \bibinfo {author} {\bibfnamefont {L.}~\bibnamefont {He}},
  \bibinfo {author} {\bibfnamefont {J.~M.}\ \bibnamefont {Mann}}, \bibinfo
  {author} {\bibfnamefont {M.}~\bibnamefont {Khafizov}}, \bibinfo {author}
  {\bibfnamefont {D.~H.}\ \bibnamefont {Hurley}},\ and\ \bibinfo {author}
  {\bibfnamefont {A.}~\bibnamefont {El-Azab}},\ }\bibfield  {title} {\bibinfo
  {title} {A combined theoretical-experimental investigation of thermal
  transport in low-dose irradiated thorium dioxide},\ }\href
  {https://doi.org/https://doi.org/10.1016/j.actamat.2022.118379} {\bibfield
  {journal} {\bibinfo  {journal} {Acta Materialia}\ }\textbf {\bibinfo {volume}
  {241}},\ \bibinfo {pages} {118379} (\bibinfo {year} {2022})}\BibitemShut
  {NoStop}%
\bibitem [{\citenamefont {Jiang}\ \emph {et~al.}(2022)\citenamefont {Jiang},
  \citenamefont {He}, \citenamefont {Dennett}, \citenamefont {Khafizov},
  \citenamefont {Mann},\ and\ \citenamefont {Hurley}}]{JIANG2022}%
  \BibitemOpen
  \bibfield  {author} {\bibinfo {author} {\bibfnamefont {C.}~\bibnamefont
  {Jiang}}, \bibinfo {author} {\bibfnamefont {L.}~\bibnamefont {He}}, \bibinfo
  {author} {\bibfnamefont {C.~A.}\ \bibnamefont {Dennett}}, \bibinfo {author}
  {\bibfnamefont {M.}~\bibnamefont {Khafizov}}, \bibinfo {author}
  {\bibfnamefont {J.~M.}\ \bibnamefont {Mann}},\ and\ \bibinfo {author}
  {\bibfnamefont {D.~H.}\ \bibnamefont {Hurley}},\ }\bibfield  {title}
  {\bibinfo {title} {Unraveling small-scale defects in irradiated tho2 using
  kinetic monte carlo simulations},\ }\href
  {https://doi.org/https://doi.org/10.1016/j.scriptamat.2022.114684} {\bibfield
   {journal} {\bibinfo  {journal} {Scripta Materialia}\ }\textbf {\bibinfo
  {volume} {214}},\ \bibinfo {pages} {114684} (\bibinfo {year}
  {2022})}\BibitemShut {NoStop}%
\bibitem [{\citenamefont {Hua}\ \emph {et~al.}(2023)\citenamefont {Hua},
  \citenamefont {Adnan}, \citenamefont {Khanolkar}, \citenamefont {Rickert},
  \citenamefont {Turner}, \citenamefont {Prusnick}, \citenamefont {Mann},
  \citenamefont {Hurley}, \citenamefont {Khafizov},\ and\ \citenamefont
  {Dennett}}]{hua2023thermal}%
  \BibitemOpen
  \bibfield  {author} {\bibinfo {author} {\bibfnamefont {Z.}~\bibnamefont
  {Hua}}, \bibinfo {author} {\bibfnamefont {S.}~\bibnamefont {Adnan}}, \bibinfo
  {author} {\bibfnamefont {A.~R.}\ \bibnamefont {Khanolkar}}, \bibinfo {author}
  {\bibfnamefont {K.}~\bibnamefont {Rickert}}, \bibinfo {author} {\bibfnamefont
  {D.~B.}\ \bibnamefont {Turner}}, \bibinfo {author} {\bibfnamefont {T.~A.}\
  \bibnamefont {Prusnick}}, \bibinfo {author} {\bibfnamefont {J.~M.}\
  \bibnamefont {Mann}}, \bibinfo {author} {\bibfnamefont {D.~H.}\ \bibnamefont
  {Hurley}}, \bibinfo {author} {\bibfnamefont {M.}~\bibnamefont {Khafizov}},\
  and\ \bibinfo {author} {\bibfnamefont {C.~A.}\ \bibnamefont {Dennett}},\
  }\href@noop {} {\bibinfo {title} {Thermal conductivity suppression in
  uranium-doped thorium dioxide due to phonon resonant scattering}} (\bibinfo
  {year} {2023}),\ \Eprint {https://arxiv.org/abs/2303.01659} {arXiv:2303.01659
  [cond-mat.mtrl-sci]} \BibitemShut {NoStop}%
\end{thebibliography}%

\end{document}


\preprint{APS/123-QED}
\title{First-principles determination of the phonon-point defect scattering and thermal transport due to fission products in ThO$_2$ }
\author{Linu Malakkal$^1$}
\email{linu.malakkal@inl.gov}
\author{Ankita Katre$^2$}
\email{ankitamkatre@gmail.com}
\author{Shuxiang Zhou$^1$}%
\author{Chao Jiang$^1$}%
\author{David H. Hurley$^3$}%
\author{Chris A. Marianetti$^4$}%
\author{Marat Khafizov$^5$}%
\email{khafizov.1@osu.edu}
\affiliation{
 $^1$Computational Mechanics and Materials Department, Idaho National Laboratory, Idaho Falls, Idaho 83415, USA
}%
\affiliation{
 $^2$Department of Scientific Computing, Modeling and Simulation, SP Pune University, Pune-411007 India}
  \affiliation{
 $^3$Idaho National Laboratory, Idaho Falls, ID, 83415, USA}
 
 \affiliation{
 $^4$Department of Applied Physics and Applied Mathematics, Columbia University, 500 West 120th Street, New York, New York 10027, United States}
 \affiliation{
 $^6$Department of Mechanical and Aerospace Engineering, The Ohio State University, 201 West 19th Ave, Columbus, Ohio 43210, United States
}
\date{\today}

\keywords{Suggested keywords}

\maketitle

\newpage
\section{\label{sec:level1}Supplementary Information \protect}

\begin{figure}[h]
    \centering
    \begin{subfigure}{0.49\textwidth}
    \centering
    \includegraphics [width=1\textwidth]{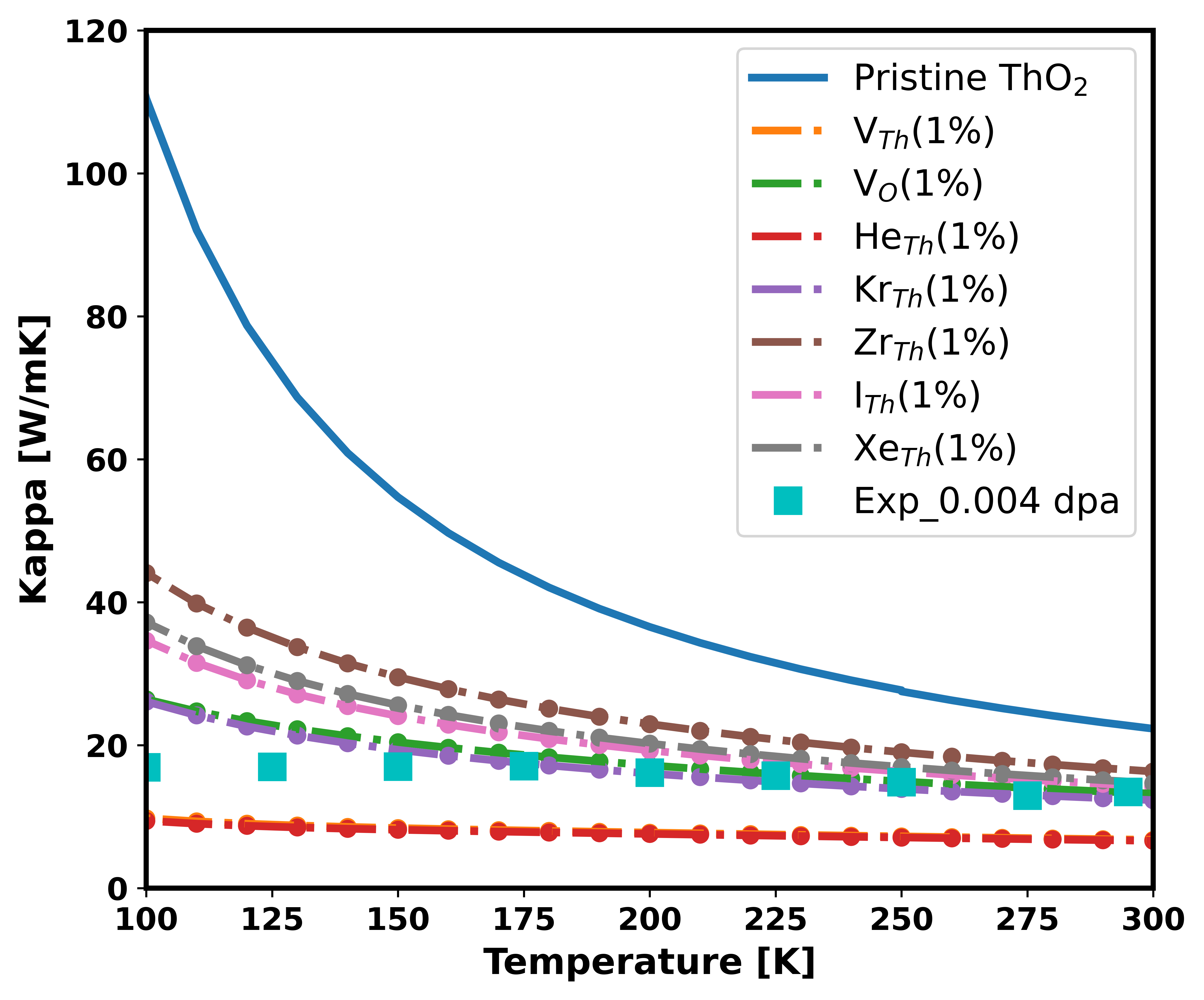}
    \caption{1$\%$ point defects (100-300 K) }
    \label{Fig8a_SI_SD_1p.png}
    \end{subfigure}
    \hfill
    \centering
    \begin{subfigure}{0.49\textwidth}
    \centering
    \includegraphics [width=1\textwidth]{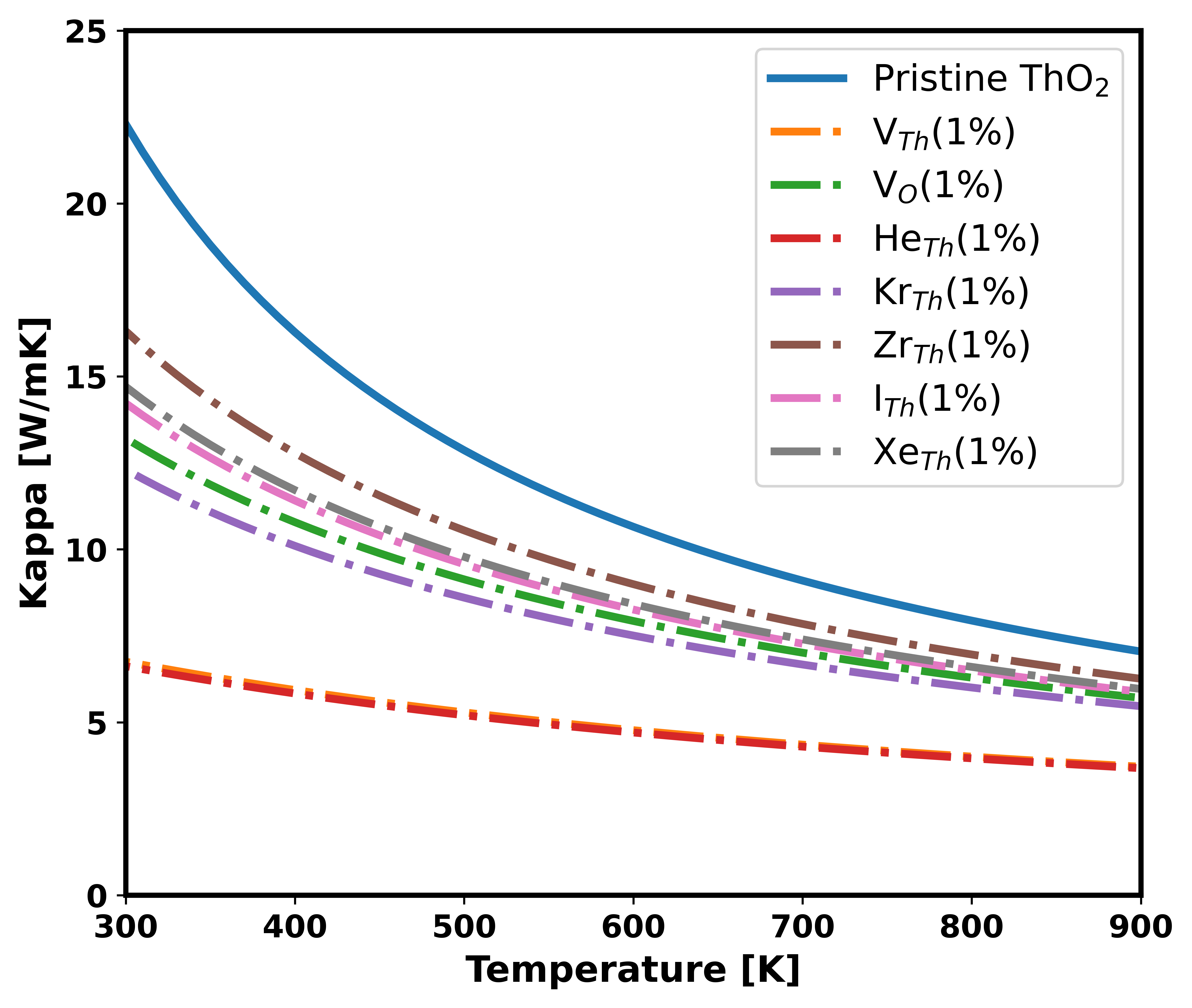}
    \caption{1$\%$ point defects (300-900 K)}
    \label{Fig8b_SI_SD_1p_ht.png}
    \end{subfigure}
    
\caption{The calculated \textit{$k_{L}$} of ThO$_{2}$ with a 1$\%$ concentration of V$_{Th}$, V$_{O}$, He$_{Th}$, Kr$_{Th}$, Zr$_{Th}$, I$_{Th}$, and Xe$_{Th}$ at (a) lower temperature range of 100--300 K and (b) higher temperature range of 300--900 K. The experiments are from Deskins \textit{et al.} \cite{DESKINS2022}, corresponding to ThO$_{2}$ single crystals irradiated with 2 MeV protons to a dose of 0.004 displacements per atom at room temperature.}
\end{figure}

Fig.~\ref{Fig8a_SI_SD_1p.png} and fig.~\ref{Fig8b_SI_SD_1p_ht.png} shows the \textit{$k_{L}$} reduction of  ThO$_{2}$ with a 1$\%$ concentration of point defects as a function of temperature from 100 to 300 K and 300 to 900 K, respectively. Though the overall trend of \textit{$k_{L}$} is similar to the 0.01$\%$ concentration, the \textit{$k_{L}$} degradation is significantly higher for a 1$\%$ concentration of all the point defects.   

\begin{figure}[h]
    \centering
    \begin{subfigure}{0.49\textwidth}
    \centering
    \includegraphics [width=0.9\textwidth]{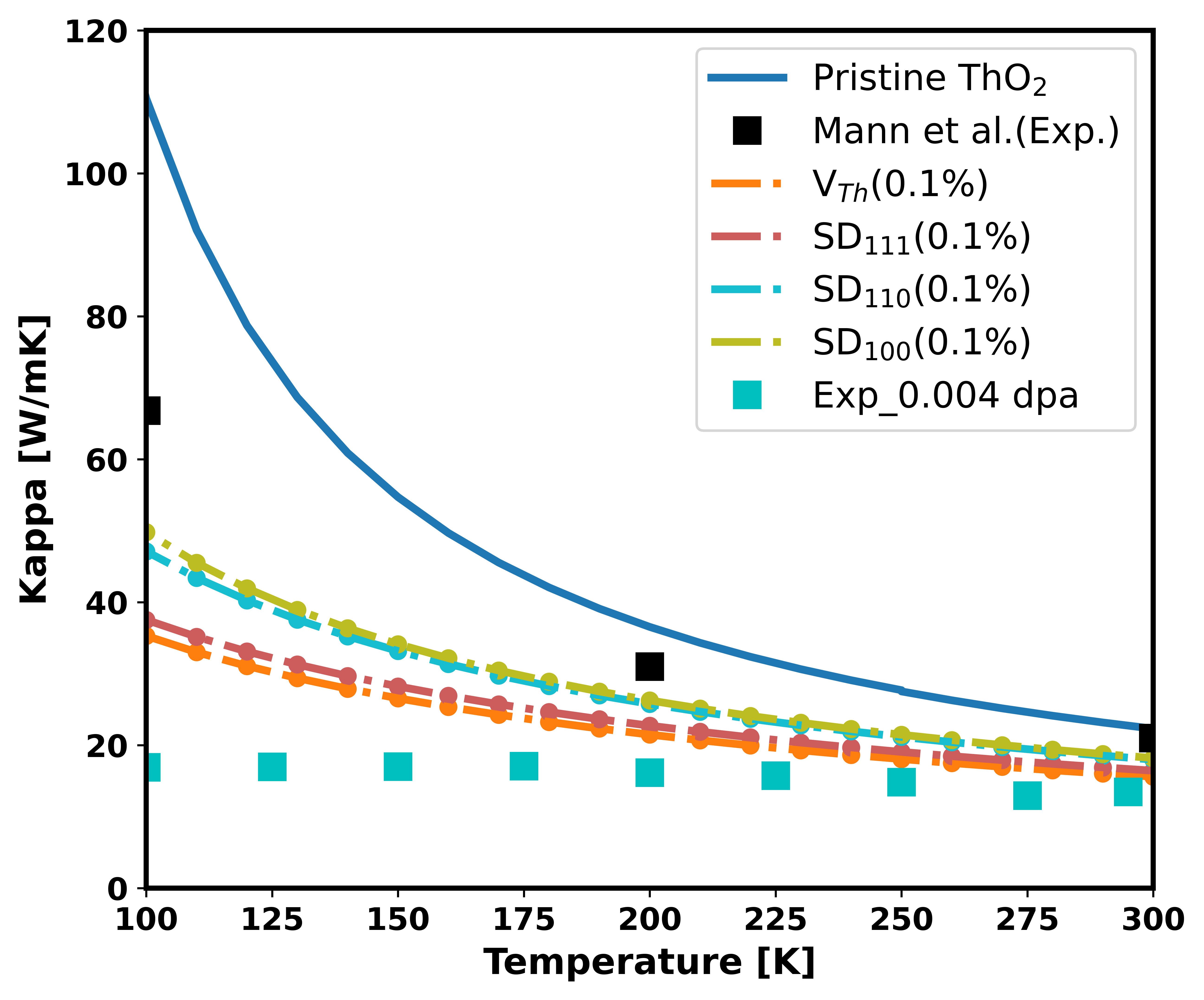}
    \caption{0.1$\%$ Schottky defects (100-300 K) }
    \label{Fig9a_SI_ScD_01p.png}
    \end{subfigure}
    \hfill
    \centering
    \begin{subfigure}{0.49\textwidth}
    \centering
    \includegraphics [width=0.9\textwidth]{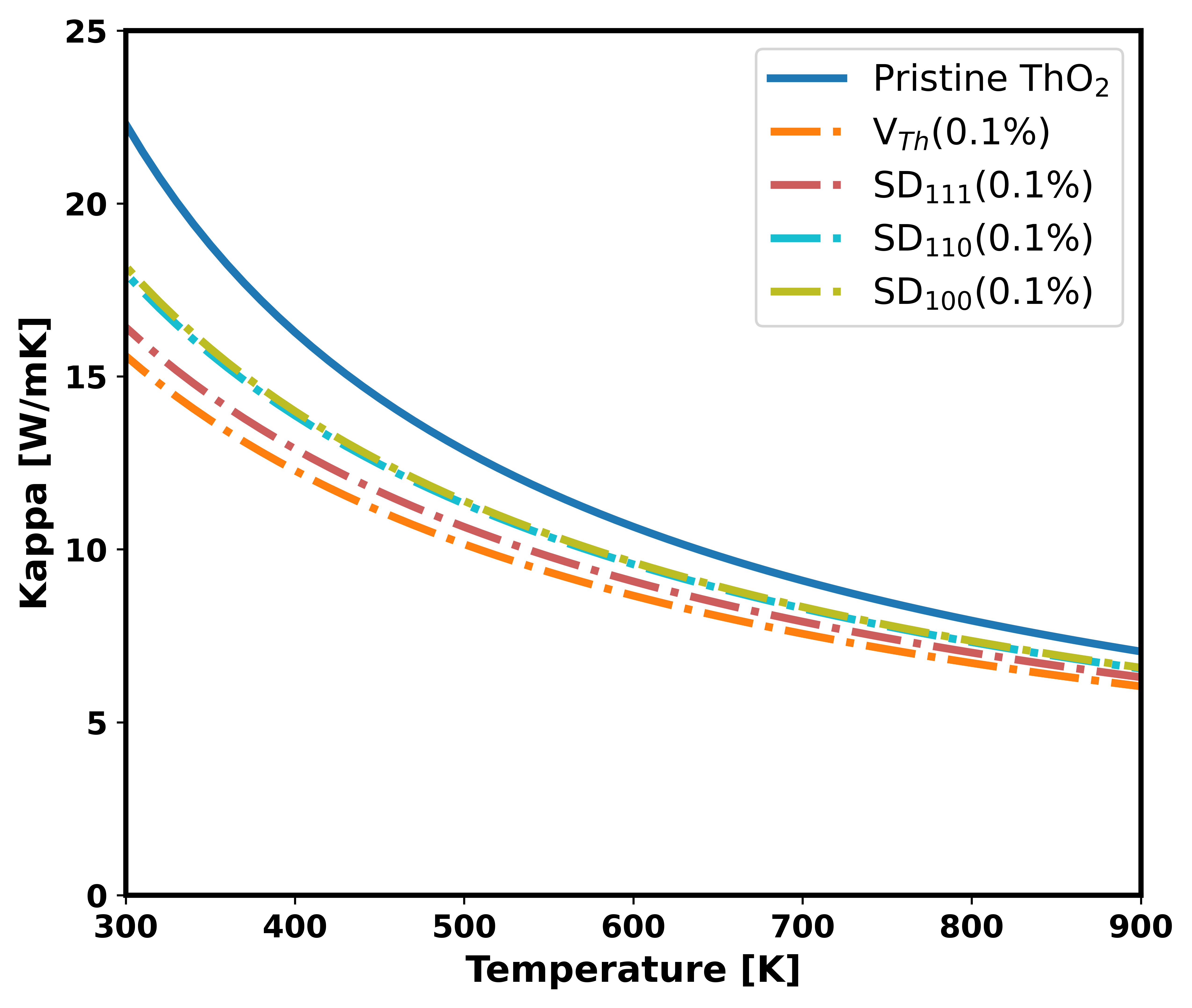}
    \caption{0.1$\%$ Schottky defects (300-900 K) }
    \label{Fig9b_SI_ScD_01p_ht.png}
    \end{subfigure}
    \centering
    \begin{subfigure}{0.49\textwidth}
    \centering
    \includegraphics [width=0.9\textwidth]{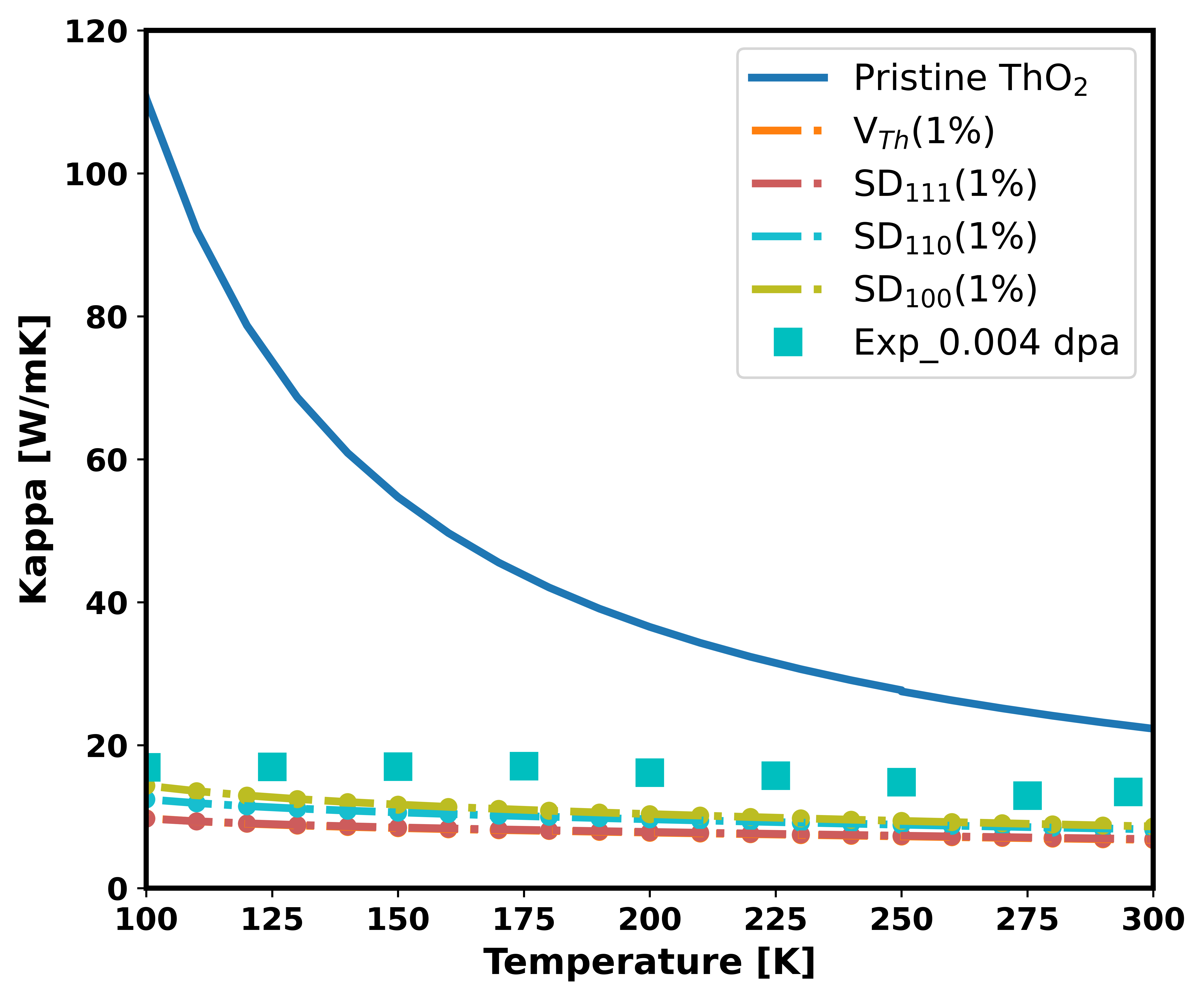}
    \caption{1$\%$ Schottky defects (100-300 K) }
    \label{Fig9c_SI_ScD_1p.png}
    \end{subfigure}
    \hfill
    \centering
    \begin{subfigure}{0.49\textwidth}
    \centering
    \includegraphics [width=0.9\textwidth]{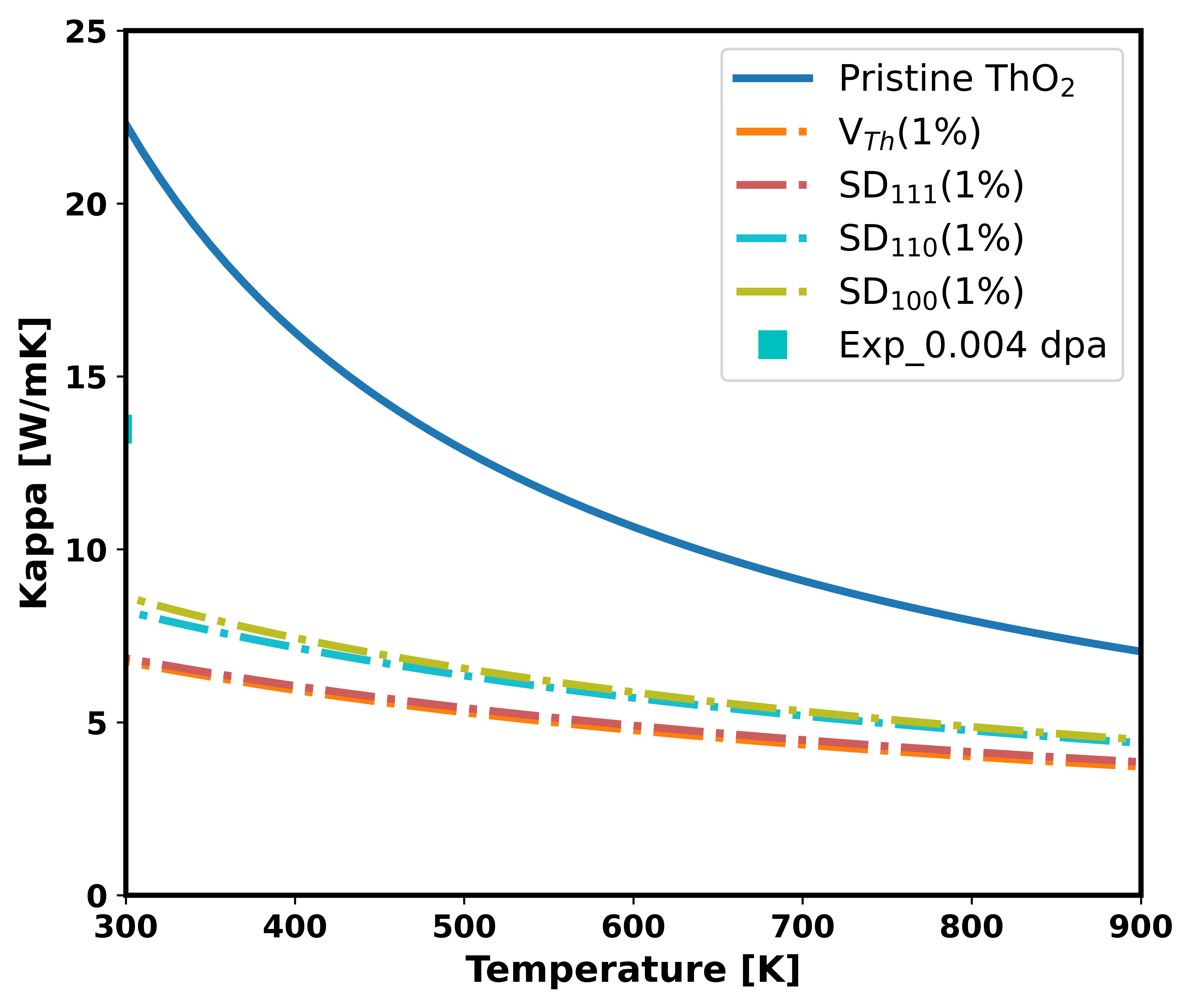}
    \caption{1$\%$ Schottky defects (300-900 K) }
    \label{Fig9d_SI_ScD_1p_ht.png}
    \end{subfigure}
\caption{The calculated \textit{$k_{L}$} of ThO$_{2}$ with three different configurations of the Schottky defects (SD$_{111}$, SD$_{110}$, and SD$_{110}$) at different concentrations and temperature range. The experiments are from Deskins \textit{et al.} \cite{DESKINS2022}, corresponding to ThO$_{2}$ single crystals irradiated with 2 MeV protons to a dose of 0.004 displacements per atom at room temperature. The experimental thermal conductivity of the pristine ThO$_{2}$ is taken from Ref. \cite{Mann_2010}.} 
\end{figure}

\newpage
Fig.~\ref{Fig9a_SI_ScD_01p.png} and fig.~\ref{Fig9b_SI_ScD_01p_ht.png} depicts the role of the 0.1$\%$ concentration of three different configurations of the Schottky defects as a function of temperature from 100 to 300 K and 300 to 900 K respectively. Whereas, fig.~\ref{Fig9c_SI_ScD_1p.png} and fig.~\ref{Fig9d_SI_ScD_1p_ht.png} respectively shows the \textit{$k_{L}$} reduction of ThO$_{2}$ with the 1$\%$ concentration of the Schottky defects at a temperature from 100 to 300 K and 300 to 900 K. The results indicate that the \textit{$k_{L}$} degradation for Schottky defects are SD$_{111}$ $>$ SD$_{110}$ $>$ SD$_{100}$. ThO$_{2}$ with an SD$_{111}$ has a similar effect to that of a Thorium vacancy.

\newpage
Fig.~\ref{Fig10_SI_2D_CD.png} Comparison of the charge density plot of various substitution defects.
\begin{figure}[h]
    \centering
    \includegraphics [width=0.5\textwidth]{Fig10_SI_2D_CD.png}
    \caption{Comparison of the charge density plot of various substitution defects. Note that positive (negative) values denote the charge accumulation (depletion) in e/Bohr$^3$.}
    \label{Fig10_SI_2D_CD.png}
\end{figure}

\begin{figure}[h]
    \centering
    \includegraphics [width=0.5\textwidth]{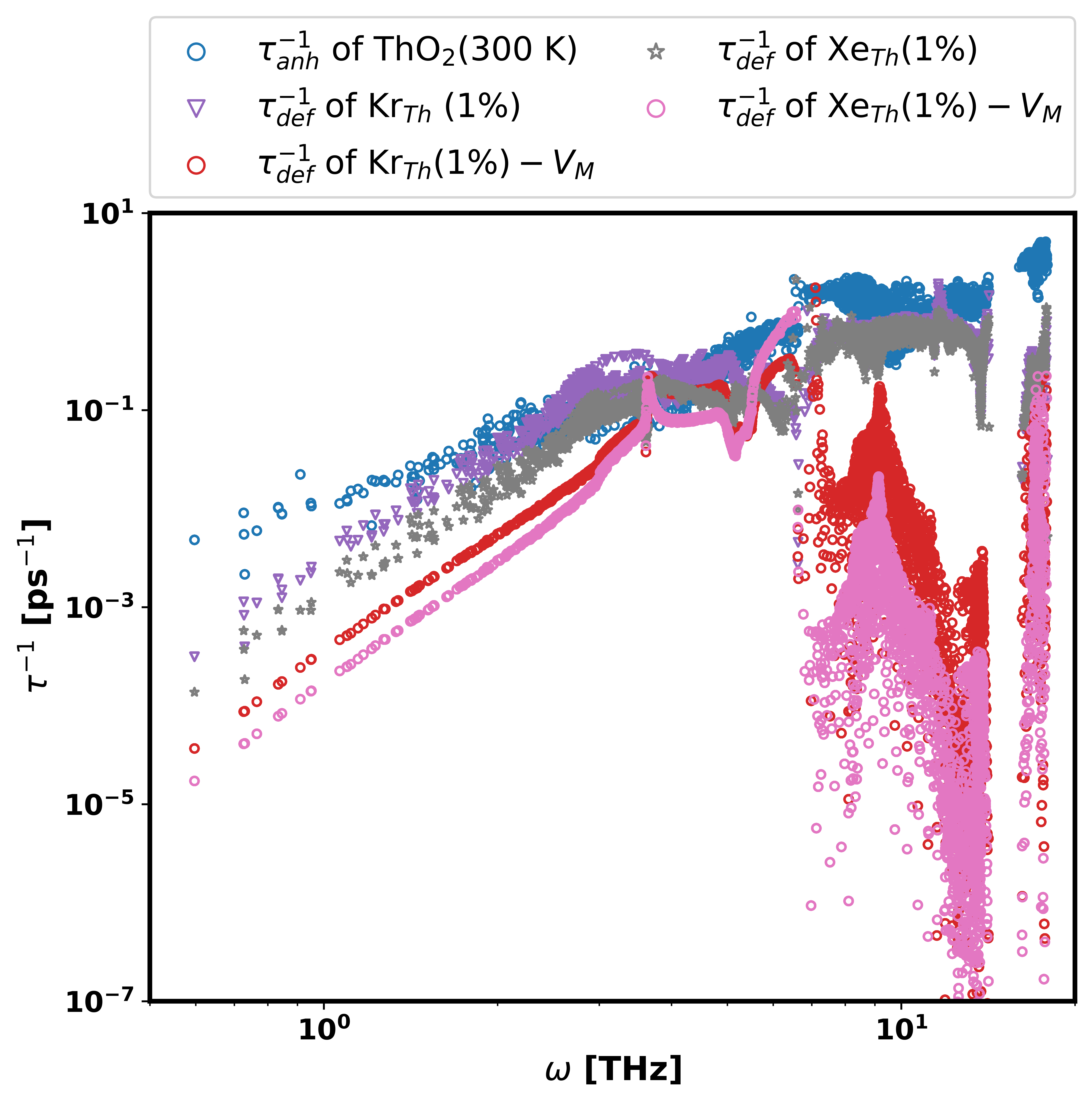}
    \caption{Comparison of the scattering rate due to the mass perturbation and force constant perturbation separately for the xenon an krypton substitution}
    \label{Fig11_SI_VM_Vk.png}
\end{figure}

Fig.~\ref{Fig11_SI_VM_Vk.png} Comparison of the scattering rate due to the mass and force constant perturbation separately for the xenon an krypton substitution. The scattering rate indicate that compared to the force constant variance the effect of mass variance is relatively smaller. 

\begin{figure}[h]
    \centering
    \includegraphics [width=0.5\textwidth]{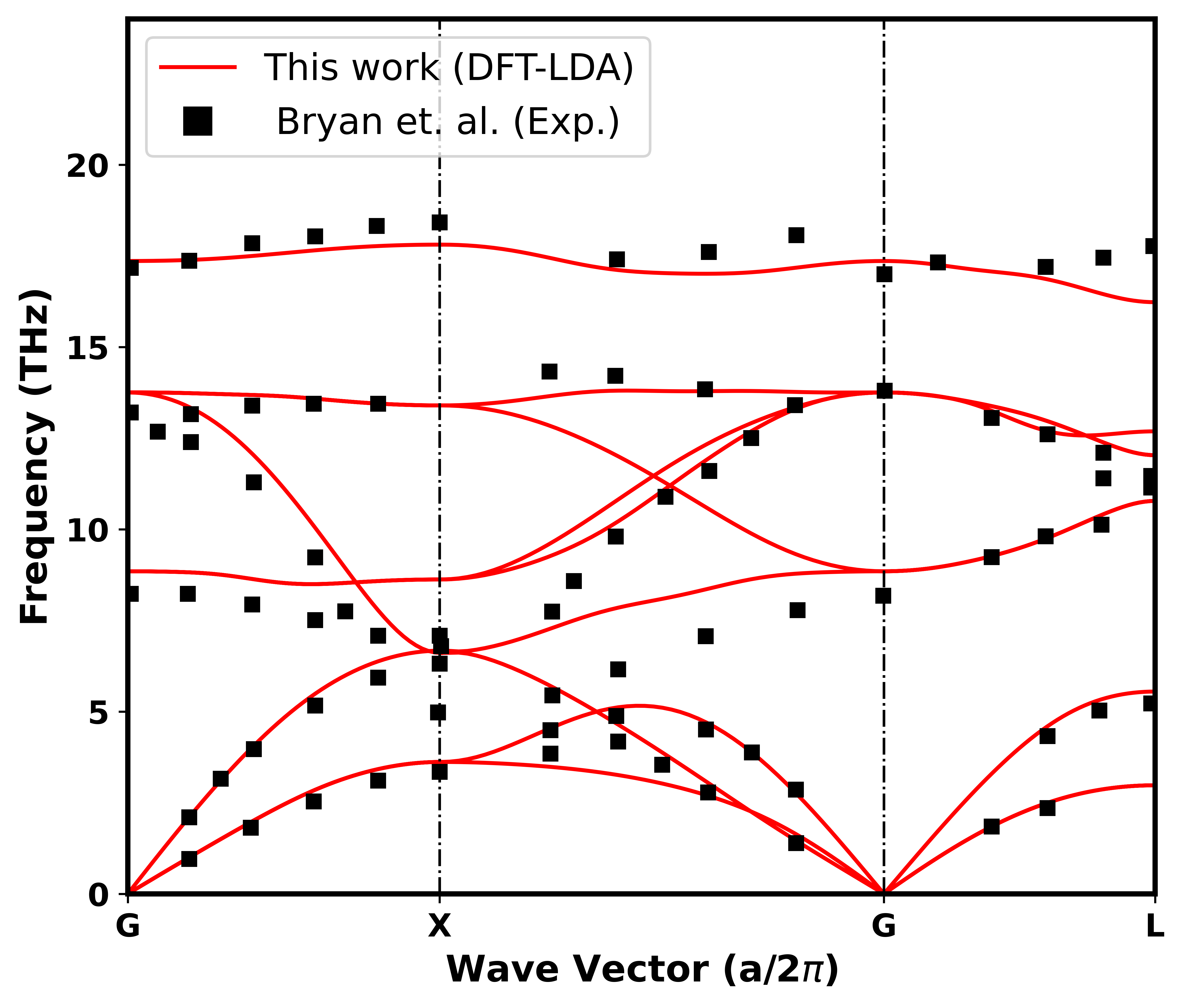}
    \caption{Phonon dispersion spectra and the total phonon density of states of ThO$_{2}$}
    \label{Fig12_SI_PD.png} 
\end{figure}

Fig.~\ref{Fig12_SI_PD.png} shows the phonon dispersion spectra of ThO$_{2}$ compared with the experimental work by Bryan et al.\cite{Bryan2020} along with the total phonon density of states 

Table ~\ref{Table1} shows the values of the Frobenius norm ratio of the IFC matrix for the different point defects. The result indicate that the force constants changes are significant when there is a thorium vacancy and helium substitution. Whereas, the force constant was least affected by the Zirconium substitution. 

\begin{table*}[h]
\caption{The frobenius norm of the force constant changes for various defects considered.}
\begin{tabular}{| >{\centering}p{1.8 cm} |  >{\centering}p{1.8 cm} | >{\centering}p{1.8 cm} |>{\centering}p{1.8 cm} |>{\centering}p{1.8 cm} |>{\centering}p{1.8 cm} |>{\centering}p{1.8 cm} |>{\centering\arraybackslash}p{1.8 cm} |}
\hline
Defects & $V_{Th}$ &$V_{O}$ &$He_{Th}$ &$Kr_{Th}$ &$Zr_{Th}$ &$I_{Th}$ &$Xe_{Th}$ \\
\hline
$\frac{\Delta K}{K}$ & 0.17& 0.05 &0.17 &0.07 &0.04 &0.05 &0.06 \\
\hline
\end{tabular}
\label{Table1}
\end{table*}

\begin{figure}[h!]
        \centering
        \begin{subfigure}[b]{0.49\textwidth}
            \centering
            \includegraphics[width=0.7\textwidth]{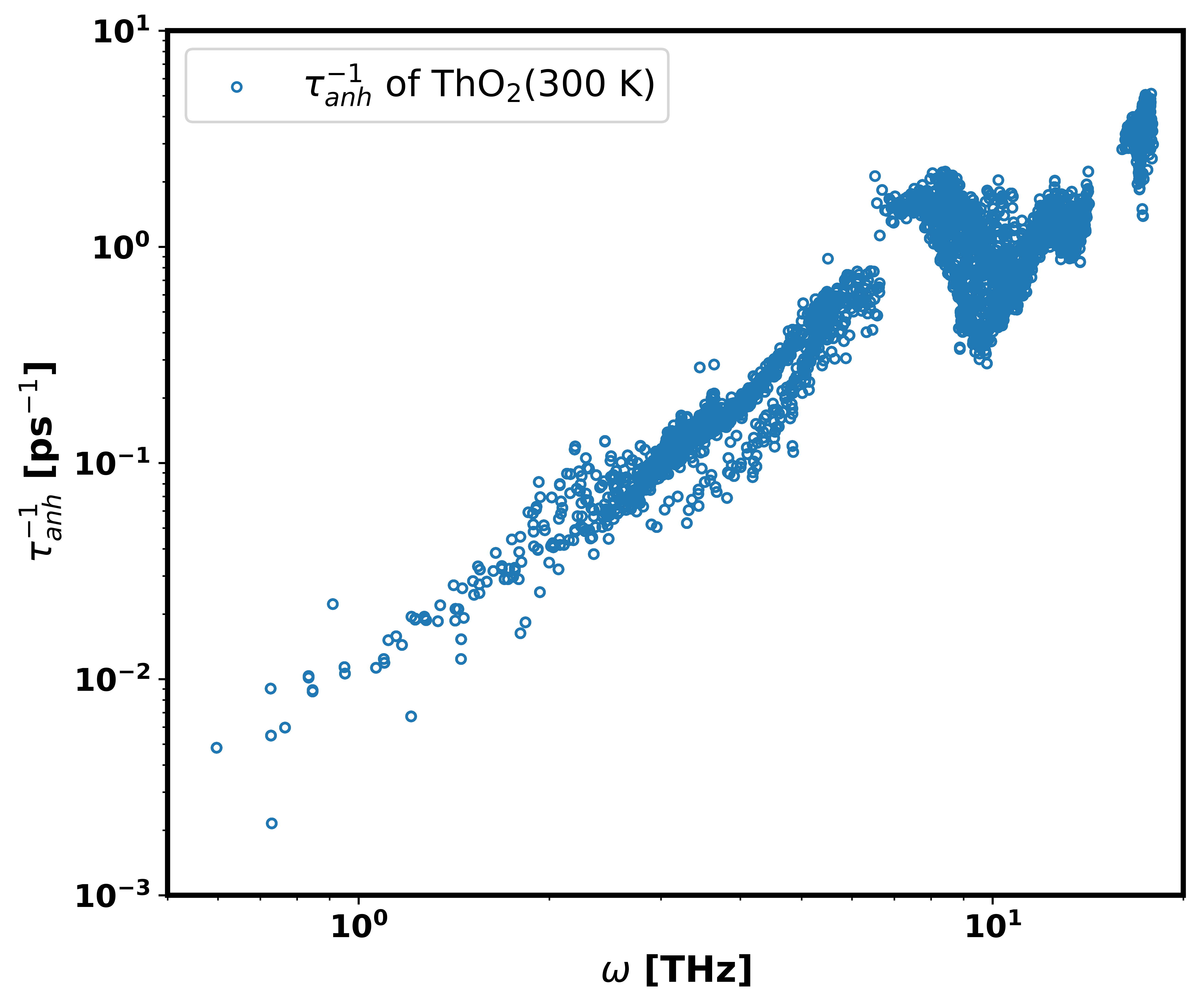}
            \caption[]%
            {{\small Anharmonic scattering of ThO$_{2}$}}    
            \label{fig:Anharmonic1}
        \end{subfigure}
        \hfill
        \begin{subfigure}[b]{0.49\textwidth}  
            \centering 
            \includegraphics[width=0.7\textwidth]{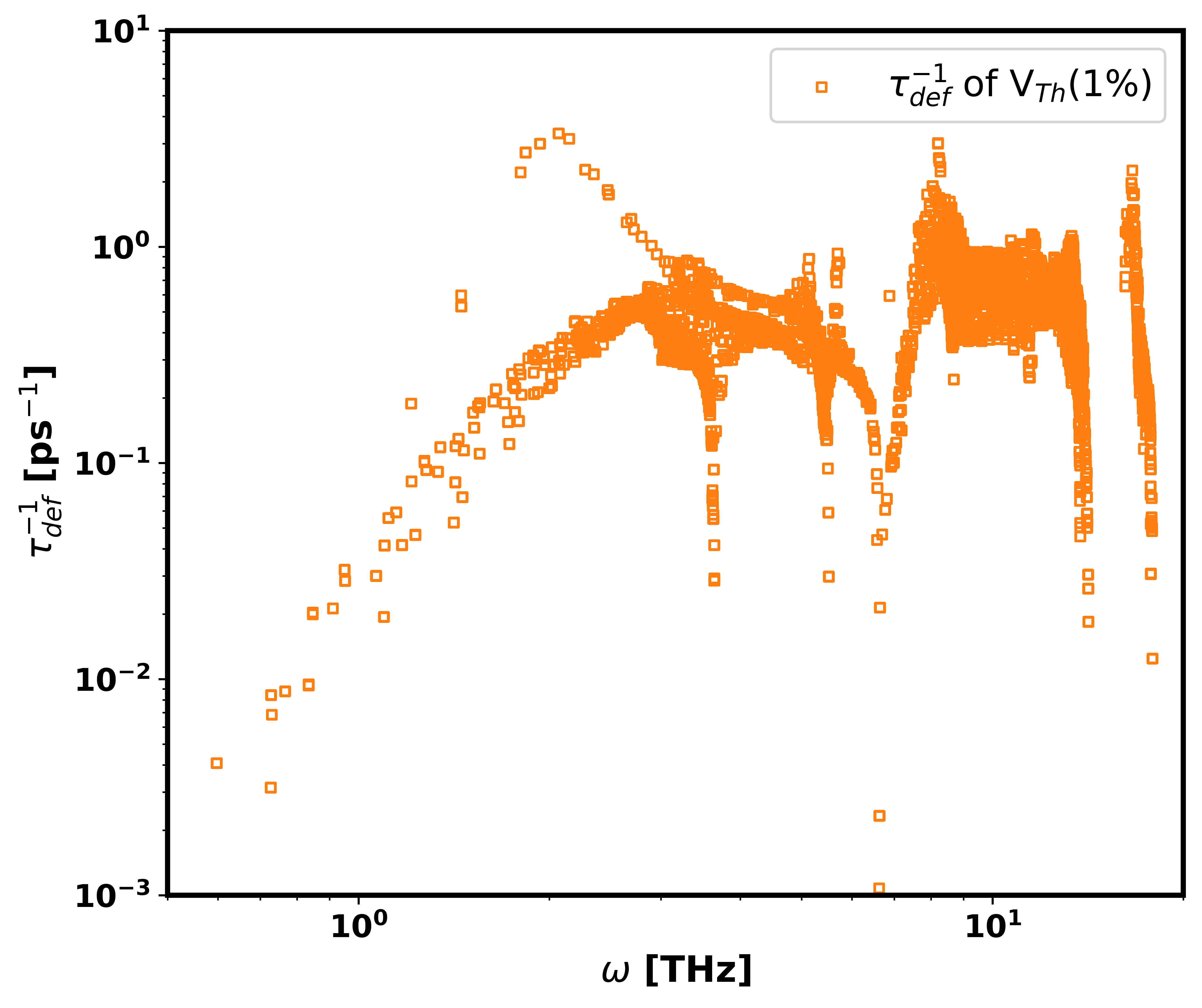}
            \caption[]%
            {{\small Thorium vacancy (V$_{Th}$})}    
            \label{fig:Thorium_vacancy}
        \end{subfigure}
        \vskip\baselineskip
        \begin{subfigure}[b]{0.49\textwidth}  
            \centering 
            \includegraphics[width=0.7\textwidth]{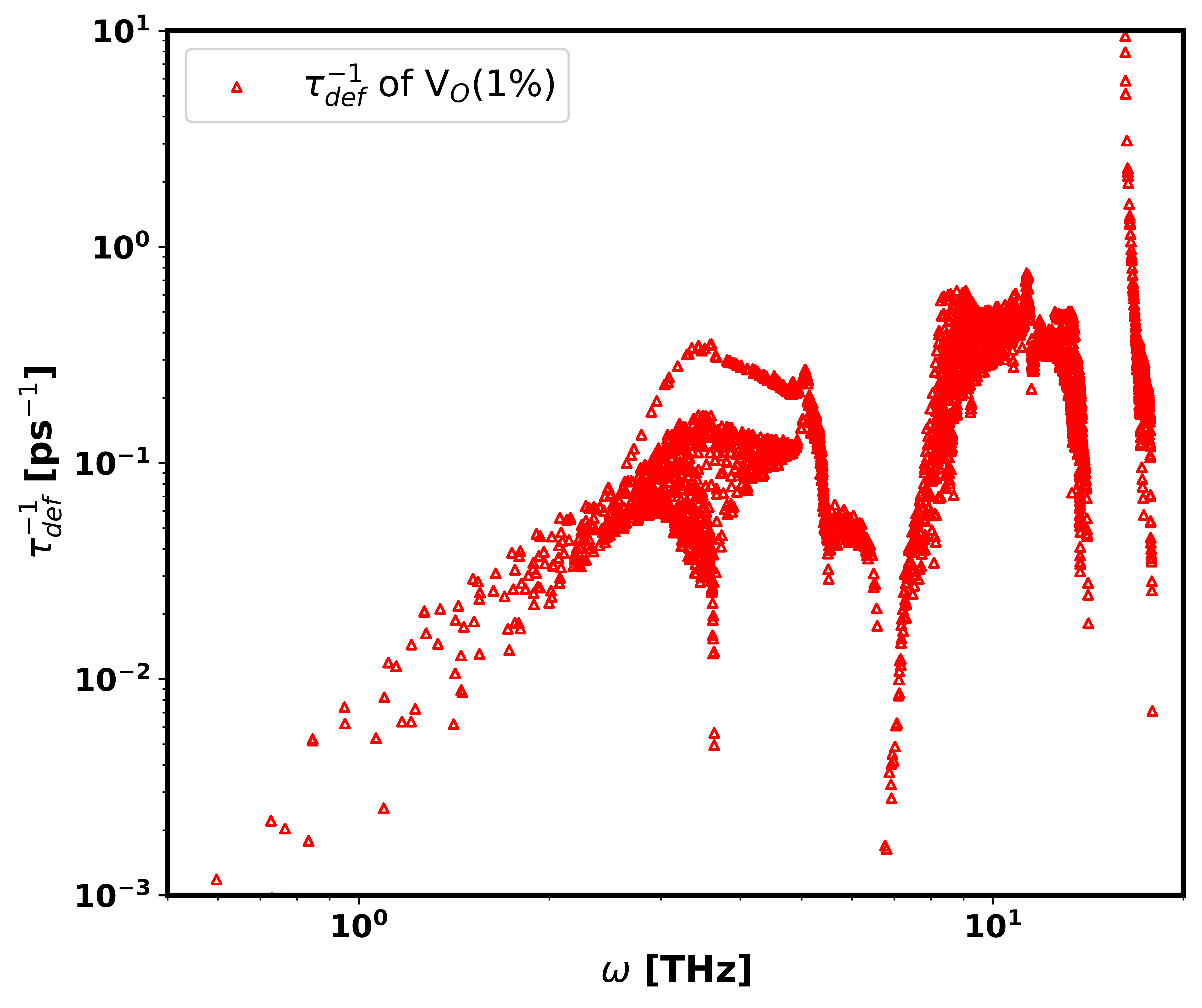}
            \caption[]%
            {{\small Oxygen vacancy (V$_{O}$})}    
            \label{fig:Oxygen_vacancy}
        \end{subfigure}
        \hfill
        \begin{subfigure}[b]{0.49\textwidth}   
            \centering 
            \includegraphics[width=0.7\textwidth]{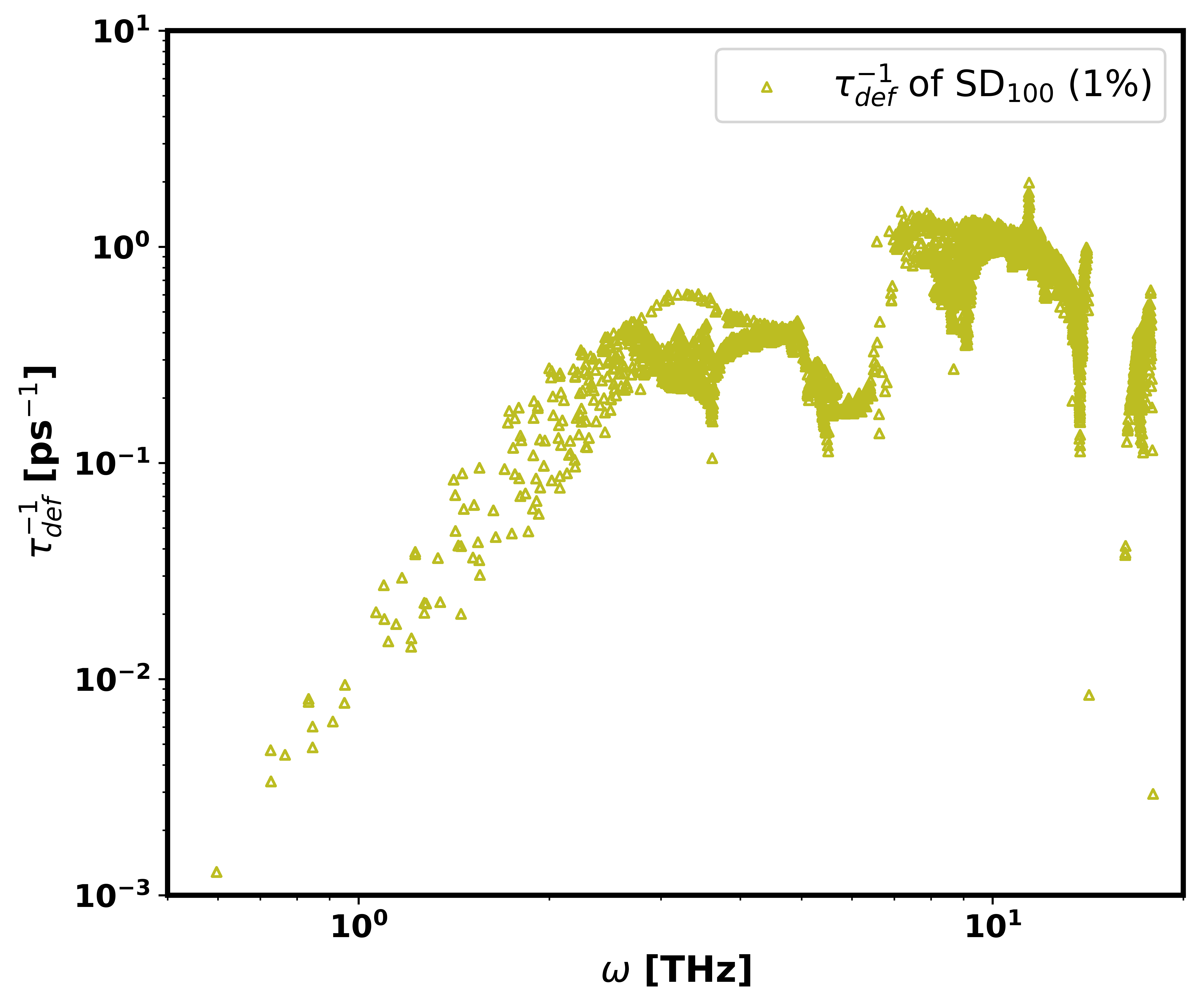}
            \caption[]%
            {{\small {100} Schottky defect (SD$_{100}$})}   
            \label{fig:Schottky_defect_100}
        \end{subfigure}
        \vskip\baselineskip
        \begin{subfigure}[b]{0.49\textwidth}   
            \centering 
            \includegraphics[width=0.7\textwidth]{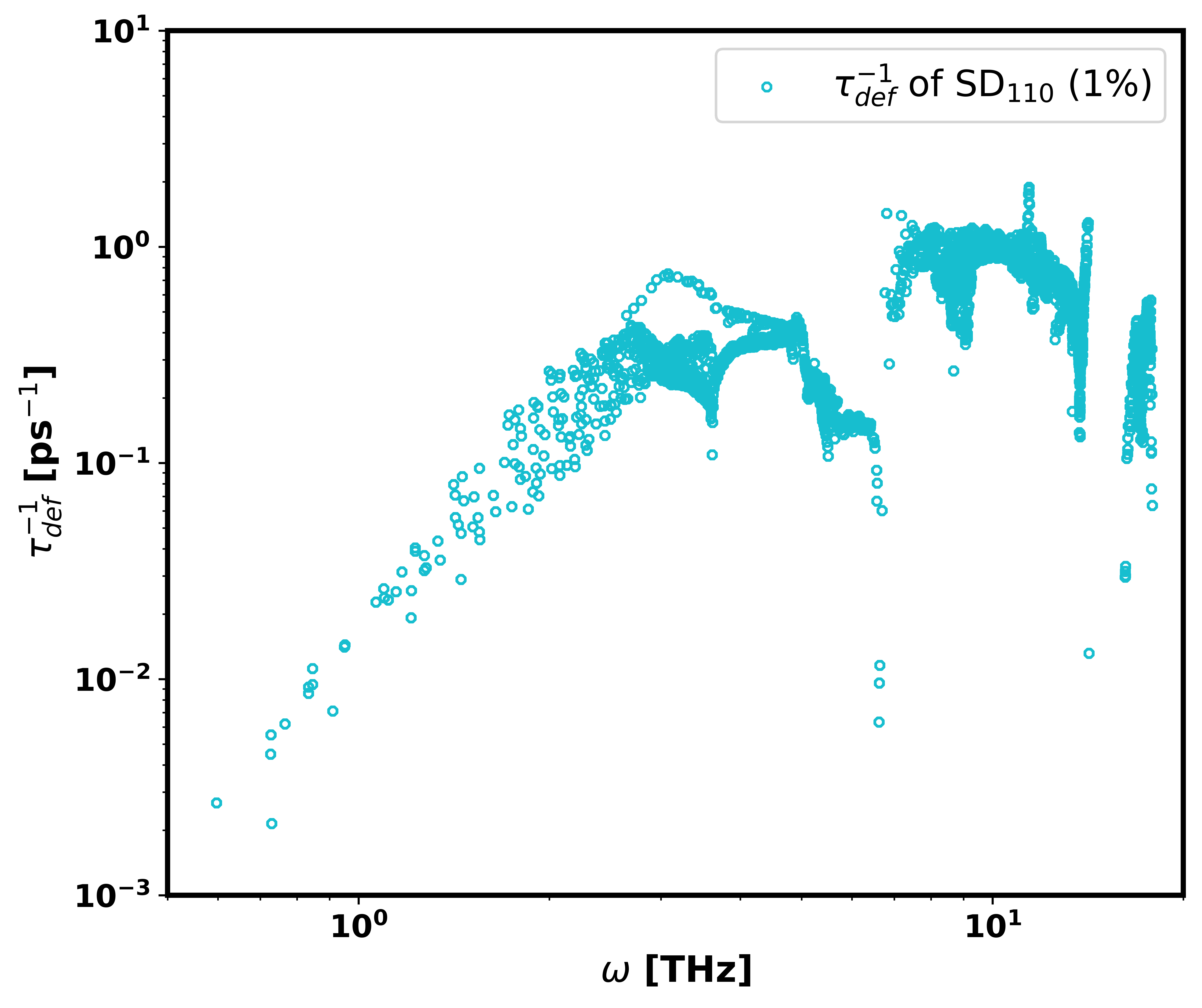}
            \caption[]%
            {{\small {110} Schottky defect (SD$_{110}$})}  
            \label{fig:Schottky_defect_110}
        \end{subfigure}
        \hfill
        \begin{subfigure}[b]{0.49\textwidth}   
            \centering 
            \includegraphics[width=0.7\textwidth]{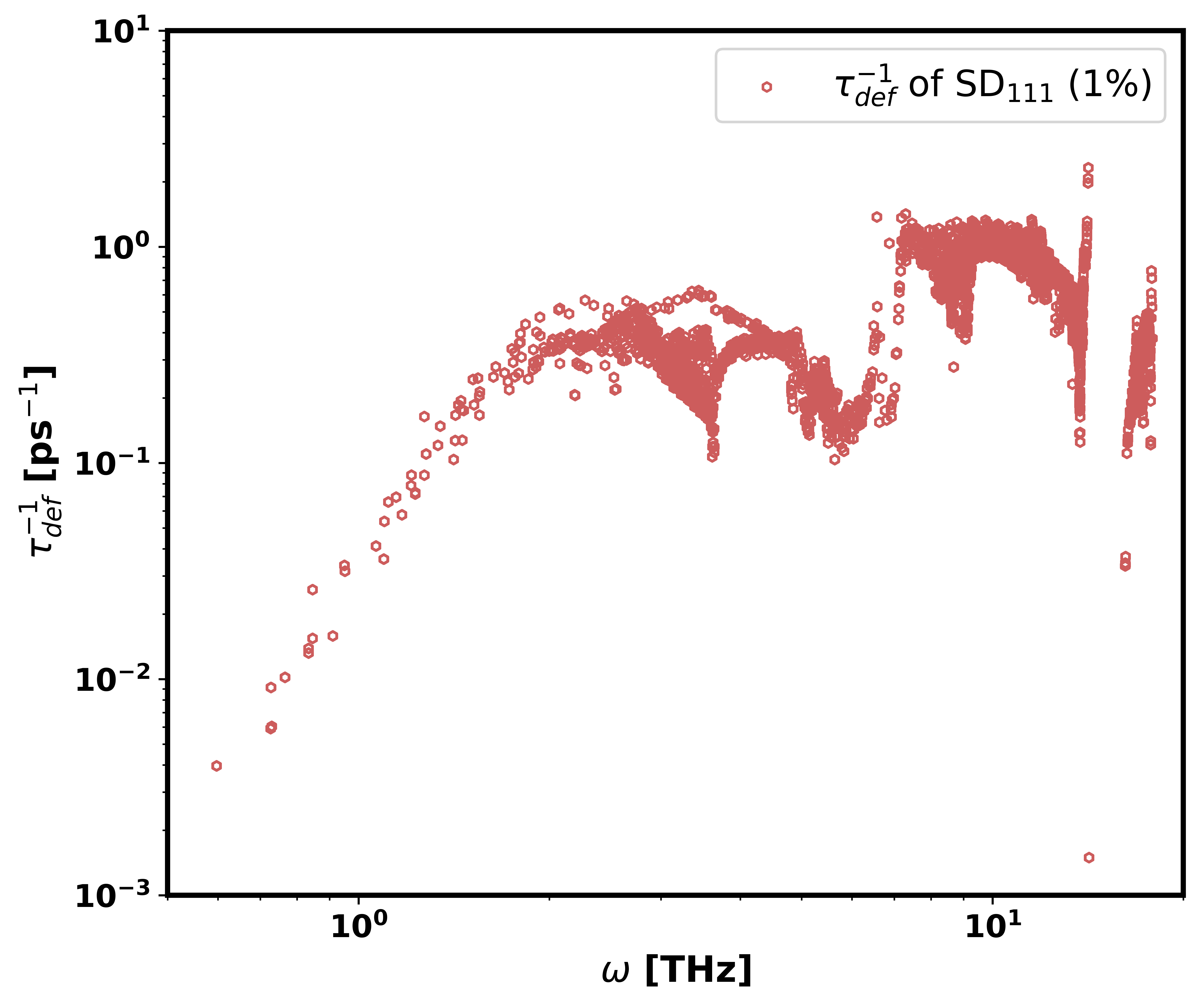}
            \caption[]%
            {{\small {111} Schottky defect (SD$_{111}$})}   
            \label{fig:Schottky_defect_111}
        \end{subfigure}
        \caption[]
        {\small Scattering rates ($\tau^{-1}_{def}$) of phonons as a function of phonon frequency $(\omega)$ due to vacancies and Schottky defects compared to $\tau^{-1}_{anh}$ of the pristine ThO$_{2}$ due to three phonon interactions at 300 K }
        \label{Fig13}
\end{figure}

\newpage
Fig.~\ref{Fig13} Shows the phonon-defect scattering of the vacancy and Schottky defects in ThO$_{2}$.

\newpage

\begin{figure}[h!]
        \centering
        \begin{subfigure}[b]{0.49\textwidth}
            \centering
            \includegraphics[width=0.7\textwidth]{Fig13a_SI_ThO2_tau.png}
            \caption[]%
            {{\small Anharmonic scattering of ThO$_{2}$}}    
            \label{fig:Anharmonic2}
        \end{subfigure}
        \hfill
        \begin{subfigure}[b]{0.49\textwidth}  
            \centering 
            \includegraphics[width=0.7\textwidth]{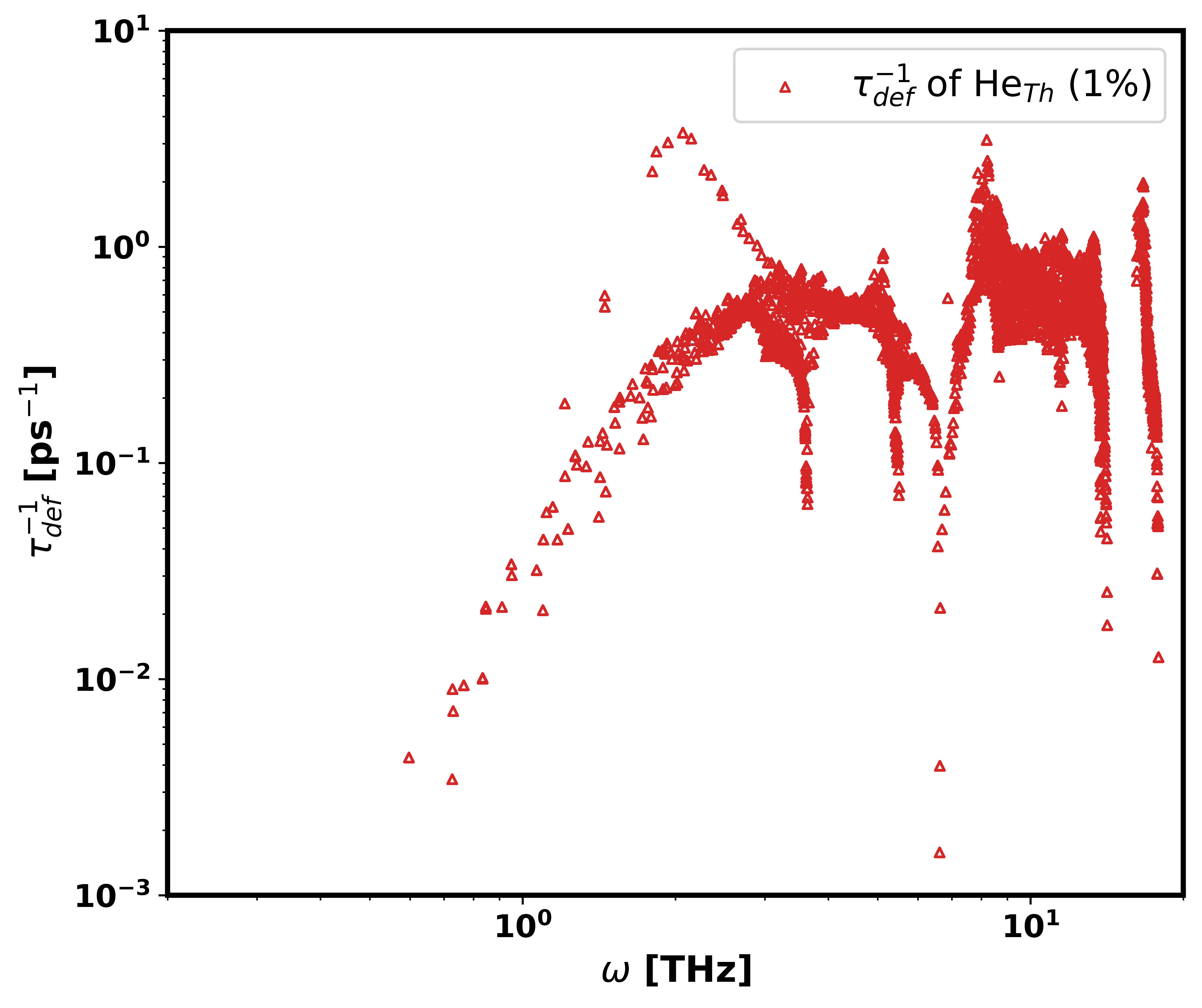}
            \caption[]%
            {{\small Helium substitution (He$_{Th}$})}    
            \label{fig:helium_substitution}
        \end{subfigure}
        \vskip\baselineskip
        \begin{subfigure}[b]{0.49\textwidth}  
            \centering 
            \includegraphics[width=0.7\textwidth]{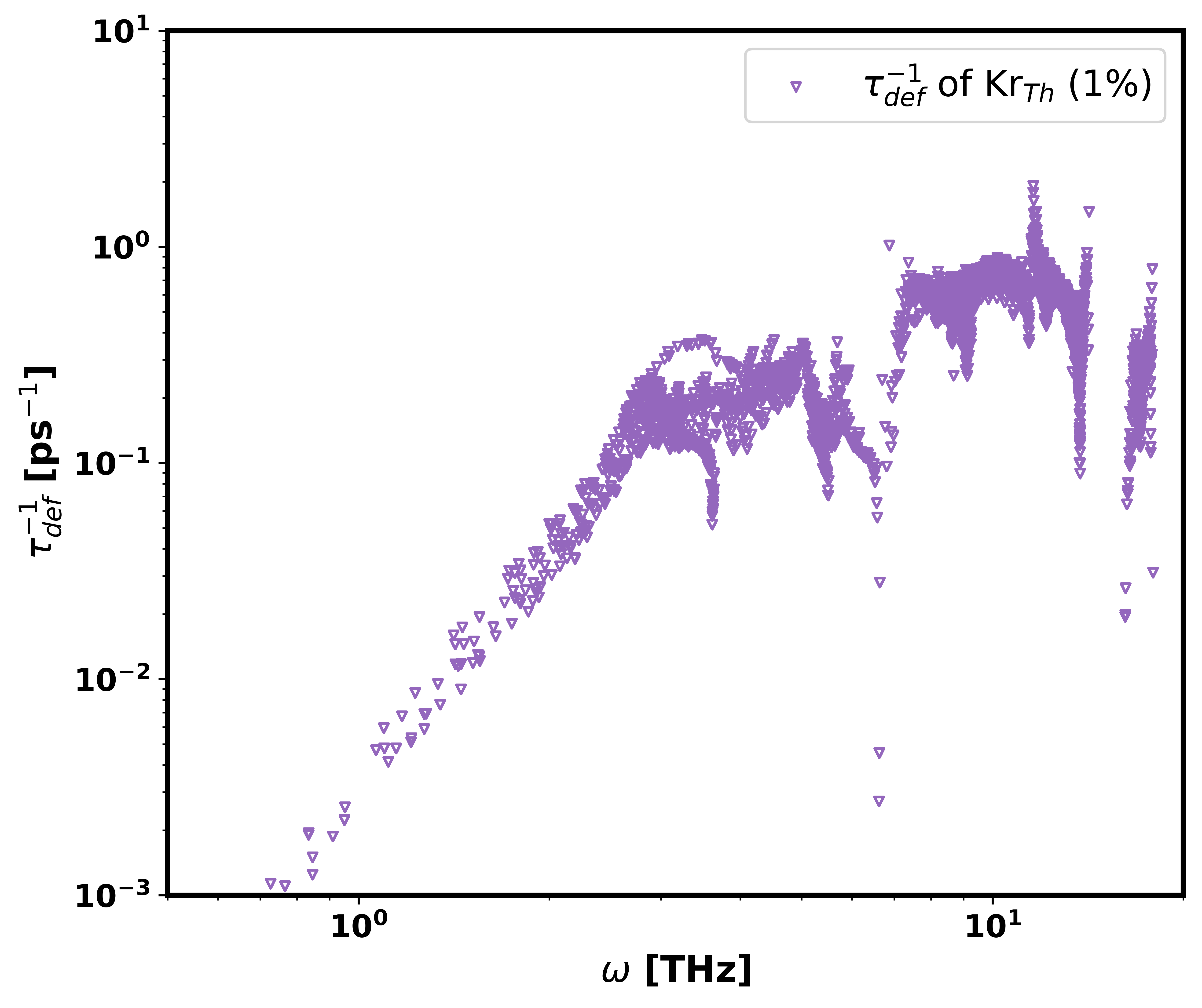}
            \caption[]%
            {{\small Krypton substitution (Kr$_{Th}$})}    
            \label{fig:krypton_substitution}
        \end{subfigure}
        \hfill
        \begin{subfigure}[b]{0.49\textwidth}   
            \centering 
            \includegraphics[width=0.7\textwidth]{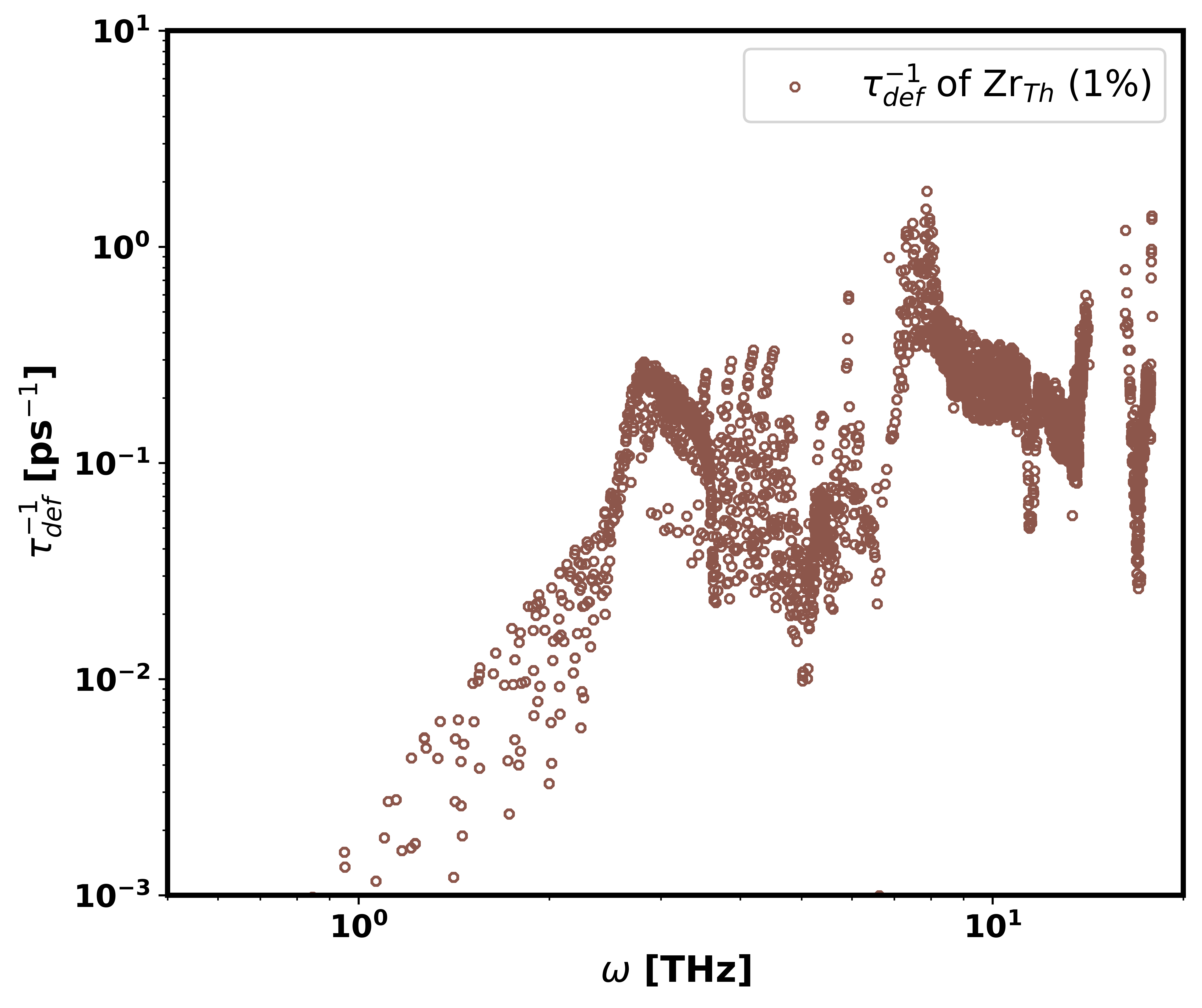}
            \caption[]%
            {{\small Zirconium substitution (Zr$_{Th}$})}    
            \label{fig:Zirconium_substitution}
        \end{subfigure}
        \vskip\baselineskip
        \begin{subfigure}[b]{0.49\textwidth}   
            \centering 
            \includegraphics[width=0.7\textwidth]{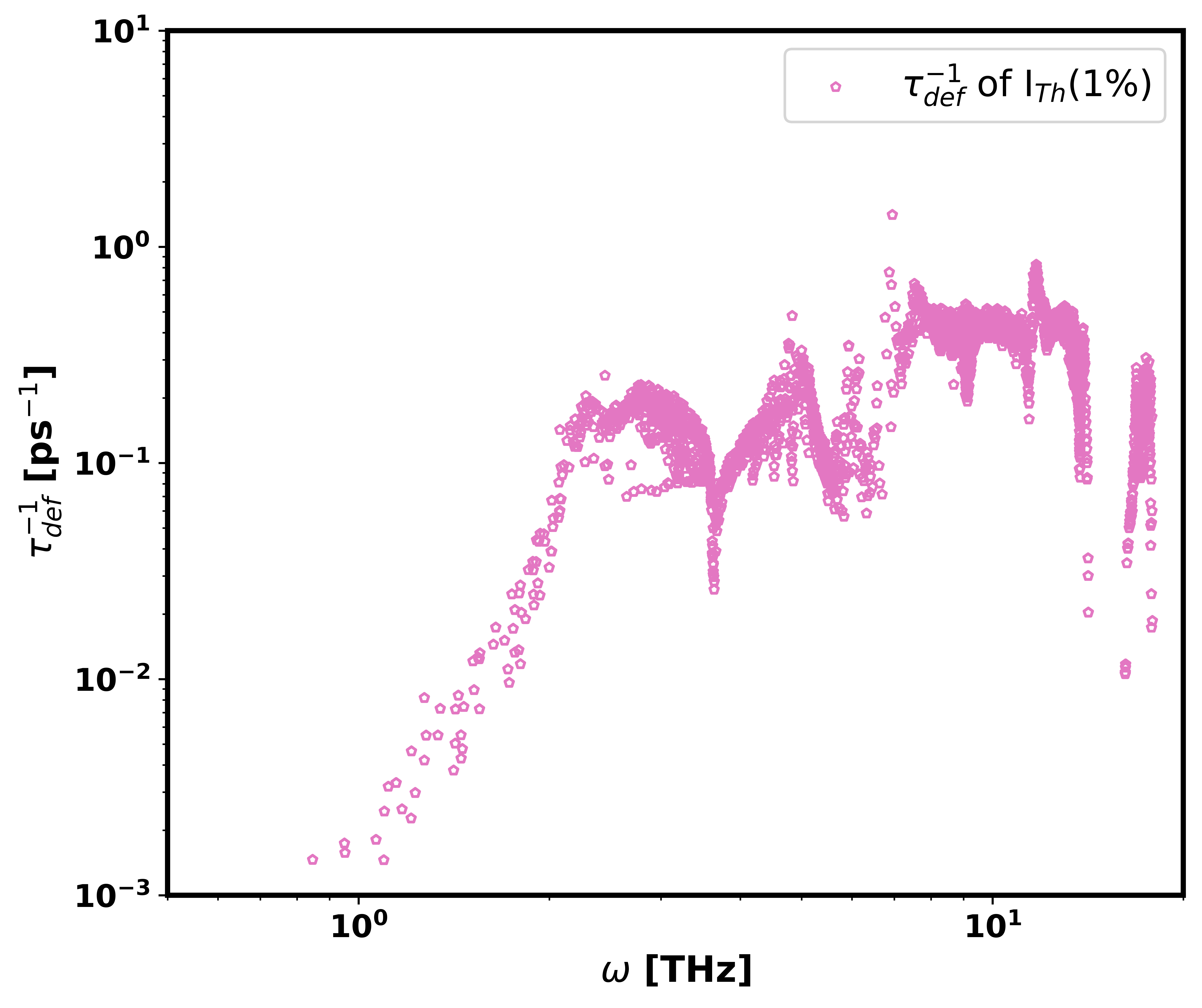}
            \caption[]%
            {{\small Iodine substitution (I$_{Th}$})}    
            \label{fig:iodine_substitution}
        \end{subfigure}
        \hfill
        \begin{subfigure}[b]{0.49\textwidth}   
            \centering 
            \includegraphics[width=0.7\textwidth]{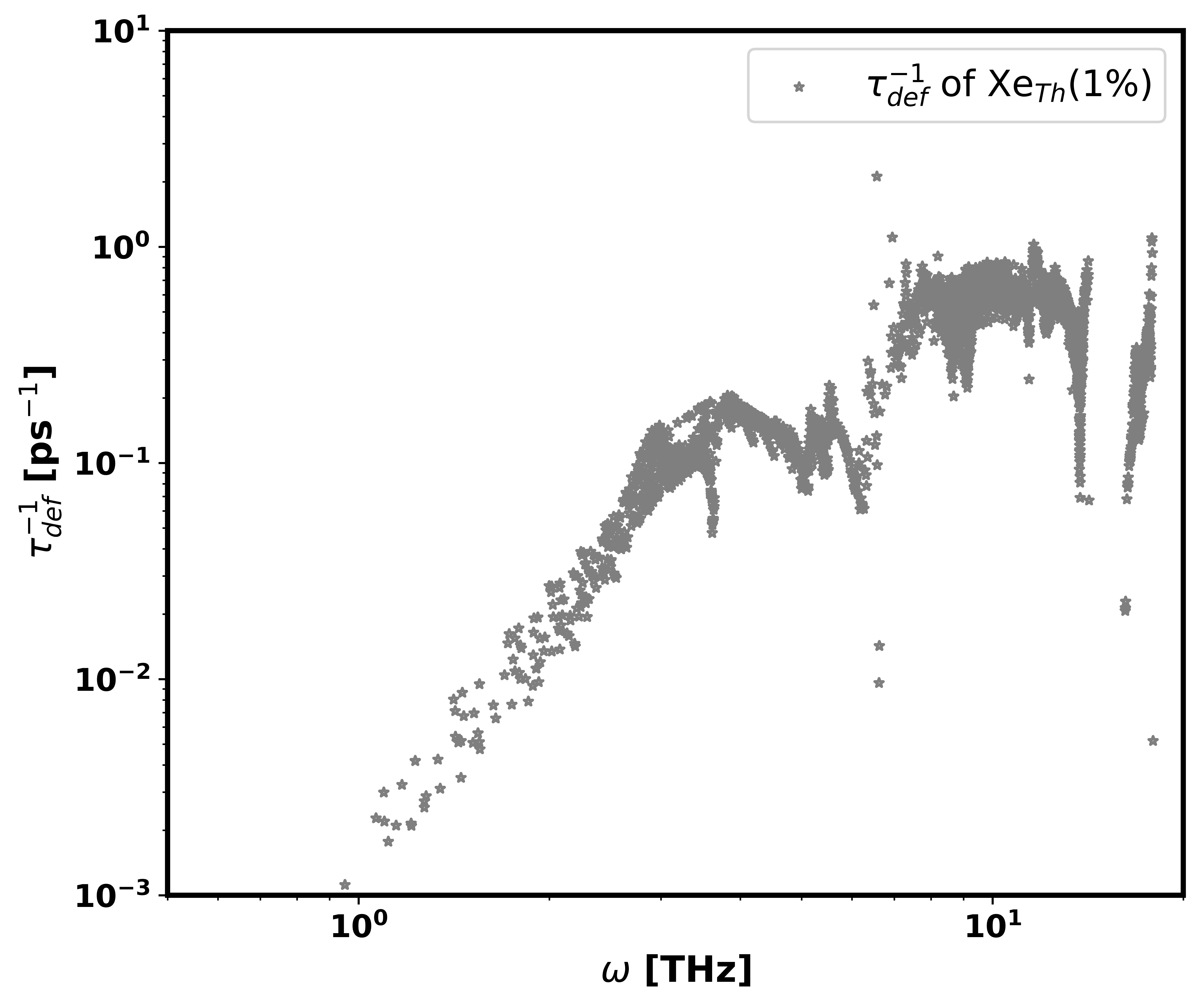}
            \caption[]%
            {{\small Xenon substitution (Xe$_{Th}$})}    
            \label{fig:xenon_substitution}
        \end{subfigure}
        \caption[]
        {\small Scattering rates ($\tau^{-1}_{def}$) of phonons as a function of phonon frequency $(\omega)$ due to substitutional point defects in thorium sites compared to $\tau^{-1}_{anh}$ of the pristine ThO$_{2}$ due to three phonon interactions at 300 K}
        \label{Fig14}
\end{figure}

Fig.~\ref{Fig14} Shows the phonon defect scattering rates of different substitutional defects.

\begin{figure}[h!]
        \centering
        \begin{subfigure}[b]{0.49\textwidth}
            \centering
            \includegraphics[width=0.6\textwidth]{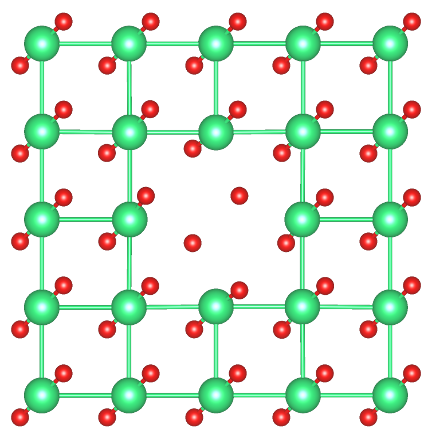}
            \caption[]%
            {{\small Thorium vacancy (V$_{Th}$})}    
            \label{fig:ThV_struc}
        \end{subfigure}
        \hfill
        \begin{subfigure}[b]{0.49\textwidth}  
            \centering 
            \includegraphics[width=0.6\textwidth]{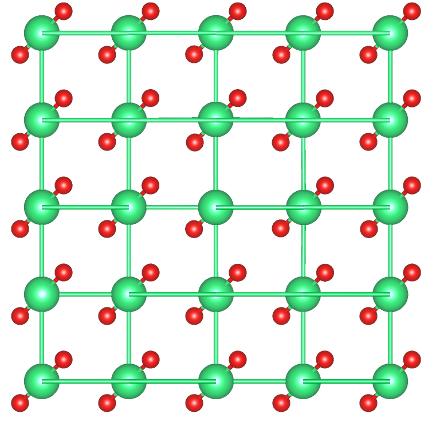}
            \caption[]%
            {{\small Oxygen vacancy (V$_{O}$)}}    
            \label{fig:OV_struc}
        \end{subfigure}
        \vskip\baselineskip
        \begin{subfigure}[b]{0.49\textwidth}  
            \centering 
            \includegraphics[width=0.6\textwidth]{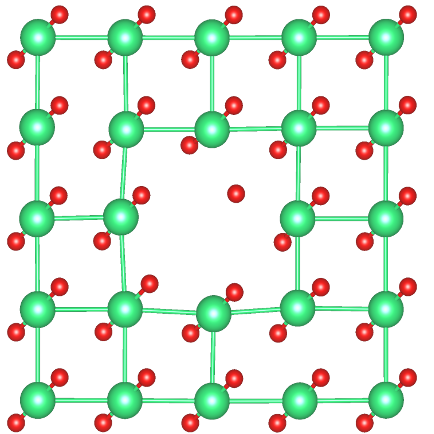}
            \caption[]%
            {{\small 100 Schottky defect (SD$_{100}$})}    
            \label{fig:SD_100}
        \end{subfigure}
        \hfill
        \begin{subfigure}[b]{0.49\textwidth}   
            \centering 
            \includegraphics[width=0.6\textwidth]{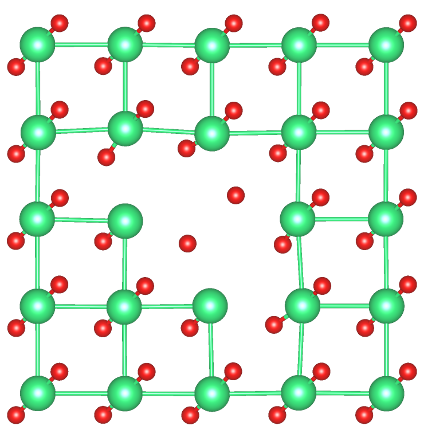}
            \caption[]%
            {{\small 110 Schottky defect (SD$_{110}$})}    
            \label{fig:SD_110}
        \end{subfigure}
        \vskip\baselineskip
        \begin{subfigure}[b]{0.49\textwidth}   
            \centering 
            \includegraphics[width=0.6\textwidth]{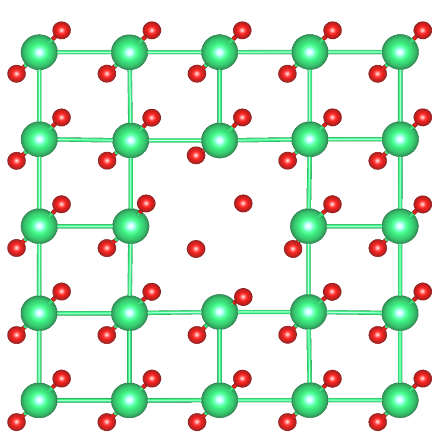}
            \caption[]%
            {{\small 111 Schottky defect (SD$_{111}$})}    
            \label{fig:SD_111}
        \end{subfigure}
        \hfill
        \begin{subfigure}[b]{0.49\textwidth}   
            \centering 
            \includegraphics[width=0.6\textwidth]{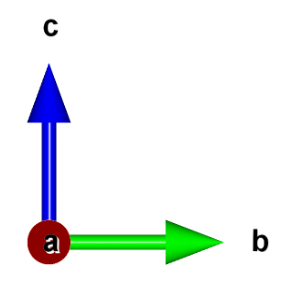}
            \label{fig:Axis}
        \end{subfigure}
        \caption[]
        {\small Relaxed structures of ThO$_2$ with vacancy defect and Schottky defects. Color codes are as follows; Th (green), and O (red).}
        \label{Fig15}
\end{figure}

\newpage
Fig.~\ref{Fig15} Shows the relaxed structure of ThO$_{2}$  near the vacancies and Schottky defects along the 100 plane.

\newpage

\begin{figure}[h!]
        \centering
        \begin{subfigure}[b]{0.49\textwidth}
            \centering
            \includegraphics[width=0.6\textwidth]{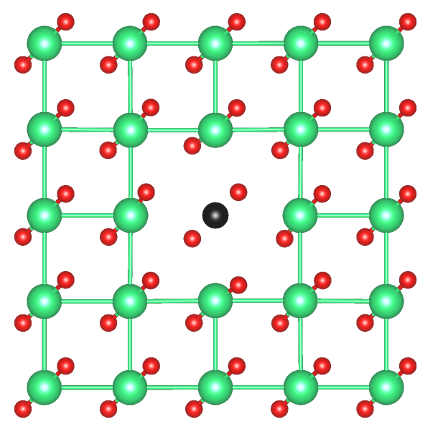}
            \caption[]%
            {{\small Helium substitution (He$_{Th}$})}    
            \label{fig:He_struc}
        \end{subfigure}
        \hfill
        \begin{subfigure}[b]{0.49\textwidth}  
            \centering 
            \includegraphics[width=0.6\textwidth]{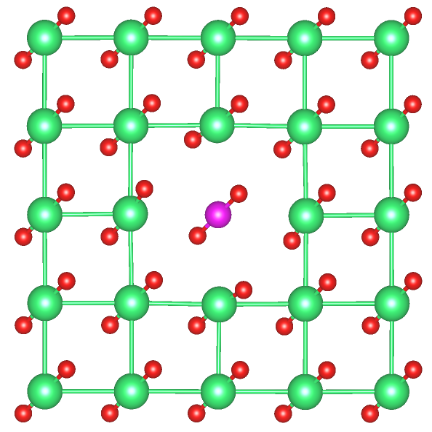}
            \caption[]%
            {{\small Krypton substitution (Kr$_{Th}$})}    
            \label{fig:Kr_struc}
        \end{subfigure}
        \vskip\baselineskip
        \begin{subfigure}[b]{0.49\textwidth}  
            \centering 
            \includegraphics[width=0.6\textwidth]{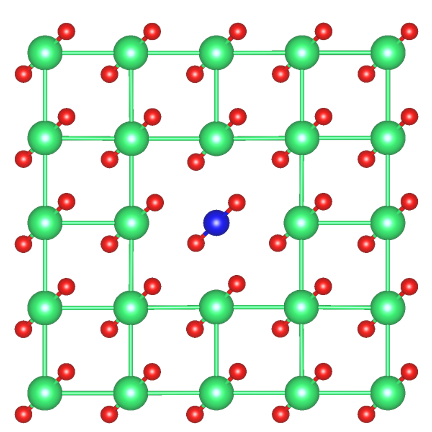}
            \caption[]%
            {{\small Zirconium substitution (Zr$_{Th}$})}    
            \label{fig:Zr_struc}
        \end{subfigure}
        \hfill
        \begin{subfigure}[b]{0.49\textwidth}   
            \centering 
            \includegraphics[width=0.6\textwidth]{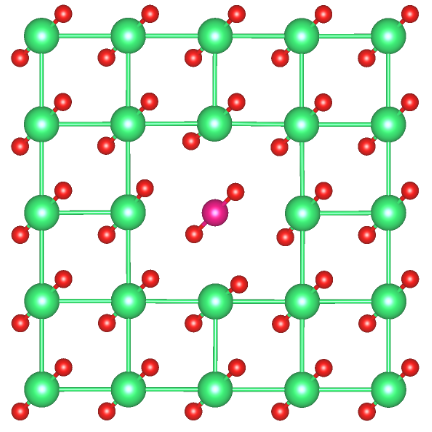}
            \caption[]%
            {{\small Iodine substitution (I$_{Th}$})}    
            \label{fig:I_struc}
        \end{subfigure}
        \vskip\baselineskip
        \begin{subfigure}[b]{0.49\textwidth}   
            \centering 
            \includegraphics[width=0.6\textwidth]{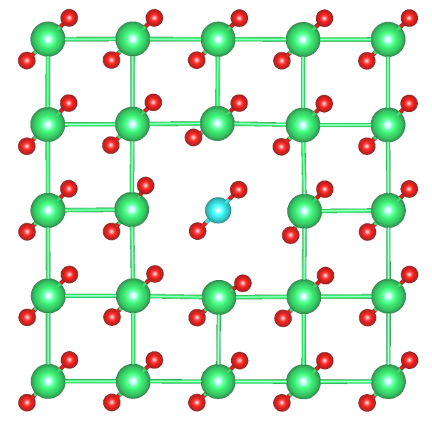}
            \caption[]%
            {{\small Xenon substitution (Xe$_{Th}$})}    
            \label{fig:Xe_struc}
        \end{subfigure}
        \hfill
        \begin{subfigure}[b]{0.49\textwidth}   
            \centering 
            \includegraphics[width=0.6\textwidth]{Fig15f_coordinates.png}
            \label{fig:Axis}
        \end{subfigure}
        \caption[]
        {\small Relaxed structures of different substitutional defects in ThO$_2$. Color codes are as follows; Th (green), O (red), He (black), Kr (magenta), Zr (blue), I (brown), and Xe (cyan).}
        \label{Fig16}
\end{figure}

Fig.~\ref{Fig16} Shows the relaxed ThO$_{2}$ structure near the substitutional defects along the 100 plane.

\newpage
Table ~\ref{Table2} shows the nearest neighbour distance changes caused due to the presence of vacancy and substitutional defects compared with the pristine. For substitution, the distances are calculated from the position of the substitutional atom after convergence. For the vacancy, the distance are calculated from the origin. Here, $d_{12}$, $d_{23}$ and $d_{34}$ are respectively, the distance between the first and second thorium atom, the distance between the second and third thorium atom and the distance between the third and fourth thorium atom.  $d1_{oxygen}$ is the distance between the defect and the nearest oxygen atom.
\begin{table*}[h]
\caption{The changes in nearest neighbour distance due to defects compared with pristine ThO$_{2}$.}
\begin{tabular}{| >{\centering}p{2.0 cm} | >{\centering}p{2.0 cm}|  >{\centering}p{2.0 cm} | >{\centering}p{2.0 cm} |>{\centering\arraybackslash}p{2.0 cm} |}

\hline 
Distance & $d_{12}$ & $d_{23}$ & $d_{34}$ & $d1_{oxygen}$\\
\hline
ThO$_{2}$ & 3.9096 & 3.9096 & 3.9096 & 2.3941\\
He$_{Th}$ & 3.8740 & 3.9404 & 3.9202 & 2.5838\\
Kr$_{Th}$ & 3.8720 & 3.9322 & 3.9322 & 2.1299\\
Xe$_{Th}$ & 3.8903 & 3.9202 & 3.9251 & 2.1842\\
I$_{Th}$  & 3.84013 & 3.9272 & 3.9317 & 2.1187\\
Zr$_{Th}$ & 3.87047 & 3.9413 & 3.9256 & 2.2832\\
V$_{Th}$ & 3.87256 & 3.9411 & 3.9207 & 2.5714\\
V$_{O}$  & 3.89489 & 3.9038 & 3.9038 & 2.3817\\
\hline
\end{tabular}
\label{Table2}
\end{table*}

\bibliography{final}